\documentclass[journal]{IEEEtran}
\usepackage{amssymb}
\usepackage{algorithm}
\usepackage{algorithmic}
\usepackage{amsfonts}
\usepackage{amsmath}
\usepackage{amssymb}
\usepackage{amsthm}
\usepackage{array}
\usepackage{bbding}
\usepackage{bigints}
\usepackage{booktabs}
\usepackage{cite}
\usepackage{cleveref}
\usepackage{color}
\usepackage{diagbox}
\usepackage{epsfig,latexsym}
\usepackage{epstopdf}
\usepackage{graphicx}
\usepackage{fancyhdr}
\usepackage{float}
\usepackage{flushend}
\usepackage{indentfirst}
\usepackage{lastpage}
\usepackage{makecell}
\usepackage{mathtools}
\usepackage{multirow}
\usepackage{pifont}
\usepackage{psfrag}
\usepackage{setspace}
\usepackage{stfloats}
\usepackage{subfloat}
\usepackage{times}
\usepackage{subfig}
\newcommand{\cmark}{\ding{52}}

\theoremstyle{remark}
\newtheorem{theorem}{\quad \textbf{Theorem}}

\allowdisplaybreaks[4]

\begin{document}
\title{Achievable Rate Analysis of the STAR-RIS Aided NOMA Uplink in the Face of Imperfect CSI and Hardware Impairments}

\author{Qingchao Li, \textit{Graduate Student Member, IEEE}, Mohammed El-Hajjar, \textit{Senior Member, IEEE}, Yanshi Sun, \textit{Member, IEEE}, Ibrahim Hemadeh, \textit{Member, IEEE}, Arman Shojaeifard, \textit{Senior Member, IEEE}, Yuanwei Liu, \textit{Senior Member, IEEE}, Lajos Hanzo, \textit{Life Fellow, IEEE}

\thanks{L. Hanzo would like to acknowledge the financial support of the Engineering and Physical Sciences Research Council projects EP/W016605/1 and EP/X01228X/1 as well as of the European Research Council¡¯s Advanced Fellow Grant QuantCom (Grant No. 789028) \textit{(Corresponding author: Lajos Hanzo.)}

Qingchao Li, Mohammed El-Hajjar and Lajos Hanzo are with the Electronics and Computer Science, University of Southampton, Southampton SO17 1BJ, U.K. (e-mail: Qingchao.Li@soton.ac.uk; meh@ecs.soton.ac.uk; lh@ecs.soton.ac.uk).

Yanshi Sun is with the School of Computer Science and Information Engineering, Hefei University of Technology, Hefei, 230009, China. (email:sys@hfut.edu.cn).

Ibrahim Hemadeh and Arman Shojaeifard are with InterDigital, London EC2A 3QR, U.K. (e-mail: Ibrahim.Hemadeh@InterDigital.com; Arman.Shojaeifard@InterDigital.com).

Yuanwei Liu is with the School of Electronic Engineering and Computer Science, Queen Mary University of London, London E1 4NS, U.K. (e-mail: Yuanwei.Liu@qmul.ac.uk).}}

\maketitle

\begin{abstract}
Reconfigurable intelligent surfaces (RIS) are capable of beneficially ameliorating the propagation environment by appropriately controlling the passive reflecting elements. To extend the coverage area, the concept of simultaneous transmitting and reflecting reconfigurable intelligent surfaces (STAR-RIS) has been proposed, yielding supporting 360$^\circ$ coverage user equipment (UE) located on both sides of the RIS. In this paper, we theoretically formulate the ergodic sum-rate of the STAR-RIS assisted non-orthogonal multiple access (NOMA) uplink in the face of channel estimation errors and hardware impairments (HWI). Specifically, the STAR-RIS phase shift is configured based on the statistical channel state information (CSI), followed by linear minimum mean square error (LMMSE) channel estimation of the equivalent channel spanning from the UEs to the access point (AP). Afterwards, successive interference cancellation (SIC) is employed at the AP using the estimated instantaneous CSI, and we derive the theoretical ergodic sum-rate upper bound for both perfect and imperfect SIC decoding algorithm. The theoretical analysis and the simulation results show that both the channel estimation and the ergodic sum-rate have performance floor at high transmit power region caused by transceiver hardware impairments.
\end{abstract}

\section{Introduction}
\IEEEPARstart{T}{he} fifth-generation (5G) systems is being rolled out across the globe and research is well under way on massive multiple-input-multiple-output (MIMO) techniques~\cite{lu2014overview,yang2016transmit}, millimeter wave (mmWave) communications~\cite{zhang2017codebook} and ultra-dense networking (UDN)~\cite{jafri2022robust}. However, driven by the ever-increasing thirst for increased data rate, the research turned to the exploration of radical next-generation concept~\cite{nguyen20216g}. Reconfigurable intelligent surfaces (RIS) are capable of enhancing the transmission reliability by harnessing passive elements for reconfiguring the phase shift of the impinging signals to construct smart radio environments~\cite{wu2019intelligent,li2022reconfigurable,li2022reconfigurable_iot,
li2023reconfigurable}. However, a conventional RIS can only support the users located at the same side of the RIS as the access point (AP) heating a 180$^\circ$ coverage. To deal with this issue, recently the simultaneous transmitting and reflecting RIS (STAR-RIS) concept was proposed for providing 360$^\circ$ coverage~\cite{zhao2022simultaneously,zhang2022secrecy}. In STAR-RIS architectures, three popular STAR-RIS protocols were proposed, including the energy splitting (ES), time switching (TS) and mode switching (MS) protocols~\cite{liu2021star}. Specifically, in the ES protocol, all the STAR-RIS elements are used for transmitting and reflecting signals simultaneously by splitting the power of the impinging waves. In the TS protocol, a pair of orthogonal time slots are employed, where all the STAR-RIS elements are switched to the transmission mode in one time slot and to the reflection mode in another time slot. By contrast, in the MS protocol, the STAR-RIS elements are divided into two parts, where the STAR-RIS elements are split into transmit and reflect mode.

Since non-orthogonal multiple access (NOMA) techniques are capable of approaching the boundary of capacity region, the beamforming design and performance analysis of STAR-RIS assisted NOMA systems was documented in~\cite{wu2021coverage,ma2022transmit,wu2022resource,
zuo2022joint,hou2021joint,aldababsa2022star,yue2022simultaneously,zhao2022ergodic,chen2022ergodic,
yang2022joint,zhang2022star,liu2022effective,wang2022outage,xie2022star}. Wu \textit{et al.}~\cite{wu2021coverage} maximized the coverage range of the STAR-RIS for both orthogonal multiple access (OMA) and NOMA systems, where the numerical results showed that the coverage range of the STAR-RIS assisted NOMA scheme is higher than that of its OMA counterpart as well as that of the conventional RIS assisted schemes. Ma \textit{et al.}~\cite{ma2022transmit} formulated the transmit power minimization problem of the STAR-RIS NOMA uplink, where the transmit power of each user and the passive beamforming weights of the STAR-RIS are jointly optimized by the popular semi-definite relaxation method. A resource allocation strategy is proposed in~\cite{wu2022resource} for multi-carrier STAR-RIS assisted NOMA systems by jointly optimizing the channel assignment, power allocation and passive beamforming at the STAR-RIS. In~\cite{zuo2022joint}, Zuo \textit{et al.} employed a two-layer iterative algorithm for maximizing the sum-rate by jointly optimizing the decoding order, power allocation, and the beamformer design at the AP and the STAR-RIS. Hou \textit{et al.}~\cite{hou2021joint} extended the STAR-RIS assisted NOMA scheme to coordinated multi-point transmission (CoMP), where both inter-cell interference cancellation and signals enhancement are attained by the passive beamforming weights of the STAR-RIS. Aldababsa \textit{et al.}~\cite{aldababsa2022star} derived the bit error rate (BER) expression of STAR-RIS NOMA networks relying on successive interference cancellation (SIC) by exploiting the central limit theorem (CLT). The ergodic rate of STAR-RIS assisted NOMA systems was theoretically analysed in~\cite{yue2022simultaneously,zhao2022ergodic,chen2022ergodic}. The closed-form expressions of the outage probability (OP) and diversity gain of STAR-RIS NOMA systems were derived in~\cite{yang2022joint,zhang2022star}, respectively. Additionally, Liu \textit{et al.}~\cite{liu2022effective} characterized the effective capacity of STAR-RIS assisted NOMA schemes in support of ultra-reliable low-latency communications. Wang \textit{et al.}~\cite{wang2022outage} employed the moment-matching method for deriving the outage probability of STAR-RIS assisted NOMA over spatially correlated channels. Furthermore, the moment-matching method was employed in~\cite{xie2022star} by Xie \textit{et al.} for analyzing the coverage probability of STAR-RIS aided multicell NOMA systems.

\begin{table*}[!t]
\footnotesize
\begin{center}
\caption{Novelty comparison with the literatures related to STAR-RIS.}\label{Table_literature}
\begin{tabular}{*{16}{l}}
\toprule
     & \makecell[c]{Our paper} & \cite{wu2021coverage} & \cite{ma2022transmit} & \cite{wu2022resource} & \cite{zuo2022joint} & \cite{hou2021joint} & \cite{aldababsa2022star} & \cite{yue2022simultaneously} & \cite{zhao2022ergodic} & \cite{chen2022ergodic} & \cite{yang2022joint} & \cite{zhang2022star} & \cite{liu2022effective} & \cite{wang2022outage} & \cite{xie2022star} \\
\midrule
\midrule
    STAR-RIS & \makecell[c]{\cmark} & \makecell[c]{\cmark} & \makecell[c]{\cmark} & \makecell[c]{\cmark} & \makecell[c]{\cmark} & \makecell[c]{\cmark} & \makecell[c]{\cmark} & \makecell[c]{\cmark} & \makecell[c]{\cmark} & \makecell[c]{\cmark} & \makecell[c]{\cmark} & \makecell[c]{\cmark} & \makecell[c]{\cmark} & \makecell[c]{\cmark} & \makecell[c]{\cmark}\\
\midrule
    NOMA & \makecell[c]{\cmark} & \makecell[c]{\cmark} & \makecell[c]{\cmark} & \makecell[c]{\cmark} & \makecell[c]{\cmark} & \makecell[c]{\cmark} & \makecell[c]{\cmark} & \makecell[c]{\cmark} & \makecell[c]{\cmark} & \makecell[c]{\cmark} & \makecell[c]{\cmark} & \makecell[c]{\cmark} & \makecell[c]{\cmark} & \makecell[c]{\cmark} & \makecell[c]{\cmark}\\
\midrule
    Downlink &  & \makecell[c]{\cmark} &  & \makecell[c]{\cmark} & \makecell[c]{\cmark} & \makecell[c]{\cmark} & \makecell[c]{\cmark} & \makecell[c]{\cmark} & \makecell[c]{\cmark} & \makecell[c]{\cmark} & \makecell[c]{\cmark} & \makecell[c]{\cmark} & \makecell[c]{\cmark} & \makecell[c]{\cmark} & \makecell[c]{\cmark}\\
\midrule
    Uplink & \makecell[c]{\cmark} &  & \makecell[c]{\cmark} &  &  &  &  &  & & & & & & & \\
\midrule
    Achievable rate & \makecell[c]{\cmark} & & & & \makecell[c]{\cmark} & \makecell[c]{\cmark} & & \makecell[c]{\cmark} & \makecell[c]{\cmark} & \makecell[c]{\cmark} & & & \makecell[c]{\cmark} & \\
\midrule
    Channel estimation & \makecell[c]{\cmark} & & &  &  &  & & & &  &  &  &  & \\
\midrule
    Imperfect CSI & \makecell[c]{\cmark} & & &  &  &  & & & &  &  &  &  & \\
\midrule
    Transceiver hardware impairments & \makecell[c]{\cmark} & & & & &  &  &  &  &  &  &  &  & \\
\midrule
    RIS phase noise & \makecell[c]{\cmark} & & & & &  &  &  &  &  &  &  &  & \\
\bottomrule
\end{tabular}
\end{center}
\end{table*}

The above treatises on the STAR-RIS aided NOMA scheme have the following limitations. Firstly, the STAR-RIS aided NOMA scheme is operated based on the assumption of perfect channel state information (CSI), ignoring the channel estimation errors. Secondly, perfect transceiver hardware is assumed, i.e. the practical signal impairments of the transceivers are ignored. However, the effect of realistic channel estimation errors and hardware impairments (HWI) may impose a performance floor at high transmit powers, both in terms of the achievable rate, outage probability and bit error ratio (BER)~\cite{yang2021performance,peng2021ris,vu2022performance,li2023performance,wang2023ris}. This means that the performance improvement of RIS-aided wireless communication systems is limited by the CSI imperfection and HWI, which cannot be compensated upon increasing the transmit power. Furthermore, the above treatises on the STAR-RIS aided NOMA scheme assume that the RIS phase shifts can be perfectly configured. However, the RIS phase noise is inevitable, which imposes a potentially significant performance erosion on RIS-aided systems~\cite{badiu2019communication,qian2020beamforming,zhi2020uplink,
papazafeiropoulos2021intelligent,dai2021statistical}.

The above-mentioned idealized simplifying assumptions are unrealistic in practical STAR-RIS aided systems. Hence, in order to deal with the above aspects, our contributions in this paper are as follows:

\begin{itemize}
    \item We conceive a two-timescale beamforming design to reduce the CSI acquisition overhead of the STAR-RIS aided systems. Specifically, the passive beamforming pattern of the STAR-RIS is designed based on the statistical CSI and then based on the resultant STAR-RIS beamformer, the equivalent channels spanning from the user equipment (UE) to the AP are estimated in each channel coherence interval using the linear minimum mean square error (LMMSE) technique. Then, the information is transmitted using the estimated instantaneous UE-AP CSI.
    \item We theoretically derive the ergodic sum-rate of the STAR-RIS assisted NOMA uplink relying on both perfect and imperfect SIC decoding algorithms, in the face of realistic RIS phase noise, imperfect CSI and transceiver HWIs. More specifically, we exploit the statistical knowledge of the RIS phase noise distribution, of the channel estimation error variance, and of the transceiver signal distortion distribution in our theoretical derivations.
    \item The theoretical analysis and the simulation results show that the ergodic sum-rate of the STAR-RIS assisted NOMA uplink is independent of the decoding order when perfect SIC is assumed. By contrast, in the imperfect SIC decoding case, the user having lower channel gain should have decoding priority to maximize the ergodic sum-rate.
    \item The theoretical analysis and the simulation results show that the achievable sum-rate performance degradation of the STAR-RIS assisted NOMA uplink caused by the RIS phase noise can be compensated upon increasing the transmit power or deploying more RIS elements. We also show that the CSI accuracy can be improved by increasing the pilot sequence length. However, the transceiver HWI can cause an error floor at high SNRs. Furthermore, the transceiver HWI has a non-negligible effect on the rate-fairness in the STAR-RIS aided NOMA uplink.
\end{itemize}

Finally, Table \ref{Table_literature} explicitly contrasts our contributions to the literature.

The rest of this paper is organized as follows. In Section \ref{System_Model}, we present the system model, while the associated channel estimation is described in Section \ref{Channel_Estimation}. Section \ref{Performance_Analysis_and_Beamforming_Design} presents the performance analysis and beamforming design of the STAR-RIS assisted NOMA uplink. Our simulation results are presented in Section \ref{Numerical_and_Simulation_Results}, while we conclude in Section \ref{Conclusion}.

\textit{Notations:} $\jmath=\sqrt{-1}$, while $(\cdot)^{*}$ and $(\cdot)^{\text{H}}$ represent the operations of conjugation and Hermitian transpose, respectively, $|a|$ represents the amplitude of the complex scalar $a$, $\mathbb{C}^{m\times n}$ denotes the space of $m\times n$ complex-valued matrices, $\mathbf{I}_{N}$ represents the $N\times N$ identity matrix, $\text{diag}\{\mathbf{a}\}$ denotes a diagonal matrix with the diagonal elements being the elements of $\mathbf{a}$ in order, $\mathcal{CN}(\boldsymbol{\mu},\mathbf{\Sigma})$ is a circularly symmetric complex Gaussian random vector with the mean $\boldsymbol{\mu}$ and the covariance matrix $\mathbf{\Sigma}$, $\mathbb{E}[x]$ represents the mean of the random variable $x$, the covariance between the random variables $x$ and $y$ is denoted by $\mathrm{C}_{xy}$.

\section{System Model}\label{System_Model}
\begin{figure}[!t]
    \centering
    \includegraphics[width=3.4in]{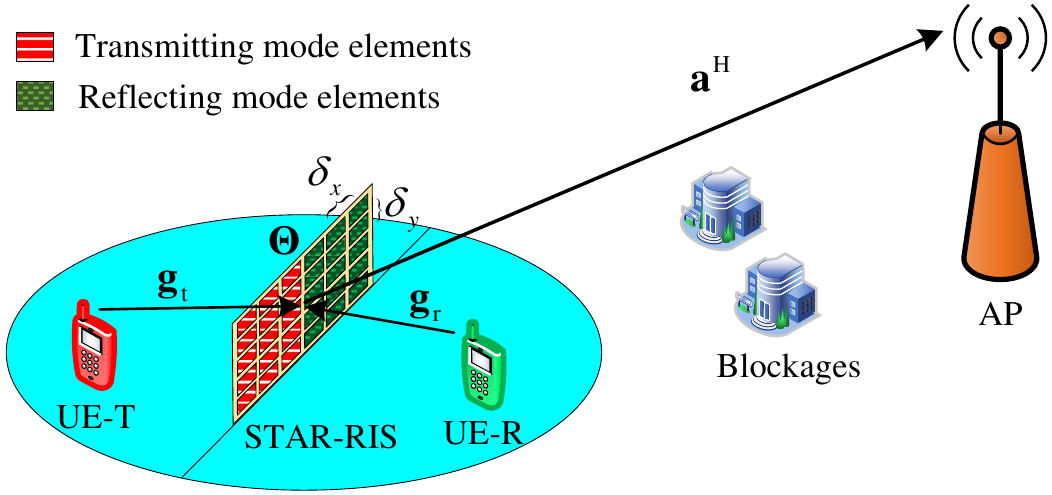}
    \caption{System model of the STAR-RIS aided uplink NOMA wireless communication system.}\label{Fig_system_model}
\end{figure}

The STAR-RIS-aided wireless communication system model of~\cite{yue2022simultaneously,yang2022joint} is shown in Fig. \ref{Fig_system_model}, including a single-antenna AP\footnote{To unveil the theoretically achievable rate limit of the STAR-RIS aided NOMA uplink, we assume that the AP is equipped with a single-antenna. Note that the alternating optimization (AO) algorithm is also applicable to the case of multiple antennas at the AP~\cite{wu2020joint}.}, a STAR-RIS and a pair of single-antenna UEs distributed at the different sides of the STAR-RIS\footnote{To support multiple UEs at each sides of the STAR-RIS, we can employ the general hybrid time division multiple access (TDMA)-NOMA technique associated with device grouping~\cite{chen2023active}. Specifically, the UEs are divided into multiple groups, where the signals of the UEs in the same group are transmitted simultaneously via NOMA, while that in different groups occupy orthogonal time slots.}. We denote the UE at the different side of the AP as UE-T and the UE at the same side of the AP as UE-R. Since the direct links between the UEs and the AP are blocked, the STAR-RIS creates additional links to support information transfer. We focus our attention on the power-domain NOMA uplink architecture, where the transmit power at the UE-T and UE-R are $\rho_\mathrm{t}$ and $\rho_\mathrm{r}$, respectively. Specifically, the popular SIC algorithm is employed for information recovery in NOMA~\cite{dai2018survey}, where the AP first detects the signal of one user, followed by remodulating it and subtracting the interference imposed by it on the composite NOMA signal. As a result, this operation leaves behind the uncontaminated signal for another user.

\subsection{STAR-RIS Architecture}\label{System_Model_RIS_Architecture}
In our work, we focus on the MS STAR-RIS architecture. Recall that for the ES protocol power splitting circuits are required and having energy leakage is inevitable, the TS protocol requires mode switching circuits for each elements. Hence, the MS protocol is deemed simplest implementation~\cite{liu2021star}. We assume that a total of $N_\mathrm{t}+N_\mathrm{r}$ STAR-RIS elements are used in a uniform rectangular planar array (URPA) containing $N_\mathrm{t}=N_{\mathrm{t},x}\times N_{\mathrm{t},y}$ elements operating in the transmit mode to support the UE-T, and a URPA containing $N_\mathrm{r}=N_{\mathrm{r},x}\times N_{\mathrm{r},y}$ element in the reflecting mode to support the UE-R. These two types of RIS elements are placed block-wise, as shown in Fig. \ref{Fig_system_model}. Furthermore, the distances between the adjacent STAR-RIS elements in the horizontal and vertical direction are represented by $\delta_x$ and $\delta_y$, respectively. We denote the phase shift of the $n_\mathrm{t}$th element in the transmit mode as $\theta^{\mathrm{t}}_{n_\mathrm{t}}$ and that of the $n_\mathrm{r}$th element in the reflect mode as $\theta^{\mathrm{r}}_{n_\mathrm{r}}$, where we have $0\leq\theta^{\mathrm{t}}_{n_\mathrm{t}}<2\pi$ for $n_\mathrm{t}=1,2,\cdots,N_\mathrm{t}$, and $0\leq\theta^{\mathrm{r}}_{n_\mathrm{r}}<2\pi$ for $n_\mathrm{r}=1,2,\cdots,N_\mathrm{r}$. Therefore, the phase shift matrix of the STAR-RIS can be characterized as $\mathbf{\Theta}=\text{diag}\{\mathbf{\Theta}^{\mathrm{t}},\mathbf{\Theta}^{\mathrm{r}}\}$, with $\mathbf{\Theta}^{\mathrm{t}}=\text{diag}\{\text{e}^{\jmath\theta^{\mathrm{t}}_1},
\text{e}^{\jmath\theta^{\mathrm{t}}_2},\cdots,
\text{e}^{\jmath\theta^{\mathrm{t}}_{N_\mathrm{t}}}\}$ and $\mathbf{\Theta}^{\mathrm{r}}=\text{diag}\{\text{e}^{\jmath\theta^{\mathrm{r}}_1},
\text{e}^{\jmath\theta^{\mathrm{r}}_2},\cdots,
\text{e}^{\jmath\theta^{\mathrm{r}}_{N_\mathrm{r}}}\}$.

In~\cite{wu2021coverage,ma2022transmit,wu2022resource,
zuo2022joint,hou2021joint,aldababsa2022star,yue2022simultaneously,zhao2022ergodic,chen2022ergodic,
yang2022joint,zhang2022star,liu2022effective,wang2022outage,xie2022star} it was assumed that the phase shift can be perfectly configured without phase noise. However, due to the realistic RIS hardware impairments, the phase shift of each reflecting element is practically modelled as~\cite{badiu2019communication}
\begin{align}\label{RIS_Architecture_3}
    \theta^{\mathrm{i}}_{n_\mathrm{i}}=\bar{\theta}^{\mathrm{i}}_{n_\mathrm{i}}+
    \tilde{\theta}^{\mathrm{i}}_{n_\mathrm{i}},\ \mathrm{i}\in\{\mathrm{t},\mathrm{r}\},
\end{align}
where $\bar{\theta}^{\mathrm{i}}_{n_\mathrm{i}}$ represents the expected phase shift, while $\tilde{\theta}^{\mathrm{i}}_{n_\mathrm{i}}$ is the phase noise of the $n_\mathrm{i}$-th element. The phase noise $\tilde{\theta}^{\mathrm{i}}_{n_\mathrm{i}}$ is modelled by identically and independently distributed (i.i.d.) random variables having a mean of 0, and following either the von Mises distribution or the uniform distribution~\cite{badiu2019communication}. These may be represented as
$\tilde{\theta}^{\mathrm{i}}_{n_\mathrm{i}}\sim\mathcal{VM}(0,\varsigma_\text{p})$ and $\tilde{\theta}^{\mathrm{i}}_{n_\mathrm{i}}\sim\mathcal{UF}(-\iota_\text{p},\iota_\text{p})$, where $\varsigma_\mathrm{p}$ is the concentration parameter of the von Mises distributed variables and $(-\iota_\mathrm{p},\iota_\mathrm{p})$ is the support interval of the uniformly distributed variables. We denote the RIS phase noise power as $\sigma_\mathrm{p}^2$. Hence, $\varsigma_\text{p}=\frac{1}{\sigma_\mathrm{p}^2}$ and $\iota_\mathrm{p}=\sqrt{3\sigma_\mathrm{p}^2}$.

\subsection{Channel Model}
As shown in Fig. \ref{Fig_system_model}, we denote the links spanning from the UE-T to the STAR-RIS transmitting elements as $\mathbf{g}_\mathrm{t}\in\mathbb{C}^{N_\mathrm{t}\times1}$, and the channel from the UE-R to the STAR-RIS reflecting elements as $\mathbf{g}_\mathrm{r}\in\mathbb{C}^{N_\mathrm{r}\times1}$. The channel impinging from the STAR-RIS upon the AP is denoted as $\mathbf{a}^{\mathrm{H}}=[\mathbf{a}_\mathrm{t}^{\mathrm{H}},\mathbf{a}_\mathrm{r}^{\mathrm{H}}]$, where $\mathbf{a}_\mathrm{t}^{\mathrm{H}}\in\mathbb{C}^{1\times{N}_\mathrm{r}}$ and $\mathbf{a}_\mathrm{r}^{\mathrm{H}}\in\mathbb{C}^{1\times{N}_\mathrm{t}}$ are the links spanning from the transmitting elements to the AP and those from the reflecting elements to the AP, respectively. Then, $\mathbf{g}_\mathrm{t}$, $\mathbf{g}_\mathrm{r}$ $\mathbf{a}_\mathrm{t}$ and $\mathbf{a}_\mathrm{r}$ are assumed to obey Rician fading, given by~\cite{hou2021joint,yue2022simultaneously}
\begin{align}\label{Channel_Model_1_1}
    \mathbf{g}_{\mathrm{i}}\sim\mathcal{CN}\Big(
    \frac{\sqrt{\kappa_{\mathrm{i}}}\bar{\mathbf{g}}_\mathrm{i}}{\sqrt{1+\kappa_{\mathrm{i}}}},
    \frac{\mathbf{I}_{N_\mathrm{i}}}{1+\kappa_{\mathrm{i}}}\Big),
\end{align}
\begin{align}\label{Channel_Model_2_1}
    \mathbf{a}_{\mathrm{i}}\sim\mathcal{CN}\Big(
    \frac{\sqrt{\kappa_a}\bar{\mathbf{a}}_\mathrm{i}}{\sqrt{1+\kappa_a}},
    \frac{\mathbf{I}_{N_\mathrm{i}}}{1+\kappa_a}\Big),
\end{align}
where $\mathrm{i}\in\{\mathrm{t},\mathrm{r}\}$, $\kappa_{\mathrm{t}}$, $\kappa_{\mathrm{r}}$ and $\kappa_a$ denote the Rician factors of the corresponding link, $\bar{\mathbf{g}}_\mathrm{t}$, $\bar{\mathbf{g}}_\mathrm{r}$, $\bar{\mathbf{a}}_\mathrm{t}$ and $\bar{\mathbf{a}}_\mathrm{r}$ represent the LoS component vectors, given by
\begin{align}\label{Channel_Model_3_1}
    \notag\bar{\mathbf{g}}_\mathrm{i}=&[1,\cdots,\mathrm{e}^{-\jmath\frac{2\pi}{\lambda}
    (\delta_{x}n_{\mathrm{i},x}\sin\phi_\mathrm{i}\cos\varphi_\mathrm{i}
    +\delta_{y}n_{\mathrm{i},y}\cos\phi_\mathrm{i})},\cdots,\\
    &\mathrm{e}^{-\jmath\frac{2\pi}{\lambda}(\delta_{x}(N_{\mathrm{i},x}-1)
    \sin\phi_\mathrm{i}\cos\varphi_\mathrm{i}
    +\delta_{y}(N_{\mathrm{i},y}-1)\cos\phi_\mathrm{i})}]^{\mathrm{H}},
\end{align}
\begin{align}\label{Channel_Model_4_1}
    \notag\bar{\mathbf{a}}_\mathrm{i}=&[1,\cdots,\mathrm{e}^{-\jmath\frac{2\pi}{\lambda}
    (\delta_{x}n_{\mathrm{i},x}\sin\omega_\mathrm{i}\cos\varpi_\mathrm{i}
    +\delta_{y}n_{\mathrm{i},y}\cos\omega_\mathrm{i})},\cdots,\\
    &\mathrm{e}^{-\jmath\frac{2\pi}{\lambda}(\delta_{x}(N_{\mathrm{i},x}-1)
    \sin\omega_\mathrm{i}\cos\varpi_\mathrm{i}
    +\delta_{y}(N_{\mathrm{i},y}-1)\cos\omega_\mathrm{i})}]^{\mathrm{H}},
\end{align}
where $\mathrm{i}\in\{\mathrm{t},\mathrm{r}\}$, $\lambda$ is the wavelength~\cite{hou2021joint,yue2022simultaneously}. $\phi_\mathrm{t}$ and $\varphi_\mathrm{t}$ are the elevation and azimuth angle of arrival (AoA) to the STAR-RIS transmitting elements, respectively, while $\phi_\mathrm{r}$ and $\varphi_\mathrm{r}$ are the elevation and azimuth AoA to the STAR-RIS reflecting elements, respectively. Furthermore, $\omega_\mathrm{t}$ and $\varpi_\mathrm{t}$ are the elevation and azimuth angle of departure (AoD) from the STAR-RIS transmitting elements, respectively, while $\omega_\mathrm{r}$ and $\varpi_\mathrm{r}$ are the elevation and azimuth AoD from the reflecting STAR-RIS elements, respectively.

We denote the path loss of the signals from the UE-T to the STAR-RIS and that from the UE-R to the STAR-RIS by $\varrho_{\mathrm{t}}$ and $\varrho_{\mathrm{r}}$, respectively. While $\varrho_{a}$ denotes the path loss of the signals from the STAR-RIS to the AP. The distance-dependent path loss model is employed~\cite{li2022reconfigurable}, which is given by $\varrho_{\mathrm{t}}=\varrho_0d_{\text{t}}^{-\alpha_{\text{t}}}$, $\varrho_{\mathrm{r}}=\varrho_0d_{\text{r}}^{-\alpha_{\text{r}}}$ and
$\varrho_a=\varrho_0d_a^{-\alpha_a}$, where $d_\mathrm{t}$, $d_\mathrm{r}$ and $d_a$ represent the length of the corresponding links, $\alpha_\mathrm{t}$, $\alpha_\mathrm{r}$ and $\alpha_a$ represent the path loss exponent of the corresponding links, and $\varrho_0$ is the path loss at the reference distance of 1 meter. The equivalent channel spanning from the UE-T to the AP is the function of $\mathbf{\Theta}^{\mathrm{t}}$, given by $h_\mathrm{t}(\mathbf{\Theta}^{\mathrm{t}})
=\sqrt{\varrho_\mathrm{t}\varrho_a}\mathbf{a}_{\mathrm{t}}^{\text{H}}
\mathbf{\Theta}^{\mathrm{t}}\mathbf{g}_{\text{t}}$. Similarly, the equivalent channel departing  from the UE-R to the AP is the function of $\mathbf{\Theta}^{\mathrm{r}}$, given by $h_\mathrm{r}(\mathbf{\Theta}^{\mathrm{r}})
=\sqrt{\varrho_\mathrm{r}\varrho_a}\mathbf{a}_{\mathrm{r}}^{\text{H}}
\mathbf{\Theta}^{\mathrm{r}}\mathbf{g}_{\text{r}}$.

\section{Channel Estimation}\label{Channel_Estimation}
\begin{figure}[!t]
    \centering
    \includegraphics[width=3.5in]{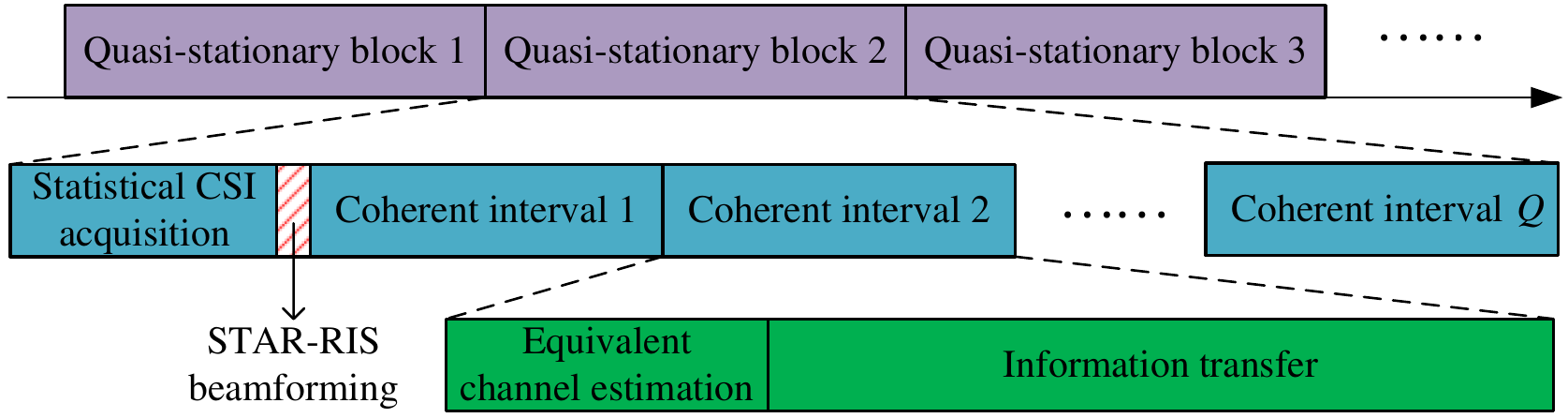}
    \caption{Illustration of the two-timescale channel estimation and information transfer protocol.}\label{Fig_two_timescale}
\end{figure}

Channel estimation is essential for RIS-aided wireless communication systems. To mitigate the CSI acquisition overhead, we employ the two-timescale beamforming protocol of~\cite{pan2022overview}. Specifically, the passive beamforming at the STAR-RIS is designed based on the statistical CSI, i.e. $\bar{\mathbf{g}}_\mathrm{t}$, $\bar{\mathbf{g}}_\mathrm{r}$, $\bar{\mathbf{a}}_\mathrm{t}$ and $\bar{\mathbf{a}}_\mathrm{r}$, which remain constant for numerous coherent intervals. Based on the passive beamforming at the RIS, the equivalent channels, i.e. $h_\mathrm{t}(\mathbf{\Theta}^{\mathrm{t}})$ and $h_\mathrm{r}(\mathbf{\Theta}^{\mathrm{r}})$, can be estimated for each coherent time. Based on the estimated equivalent channels, the SIC detection may be used at the AP. The two-timescale channel estimation and information transfer protocol conceived is illustrated in Fig.~\ref{Fig_two_timescale}. In each quasi-stationary block, the statistical CSI is fixed, which is estimated at the beginning of the block. The rest of the quasi-stationary block includes $Q$ coherence intervals. Since the statistical CSI changes slowly, the pilot overhead of the statistical CSI acquisition remains modest~\cite{zhi2022power}. Each coherence interval is composed of several symbol slots, which are assumed to have the same instantaneous CSI. The STAR-RIS beamforming is designed based on the statistical CSI. Thus, the STAR-RIS phase shift is fixed within each quasi-stationary block as seen in Fig.~\ref{Fig_two_timescale}, and the equivalent instantaneous UE-AP channel is estimated at the beginning of each coherence interval. Then, the information delivery and recovery are performed based on the corresponding instantaneous UE-AP channel.

To realize the LMMSE estimator, both the mean and the second moment of the equivalent channels are required.

\begin{theorem}\label{theorem_1}
Based on a specific STAR-RIS phase shift, the mean and the second moment of the equivalent channels $h_\mathrm{t}(\mathbf{\Theta}^{\mathrm{t}})$ and $h_\mathrm{r}(\mathbf{\Theta}^{\mathrm{r}})$ are given by
\begin{align}\label{Channel_Estimation_1_1}
    \mathbb{E}[h_\mathrm{t}(\mathbf{\Theta}^{\mathrm{t}})]
    =\sqrt{\frac{\varrho_\mathrm{t}\varrho_a\kappa_{\mathrm{t}}\kappa_a}
    {(1+\kappa_{\mathrm{t}})(1+\kappa_a)}}\xi
    \sum_{n=1}^{N_\mathrm{t}}\bar{a}_{\mathrm{t},n}^*\bar{g}_{\mathrm{t},n}\mathrm{e}^
    {\jmath\bar{\theta}_{n}^{\mathrm{t}}},
\end{align}
\begin{align}\label{Channel_Estimation_1_2}
    \mathbb{E}[h_\mathrm{r}(\mathbf{\Theta}^{\mathrm{r}})]
    =\sqrt{\frac{\varrho_\mathrm{r}\varrho_a\kappa_{\mathrm{r}}\kappa_a}
    {(1+\kappa_{\mathrm{r}})(1+\kappa_a)}}\xi
    \sum_{n=1}^{N_\mathrm{r}}\bar{a}_{\mathrm{r},n}^*\bar{g}_{\mathrm{r},n}\mathrm{e}^
    {\jmath\bar{\theta}_{n}^{\mathrm{r}}},
\end{align}
\begin{align}\label{Channel_Estimation_1_2_1}
    \notag\mathbb{E}[|h_\mathrm{t}(\mathbf{\Theta}^{\mathrm{t}})|^2]
    =&\frac{\varrho_\mathrm{t}\varrho_a}{(1+\kappa_{\mathrm{t}})(1+\kappa_a)}
    \Big(\kappa_{\mathrm{t}}\kappa_a\xi^2\Big|\sum_{n=1}^{N_\mathrm{t}}
    \bar{a}_{\mathrm{t},n}^*\bar{g}_{\mathrm{t},n}\mathrm{e}^
    {\jmath\bar{\theta}_{n}^{\mathrm{t}}}\Big|^2\\
    &+(\kappa_{\mathrm{t}}+\kappa_a+1)N_\mathrm{t}\Big),
\end{align}
\begin{align}\label{Channel_Estimation_1_2_2}
    \notag\mathbb{E}[|h_\mathrm{r}(\mathbf{\Theta}^{\mathrm{r}})|^2]
    =&\frac{\varrho_\mathrm{r}\varrho_a}{(1+\kappa_{\mathrm{r}})(1+\kappa_a)}
    \Big(\kappa_{\mathrm{r}}\kappa_a\xi^2\Big|\sum_{n=1}^{N_\mathrm{r}}
    \bar{a}_{\mathrm{r},n}^*\bar{g}_{\mathrm{r},n}\mathrm{e}^
    {\jmath\bar{\theta}_{n}^{\mathrm{r}}}\Big|^2\\
    &+(\kappa_{\mathrm{r}}+\kappa_a+1)N_\mathrm{r}\Big),
\end{align}
where $\xi=\frac{I_{1}(\varsigma_\text{p})}{I_{0}(\varsigma_\text{p})}$ when $\tilde{\theta}^{\mathrm{i}}_{n_\mathrm{i}}\sim\mathcal{VM}(0,\varsigma_\text{p})$ or $\xi=\frac{\sin(\iota_\mathrm{p})}{\iota_\mathrm{p}}$ when $\tilde{\theta}^{\mathrm{i}}_{n_\mathrm{i}}\sim\mathcal{UF}(-\iota_\mathrm{p},\iota_\mathrm{p})$, with $I_m(\cdot)$ representing the modified Bessel functions of the first kind of order $m$. Furthermore, the variance of the equivalent channels $h_\mathrm{t}(\mathbf{\Theta}^{\mathrm{t}})$ and $h_\mathrm{r}(\mathbf{\Theta}^{\mathrm{r}})$ are
\begin{align}\label{Channel_Estimation_2_1}
    \mathrm{C}_{h_\mathrm{t}(\mathbf{\Theta}^{\mathrm{t}})h_\mathrm{t}
    (\mathbf{\Theta}^{\mathrm{t}})}
    =\frac{\varrho_\mathrm{t}\varrho_a(\kappa_{\mathrm{t}}+\kappa_a+1)N_\mathrm{t}}
    {(1+\kappa_{\mathrm{t}})(1+\kappa_a)},
\end{align}
\begin{align}\label{Channel_Estimation_2_2}
    \mathrm{C}_{h_\mathrm{r}(\mathbf{\Theta}^{\mathrm{r}})h_\mathrm{r}
    (\mathbf{\Theta}^{\mathrm{r}})}
    =\frac{\varrho_\mathrm{r}\varrho_a(\kappa_{\mathrm{r}}+\kappa_a+1)N_\mathrm{r}}
    {(1+\kappa_{\mathrm{r}})(1+\kappa_a)}.
\end{align}
\end{theorem}
\begin{IEEEproof}
    See Appendix \ref{Appendix_A}.
\end{IEEEproof}

To eliminate the pilot contamination between the UE-T and UE-R, a pair of orthogonal pilot sequences of length $K\geq2$ are deployed, denoted as $[\tau_{\mathrm{t}}^{(1)};\tau_{\mathrm{t}}^{(2)};\cdots;\tau_{\mathrm{t}}^{(K)}]$
and $[\tau_{\mathrm{r}}^{(1)};\tau_{\mathrm{r}}^{(2)};\cdots;\tau_{\mathrm{r}}^{(K)}]$. In the $k$th ($k=1,2,\cdots,K$) symbol slot, the pilot $\tau_{\mathrm{t}}^{(k)}$ at the UE-T and the pilot $\tau_{\mathrm{r}}^{(k)}$ at the UE-R are transmitted simultaneously, leading to the signal received at the AP formulated as
\begin{align}\label{Channel_Estimation_3}
    \notag&x^{(k)}(\mathbf{\Theta})\\
    \notag=&\underbrace{\sqrt{P_\mathrm{t}\varepsilon_v\varepsilon_{u_\mathrm{t}}}
    h_\mathrm{t}(\mathbf{\Theta}^{\mathrm{t}})\tau_{\mathrm{t}}^{(k)}}_{\text{Desired UE-T pilot}}+\underbrace{\sqrt{P_\mathrm{r}\varepsilon_v\varepsilon_{u_\mathrm{r}}}
    h_\mathrm{r}(\mathbf{\Theta}^{\mathrm{r}})
    \tau_{\mathrm{r}}^{(k)}}_{\text{Desired UE-R pilot}}\\
    \notag&+\underbrace{\sqrt{P_\mathrm{t}(1-\varepsilon_v)}
    h_\mathrm{t}(\mathbf{\Theta}^{\mathrm{t}})
    v_\mathrm{t}^{(k)}+\sqrt{P_\mathrm{r}(1-\varepsilon_v)}
    h_\mathrm{r}(\mathbf{\Theta}^{\mathrm{r}})
    v_\mathrm{r}^{(k)}}_{\text{HWI distortion at AP}}\\
    \notag&+\underbrace{\sqrt{P_\mathrm{t}\varepsilon_v(1-\varepsilon_{u_\mathrm{t}})}
    h_\mathrm{t}(\mathbf{\Theta}^{\mathrm{t}})u_\mathrm{t}^{(k)}}_{\text{HWI distortion at UE-T}}\\
    &+\underbrace{\sqrt{P_\mathrm{r}\varepsilon_v(1-\varepsilon_{u_\mathrm{r}})}
    h_\mathrm{r}(\mathbf{\Theta}^{\mathrm{r}})u_\mathrm{r}^{(k)}}_{\text{HWI distortion at UE-R}}+\underbrace{w^{(k)}}_{\text{Noise}},
\end{align}
where $P_\mathrm{t}$ and $P_\mathrm{r}$ are the power of pilot sequences at UE-T and UE-R respectively, $\varepsilon_{u_\mathrm{t}}$, $\varepsilon_{u_\mathrm{r}}$ and $\varepsilon_v$ represent the hardware quality factor of the UE-T, UE-R and AP, respectively, satisfying $0\leq\varepsilon_v\leq1$, $0\leq\varepsilon_{u_\mathrm{t}}\leq1$ and $0\leq\varepsilon_{u_\mathrm{r}}\leq1$~\cite{bjornson2017massive}. Explicitly, a hardware quality factor of 1 indicates that the hardware is ideal, while 0 means the hardware is completely inadequate, where $u_\mathrm{t}^{(k)}\sim\mathcal{CN}(0,1)$, $u_\mathrm{r}^{(k)}\sim\mathcal{CN}(0,1)$, $v_\mathrm{t}^{(k)}\sim\mathcal{CN}(0,1)$, and $v_\mathrm{r}^{(k)}\sim\mathcal{CN}(0,1)$ are the HWIs of the $k$th pilot. Furthermore, $w^{(k)}\sim\mathcal{CN}(0,\sigma_w^2)$ is the additive noise at the $k$th symbol interval with $\sigma_w^2$ being the power spectral density.

Firstly, to estimate $h_\mathrm{t}(\mathbf{\Theta}^{\mathrm{t}})$, the conjugate transpose of the UE-T pilot sequence, i.e. $[\tau_{\mathrm{t}}^{(1)*},\tau_{\mathrm{t}}^{(2)*},\cdots,\tau_{\mathrm{t}}^{(K)*}]$, is employed to combine the AP observations $x^{(1)}(\mathbf{\Theta}),x^{(2)}(\mathbf{\Theta}),
\cdots,x^{(K)}(\mathbf{\Theta})$. Then we arrive at the processed received observation formulated as
\begin{align}\label{Channel_Estimation_4}
    \notag{x}_\mathrm{t}(\mathbf{\Theta})
    =&\frac{1}{\sqrt{K}}\sum_{k=1}^{K}x^{(k)}(\mathbf{\Theta})
    \tau_{\mathrm{t}}^{(k)*}\\
    \notag=&\sqrt{KP_\mathrm{t}\varepsilon_v\varepsilon_{u_\mathrm{t}}}h_\mathrm{t}
    (\mathbf{\Theta}^{\mathrm{t}})
    +\sqrt{P_\mathrm{t}\varepsilon_v(1-\varepsilon_{u_\mathrm{t}})}h_\mathrm{t}
    (\mathbf{\Theta}^{\mathrm{t}})u'_\mathrm{t}\\
    \notag&+\sqrt{P_\mathrm{r}\varepsilon_v(1-\varepsilon_{u_\mathrm{r}})}h_\mathrm{r}
    (\mathbf{\Theta}^{\mathrm{r}})u'_\mathrm{r}\\
    \notag&+\sqrt{P_\mathrm{t}(1-\varepsilon_v)}
    h_\mathrm{t}(\mathbf{\Theta}^{\mathrm{t}})v'_\mathrm{t}
    +\sqrt{P_\mathrm{r}(1-\varepsilon_v)}h_\mathrm{r}
    (\mathbf{\Theta}^{\mathrm{r}})v'_\mathrm{r}\\
    &+w',
\end{align}
where $u'_\mathrm{t}=\frac{1}{\sqrt{K}}\sum_{k=1}^{K}u_\mathrm{t}^{(k)}\tau_{\mathrm{t}}^{(k)*}$, $u'_\mathrm{r}=\frac{1}{\sqrt{K}}\sum_{k=1}^{K}u_\mathrm{r}^{(k)}\tau_{\mathrm{t}}^{(k)*}$, $v'_\mathrm{t}=\frac{1}{\sqrt{K}}\sum_{k=1}^{K}v_\mathrm{t}^{(k)}\tau_{\mathrm{t}}^{(k)*}$, $v'_\mathrm{r}=\frac{1}{\sqrt{K}}\sum_{k=1}^{K}v_\mathrm{r}^{(k)}\tau_{\mathrm{t}}^{(k)*}$ and $w'=\frac{1}{\sqrt{K}}\sum_{k=1}^{K}w^{(k)}\tau_{\mathrm{t}}^{(k)*}$. Due to the independence of $u_\mathrm{t}^{(k)}$, $u_\mathrm{r}^{(k)}$, $v_\mathrm{t}^{(k)}$, $v_\mathrm{r}^{(k)}$ and $w^{(k)}$ for $k=1,2,\cdots,K$, it satisfies $u'_\mathrm{t}\sim\mathcal{CN}(0,1)$, $u'_\mathrm{r}\sim\mathcal{CN}(0,1)$, $v'_\mathrm{t}\sim\mathcal{CN}(0,1)$, $v'_\mathrm{r}\sim\mathcal{CN}(0,1)$ and $w'\sim\mathcal{CN}(0,\sigma_w^2)$.

Based on (\ref{Channel_Estimation_4}), we can express the mean of $x_\mathrm{t}(\mathbf{\Theta})$, the covariance of $h_\mathrm{t}(\mathbf{\Theta}^{\mathrm{t}})$ and $x_\mathrm{t}(\mathbf{\Theta})$, and the variance of $x_\mathrm{t}(\mathbf{\Theta})$ as
\begin{align}\label{Channel_Estimation_5}
    \notag&\mathbb{E}[x_\mathrm{t}(\mathbf{\Theta})]\\
    \notag=&\sqrt{KP_\mathrm{t}\varepsilon_v\varepsilon_{u_\mathrm{t}}}
    \mathbb{E}[{h}_\mathrm{t}(\mathbf{\Theta}^{\mathrm{t}})]\\
    =&\sqrt{KP_\mathrm{t}\varepsilon_v\varepsilon_{u_\mathrm{t}}}
    \sqrt{\frac{\varrho_\mathrm{t}\varrho_a\kappa_{\mathrm{t}}\kappa_a}
    {(1+\kappa_{\mathrm{t}})(1+\kappa_a)}}\xi
    \sum_{n=1}^{N_\mathrm{t}}\bar{a}_{\mathrm{t},n}^*\bar{g}_{\mathrm{t},n}\mathrm{e}^
    {\jmath\bar{\theta}_{n}^{\mathrm{t}}},
\end{align}
\begin{align}\label{Channel_Estimation_6}
    \notag\mathrm{C}_{h_\mathrm{t}(\mathbf{\Theta}^{\mathrm{t}})
    x_\mathrm{t}(\mathbf{\Theta})}
    =&\sqrt{KP_\mathrm{t}\varepsilon_v\varepsilon_{u_\mathrm{t}}}
    \mathrm{C}_{h_\mathrm{t}(\mathbf{\Theta}^{\mathrm{t}})h_\mathrm{t}
    (\mathbf{\Theta}^{\mathrm{t}})}\\
    =&\sqrt{KP_\mathrm{t}\varepsilon_v\varepsilon_{u_\mathrm{t}}}
    \frac{\varrho_\mathrm{t}\varrho_a(\kappa_{\mathrm{t}}+\kappa_a+1)N_\mathrm{t}}
    {(1+\kappa_{\mathrm{t}})(1+\kappa_a)},
\end{align}
\begin{align}\label{Channel_Estimation_7}
    \notag&\mathrm{C}_{x_\mathrm{t}(\mathbf{\Theta})x_\mathrm{t}(\mathbf{\Theta})}\\
    \notag=&KP_\mathrm{t}\varepsilon_v\varepsilon_{u_\mathrm{t}}
    \mathrm{C}_{h_\mathrm{t}(\mathbf{\Theta}^{\mathrm{t}})
    h_\mathrm{t}(\mathbf{\Theta}^{\mathrm{t}})}
    +P_\mathrm{t}\varepsilon_v(1-\varepsilon_{u_\mathrm{t}})
    \mathbb{E}[|h_\mathrm{t}(\mathbf{\Theta}^{\mathrm{t}})|^2]\\
    \notag&+P_\mathrm{r}\varepsilon_v(1-\varepsilon_{u_\mathrm{r}})
    \mathbb{E}[|h_\mathrm{r}(\mathbf{\Theta}^{\mathrm{r}})|^2]
    +P_\mathrm{t}(1-\varepsilon_v)\mathbb{E}[|h_\mathrm{t}(\mathbf{\Theta}^{\mathrm{t}})|^2]\\
    \notag&+P_\mathrm{r}(1-\varepsilon_v)
    \mathbb{E}[|h_\mathrm{r}(\mathbf{\Theta}^{\mathrm{r}})|^2]+\sigma_w^2\\
    =&KP_\mathrm{t}\varepsilon_v\varepsilon_{u_\mathrm{t}}
    \frac{\varrho_\mathrm{t}\varrho_a(\kappa_{\mathrm{t}}+\kappa_a+1)N_\mathrm{t}}
    {(1+\kappa_{\mathrm{t}})(1+\kappa_a)}+\zeta+\sigma_w^2.
\end{align}
where $\zeta$ is given by
\begin{align}\label{Channel_Estimation_7_0}
    \notag\zeta=&\sum\limits_{\mathrm{i}\in\{\mathrm{t},\mathrm{r}\}}
    \frac{P_\mathrm{i}(1-\varepsilon_v\varepsilon_{u_\mathrm{i}})\varrho_\mathrm{i}\varrho_a}
    {(1+\kappa_{\mathrm{i}})(1+\kappa_a)}\big(\kappa_{\mathrm{i}}\kappa_a\xi^2
    |\sum\nolimits_{n=1}^{N_\mathrm{i}}\bar{a}_{\mathrm{i},n}^*
    \bar{g}_{\mathrm{i},n}\mathrm{e}^{\jmath\bar{\theta}_{n}^{\mathrm{i}}}|^2\\
    &+(\kappa_{\mathrm{i}}+\kappa_a+1)N_\mathrm{i}\big).
\end{align}

According to (\ref{Channel_Estimation_1_1}), (\ref{Channel_Estimation_4}), (\ref{Channel_Estimation_5}), (\ref{Channel_Estimation_6}) and (\ref{Channel_Estimation_7}), the LMMSE estimator of $h_\mathrm{t}(\mathbf{\Theta}^{\mathrm{t}})$, denoted as $\hat{h}_\mathrm{t}(\mathbf{\Theta}^{\mathrm{t}})$, is given by
\begin{align}\label{Channel_Estimation_8}
    \notag\hat{h}_\mathrm{t}(\mathbf{\Theta}^{\mathrm{t}})
    =&\mathbb{E}[h_\mathrm{t}(\mathbf{\Theta}^{\mathrm{t}})]+
    \mathrm{C}_{h_\mathrm{t}(\mathbf{\Theta}^{\mathrm{t}})
    x_\mathrm{t}(\mathbf{\Theta})}\mathrm{C}_{x_\mathrm{t}(\mathbf{\Theta})
    x_\mathrm{t}(\mathbf{\Theta})}^{-1}\cdot\\
    &(x_\mathrm{t}(\mathbf{\Theta})-\mathbb{E}[x_\mathrm{t}(\mathbf{\Theta})]),
\end{align}
and its corresponding variance is
\begin{align}\label{Channel_Estimation_9}
    \notag\mathrm{C}_{\hat{h}_\mathrm{t}(\mathbf{\Theta}^{\mathrm{t}})
    \hat{h}_\mathrm{t}(\mathbf{\Theta}^{\mathrm{t}})}
    =&\mathrm{C}_{h_\mathrm{t}(\mathbf{\Theta}^{\mathrm{t}})
    x_\mathrm{t}(\mathbf{\Theta})}\mathrm{C}_{x_\mathrm{t}(\mathbf{\Theta})
    x_\mathrm{t}(\mathbf{\Theta})}^{-1}\mathrm{C}_{h_\mathrm{t}(\mathbf{\Theta}^{\mathrm{t}})
    x_\mathrm{t}(\mathbf{\Theta})}^{\mathrm{H}}\\
    =&\frac{KP_\mathrm{t}\varepsilon_v\varepsilon_{u_\mathrm{t}}
    \big(\frac{\varrho_\mathrm{t}\varrho_a(\kappa_{\mathrm{t}}+\kappa_a+1)N_\mathrm{t}}
    {(1+\kappa_{\mathrm{t}})(1+\kappa_a)}\big)^2}
    {KP_\mathrm{t}\varepsilon_v\varepsilon_{u_\mathrm{t}}
    \frac{\varrho_\mathrm{t}\varrho_a(\kappa_{\mathrm{t}}+\kappa_a+1)N_\mathrm{t}}
    {(1+\kappa_{\mathrm{t}})(1+\kappa_a)}+\zeta+\sigma_w^2}.
\end{align}
Thus, the estimation error is $\check{h}_\mathrm{t}(\mathbf{\Theta}^{\mathrm{t}})
=h_\mathrm{t}(\mathbf{\Theta}^{\mathrm{t}})-\hat{h}_\mathrm{t}(\mathbf{\Theta}^{\mathrm{t}})$, and the estimation error variance is
\begin{align}\label{Channel_Estimation_10}
    \notag\mathrm{C}_{\check{h}_\mathrm{t}(\mathbf{\Theta}^{\mathrm{t}})
    \check{h}_\mathrm{t}(\mathbf{\Theta}^{\mathrm{t}})}
    =&\mathrm{C}_{h_\mathrm{t}(\mathbf{\Theta}^{\mathrm{t}})
    h_\mathrm{t}(\mathbf{\Theta}^{\mathrm{t}})}-\mathrm{C}_{\hat{h}_\mathrm{t}
    (\mathbf{\Theta}^{\mathrm{t}})\hat{h}_\mathrm{t}(\mathbf{\Theta}^{\mathrm{t}})}\\
    =&\frac{\frac{\varrho_\mathrm{t}\varrho_a(\kappa_{\mathrm{t}}+\kappa_a+1)N_\mathrm{t}}
    {(1+\kappa_{\mathrm{t}})(1+\kappa_a)}(\zeta+\sigma_w^2)}
    {KP_\mathrm{t}\varepsilon_v\varepsilon_{u_\mathrm{t}}
    \frac{\varrho_\mathrm{t}\varrho_a(\kappa_{\mathrm{t}}+\kappa_a+1)N_\mathrm{t}}
    {(1+\kappa_{\mathrm{t}})(1+\kappa_a)}+\zeta+\sigma_w^2}.
\end{align}

Similarly, to estimate $h_\mathrm{r}(\mathbf{\Theta}^{\mathrm{r}})$, the conjugate transpose of the UE-R pilot sequence, i.e. $[\tau_{\mathrm{r}}^{(1)*},\tau_{\mathrm{r}}^{(2)*},\cdots,\tau_{\mathrm{r}}^{(K)*}]$, is employed for combining the AP observations $x^{(1)}(\mathbf{\Theta}),x^{(2)}(\mathbf{\Theta}),
\cdots,x^{(K)}(\mathbf{\Theta})$, and then the LMMSE estimator of $h_\mathrm{r}(\mathbf{\Theta}^{\mathrm{r}})$, denoted as $\hat{h}_\mathrm{r}(\mathbf{\Theta}^{\mathrm{r}})$, is given by
\begin{align}\label{Channel_Estimation_11}
    \notag\hat{h}_\mathrm{r}(\mathbf{\Theta}^{\mathrm{r}})
    =&\mathbb{E}[h_\mathrm{r}(\mathbf{\Theta}^{\mathrm{r}})]+
    \mathrm{C}_{h_\mathrm{r}(\mathbf{\Theta}^{\mathrm{r}})
    x_\mathrm{r}(\mathbf{\Theta})}\mathrm{C}_{x_\mathrm{r}(\mathbf{\Theta})
    x_\mathrm{r}(\mathbf{\Theta})}^{-1}\cdot\\
    &(x_\mathrm{r}(\mathbf{\Theta})-\mathbb{E}[x_\mathrm{r}(\mathbf{\Theta})]),
\end{align}
and the corresponding estimation variance becomes
\begin{align}\label{Channel_Estimation_12}
    \notag\mathrm{C}_{\hat{h}_\mathrm{r}(\mathbf{\Theta}^{\mathrm{r}})
    \hat{h}_\mathrm{r}(\mathbf{\Theta}^{\mathrm{r}})}
    =&\mathrm{C}_{h_\mathrm{r}(\mathbf{\Theta}^{\mathrm{r}})
    x_\mathrm{r}(\mathbf{\Theta})}\mathrm{C}_{x_\mathrm{r}(\mathbf{\Theta})
    x_\mathrm{r}(\mathbf{\Theta})}^{-1}\mathrm{C}_{h_\mathrm{r}(\mathbf{\Theta}^{\mathrm{r}})
    x_\mathrm{r}(\mathbf{\Theta})}^{\mathrm{H}}\\
    =&\frac{KP_\mathrm{r}\varepsilon_v\varepsilon_{u_\mathrm{r}}
    \big(\frac{\varrho_\mathrm{r}\varrho_a(\kappa_{\mathrm{r}}+\kappa_a+1)N_\mathrm{r}}
    {(1+\kappa_{\mathrm{r}})(1+\kappa_a)}\big)^2}
    {KP_\mathrm{r}\varepsilon_v\varepsilon_{u_\mathrm{r}}
    \frac{\varrho_\mathrm{r}\varrho_a(\kappa_{\mathrm{r}}+\kappa_a+1)N_\mathrm{r}}
    {(1+\kappa_{\mathrm{r}})(1+\kappa_a)}+\zeta+\sigma_w^2}.
\end{align}
Thus, the estimation error is $\check{h}_\mathrm{r}(\mathbf{\Theta}^{\mathrm{r}})=h_\mathrm{r}(\mathbf{\Theta}^{\mathrm{r}})
-\hat{h}_\mathrm{r}(\mathbf{\Theta}^{\mathrm{r}})$, and the estimation error variance is
\begin{align}\label{Channel_Estimation_13}
    \notag\mathrm{C}_{\check{h}_\mathrm{r}(\mathbf{\Theta}^{\mathrm{r}})
    \check{h}_\mathrm{r}(\mathbf{\Theta}^{\mathrm{r}})}
    =&\mathrm{C}_{h_\mathrm{r}(\mathbf{\Theta}^{\mathrm{r}})h_\mathrm{r}
    (\mathbf{\Theta}^{\mathrm{r}})}-\mathrm{C}_{\hat{h}_\mathrm{r}(\mathbf{\Theta}^{\mathrm{r}})
    \hat{h}_\mathrm{r}(\mathbf{\Theta}^{\mathrm{r}})}\\
    =&\frac{\frac{\varrho_\mathrm{r}\varrho_a(\kappa_{\mathrm{r}}+\kappa_a+1)N_\mathrm{r}}
    {(1+\kappa_{\mathrm{r}})(1+\kappa_a)}(\zeta+\sigma_w^2)}
    {KP_\mathrm{r}\varepsilon_v\varepsilon_{u_\mathrm{r}}
    \frac{\varrho_\mathrm{r}\varrho_a(\kappa_{\mathrm{r}}+\kappa_a+1)N_\mathrm{r}}
    {(1+\kappa_{\mathrm{r}})(1+\kappa_a)}+\zeta+\sigma_w^2}.
\end{align}

Therefore, the normalized mean square error (N-MSE) of the estimated channel links, denoted as $\Xi_{\mathrm{LMMSE}}(\mathbf{\Theta})$, is given by
\begin{align}\label{Channel_Estimation_15}
    \Xi_{\mathrm{LMMSE}}(\mathbf{\Theta})
    =\frac{1}{2}\Big(\frac{\mathrm{C}_{\check{h}_\mathrm{t}(\mathbf{\Theta}^{\mathrm{t}})
    \check{h}_\mathrm{t}(\mathbf{\Theta}^{\mathrm{t}})}}
    {\mathrm{C}_{h_\mathrm{t}(\mathbf{\Theta}^{\mathrm{t}})
    h_\mathrm{t}(\mathbf{\Theta}^{\mathrm{t}})}}
    +\frac{\mathrm{C}_{\check{h}_\mathrm{r}(\mathbf{\Theta}^{\mathrm{r}})
    \check{h}_\mathrm{r}(\mathbf{\Theta}^{\mathrm{r}})}}
    {\mathrm{C}_{h_\mathrm{r}(\mathbf{\Theta}^{\mathrm{r}})
    h_\mathrm{r}(\mathbf{\Theta}^{\mathrm{r}})}}\Big).
\end{align}
Based on (\ref{Channel_Estimation_2_1}), (\ref{Channel_Estimation_2_2}), (\ref{Channel_Estimation_10}), (\ref{Channel_Estimation_13}) and (\ref{Channel_Estimation_15}), when the power of the pilot sequences obeys $P_\mathrm{t}\rightarrow\infty$ and $P_\mathrm{r}\rightarrow\infty$, the N-MSE tends to a constant, denoted as $\ddot{\Xi}_{\mathrm{LMMSE}}(\mathbf{\Theta})$, which is given by
\begin{align}\label{Channel_Estimation_16}
    \notag\ddot{\Xi}_{\mathrm{LMMSE}}(\mathbf{\Theta})
    =&\frac{1}{2}\Big(\frac{\zeta}{KP_\mathrm{t}\varepsilon_v\varepsilon_{u_\mathrm{t}}
    \frac{\varrho_\mathrm{t}\varrho_a(\kappa_{\mathrm{t}}+\kappa_a+1)N_\mathrm{t}}
    {(1+\kappa_{\mathrm{t}})(1+\kappa_a)}+\zeta}\\
    &+\frac{\zeta}{KP_\mathrm{r}\varepsilon_v\varepsilon_{u_\mathrm{r}}
    \frac{\varrho_\mathrm{r}\varrho_a(\kappa_{\mathrm{r}}+\kappa_a+1)N_\mathrm{r}}
    {(1+\kappa_{\mathrm{r}})(1+\kappa_a)}+\zeta}\Big).
\end{align}
It shows that when the transceiver hardware is ideal, i.e. $\varepsilon_v=\varepsilon_{u_\mathrm{t}}=\varepsilon_{u_\mathrm{r}}=1$, the N-MSE tends to 0 when $P_\mathrm{t}=P_\mathrm{r}\rightarrow\infty$. By contrast, when the transceiver hardware is non-ideal, i.e. $\varepsilon_v\varepsilon_{u_\mathrm{t}}\varepsilon_{u_\mathrm{r}}<1$, the N-MSE has a performance floor as $P_\mathrm{t}=P_\mathrm{r}\rightarrow\infty$, given by
\begin{align}\label{Channel_Estimation_17}
    \notag\ddot{\Xi}_{\mathrm{LMMSE}}(\mathbf{\Theta})
    \rightarrow&\frac{1}{2}\Big(\frac{\zeta'}
    {K\varepsilon_v\varepsilon_{u_\mathrm{t}}
    \frac{\varrho_\mathrm{t}\varrho_a(\kappa_{\mathrm{t}}+\kappa_a+1)N_\mathrm{t}}
    {(1+\kappa_{\mathrm{t}})(1+\kappa_a)}+\zeta'}\\
    \notag&+\frac{\zeta'}
    {K\varepsilon_v\varepsilon_{u_\mathrm{r}}
    \frac{\varrho_\mathrm{r}\varrho_a(\kappa_{\mathrm{r}}+\kappa_a+1)N_\mathrm{r}}
    {(1+\kappa_{\mathrm{r}})(1+\kappa_a)}+\zeta'}\Big),\\
    &P_\mathrm{t}=P_\mathrm{r}\rightarrow\infty,
\end{align}
where $\zeta'$ is
\begin{align}\label{Channel_Estimation_7_1}
    \notag\zeta'=&\sum\limits_{\mathrm{i}\in\{\mathrm{t},\mathrm{r}\}}
    \frac{(1-\varepsilon_v\varepsilon_{u_\mathrm{i}})\varrho_\mathrm{i}\varrho_a}
    {(1+\kappa_{\mathrm{i}})(1+\kappa_a)}\big(\kappa_{\mathrm{i}}\kappa_a\xi^2
    |\sum\nolimits_{n=1}^{N_\mathrm{i}}\bar{a}_{\mathrm{i},n}^*
    \bar{g}_{\mathrm{i},n}\mathrm{e}^{\jmath\bar{\theta}_{n}^{\mathrm{i}}}|^2\\
    &+(\kappa_{\mathrm{i}}+\kappa_a+1)N_\mathrm{i}\big).
\end{align}
Furthermore, when the pilot sequence length is large enough, i.e. $K\gg1$, the performance floor of the N-MSE is given by
\begin{align}\label{Channel_Estimation_18}
    \notag\ddot{\Xi}_{\mathrm{LMMSE}}(\mathbf{\Theta})
    \rightarrow&\frac{1}{2}\Big(\frac{\zeta'}{K\varepsilon_v\varepsilon_{u_\mathrm{t}}
    \frac{\varrho_\mathrm{t}\varrho_a(\kappa_{\mathrm{t}}+\kappa_a+1)N_\mathrm{t}}
    {1+\kappa_{\mathrm{t}}}}\\
    \notag&+\frac{\zeta'}{K\varepsilon_v\varepsilon_{u_\mathrm{r}}
    \frac{\varrho_\mathrm{r}\varrho_a(\kappa_{\mathrm{r}}+\kappa_a+1)N_\mathrm{r}}
    {1+\kappa_{\mathrm{r}}}}\Big),\\
    &P_\mathrm{t}=P_\mathrm{r}\rightarrow\infty,\ K\gg1.
\end{align}

When the statistical information of the links, i.e. $\mathbb{E}[h_\mathrm{t}(\mathbf{\Theta}^{\mathrm{t}})]$, $\mathbb{E}[x_\mathrm{t}(\mathbf{\Theta})]$, $\mathrm{C}_{h_\mathrm{t}(\mathbf{\Theta}^{\mathrm{t}})x_\mathrm{t}(\mathbf{\Theta})}$,
$\mathrm{C}_{x_\mathrm{t}(\mathbf{\Theta})x_\mathrm{t}(\mathbf{\Theta})}$, $\mathbb{E}[h_\mathrm{r}(\mathbf{\Theta}^{\mathrm{r}})]$, $\mathbb{E}[x_\mathrm{r}(\mathbf{\Theta})]$, $\mathrm{C}_{h_\mathrm{r}(\mathbf{\Theta}^{\mathrm{r}})x_\mathrm{r}(\mathbf{\Theta})}$ and
$\mathrm{C}_{x_\mathrm{r}(\mathbf{\Theta})x_\mathrm{r}(\mathbf{\Theta})}$, is not available, the least square (LS) estimator can be employed. Specifically, the LS estimators of $h_\mathrm{t}(\mathbf{\Theta}^{\mathrm{t}})$ and $h_\mathrm{r}(\mathbf{\Theta}^{\mathrm{r}})$ are given by $\frac{{x}_\mathrm{t}(\mathbf{\Theta})}
{\sqrt{KP_\mathrm{t}\varepsilon_v\varepsilon_{u_\mathrm{t}}}}$ and $\frac{{x}_\mathrm{r}(\mathbf{\Theta})}
{\sqrt{KP_\mathrm{r}\varepsilon_v\varepsilon_{u_\mathrm{r}}}}$, respectively.

\section{Performance Analysis and Beamforming Design}\label{Performance_Analysis_and_Beamforming_Design}
Based on the estimated CSI, we derive the ergodic sum-rate of the STAR-RIS aided NOMA systems with perfect SIC decoding algorithms considering both the channel estimation error and the HWI of the transceivers, compared to that of the OMA systems. Furthermore, we also present the optimal phase shift design of the STAR-RIS elements for maximizing the ergodic sum-rate. Finally, we discuss the effect of the imperfect SIC decoding algorithms.

\subsection{Ergodic Sum-rate Analysis}
We denote the information symbols at the UE-T and UE-R as $s_\mathrm{t}\in\mathbb{C}^{1\times1}$ and $s_\mathrm{r}\in\mathbb{C}^{1\times1}$ respectively, with a mean of 0 and variance of 1. The observation ${y}(\mathbf{\Theta})$ at the AP is given by
\begin{align}\label{NOMA_1}
    \notag\notag{y}(\mathbf{\Theta})
    =&\underbrace{\sqrt{\rho_\mathrm{t}\varepsilon_{v}\varepsilon_{u_\mathrm{t}}}
    \hat{h}_\mathrm{t}(\mathbf{\Theta}^{\mathrm{t}})s_\mathrm{t}
    +\sqrt{\rho_\mathrm{r}\varepsilon_{v}\varepsilon_{u_\mathrm{r}}}
    \hat{h}_\mathrm{r}(\mathbf{\Theta}^{\mathrm{r}})s_\mathrm{r}}
    _{\text{Desired signal of UE-T and UE-R over estimated channel}}\\
    \notag&+\underbrace{\sqrt{\rho_\mathrm{t}\varepsilon_{v}\varepsilon_{u_\mathrm{t}}}
    \check{h}_\mathrm{t}(\mathbf{\Theta}^{\mathrm{t}})s_\mathrm{t}
    +\sqrt{\rho_\mathrm{r}\varepsilon_{v}\varepsilon_{u_\mathrm{r}}}
    \check{h}_\mathrm{r}(\mathbf{\Theta}^{\mathrm{r}})s_\mathrm{r}}
    _{\text{Desired signal of UE-T and UE-R over unknown channel}}\\
    \notag&+\underbrace{\sqrt{\rho_\mathrm{t}(1-\varepsilon_{v})}
    h_\mathrm{t}(\mathbf{\Theta}^{\mathrm{t}})v_\mathrm{t}
    +\sqrt{\rho_\mathrm{r}(1-\varepsilon_{v})}
    h_\mathrm{r}(\mathbf{\Theta}^{\mathrm{r}})v_\mathrm{r}}_{\text{AP HWI distortion}}\\
    \notag&+\underbrace{\sqrt{\rho_\mathrm{t}\varepsilon_{v}(1-\varepsilon_{u_\mathrm{t}})}
    h_\mathrm{t}(\mathbf{\Theta}^{\mathrm{t}})u_\mathrm{t}}_{\text{UE-T HWI distortion}}\\
    &+\underbrace{\sqrt{\rho_\mathrm{r}\varepsilon_{v}(1-\varepsilon_{u_\mathrm{r}})}
    h_\mathrm{r}(\mathbf{\Theta}^{\mathrm{r}})u_\mathrm{r}}_{\text{UE-R HWI distortion}}
    +\underbrace{w}_{\text{Noise}},
\end{align}
where $w\sim\mathcal{CN}(0,\sigma_w^2)$ is the additive noise at the AP, $u_\mathrm{t}\sim\mathcal{CN}(0,1)$ and $v_\mathrm{t}\sim\mathcal{CN}(0,1)$ represent the distortion of the information symbol $s_\mathrm{t}$ at the UE-T and the AP, respectively. Furthermore, $u_\mathrm{r}\sim\mathcal{CN}(0,1)$ and $v_\mathrm{r}\sim\mathcal{CN}(0,1)$ represent the distortion of the information symbol $s_\mathrm{r}$ at the UE-R and the AP, respectively.

When we have a perfect SIC algorithm for information recovery~\cite{dai2018survey}, and the instantaneous achievable rate is presented in the following.

\begin{theorem}\label{theorem_2}
Based on the perfect SIC algorithm used for the NOMA information recovery, when the AP detects the UE-T signal first, followed by the detection of the UE-R signal after removing the interference imposed by the UE-T signal from the composite received observation, let us denote this as the $\mathrm{t}\rightarrow\mathrm{r}$ SIC order. The instantaneous achievable rate of the UE-T and that of the UE-R, denoted as $R_\mathrm{t}^{\mathrm{t}\rightarrow\mathrm{r}}(\mathbf{\Theta})$ and $R_\mathrm{r}^{\mathrm{t}\rightarrow\mathrm{r}}(\mathbf{\Theta})$ respectively, are expressed as
\begin{align}\label{NOMA_2}
    R_\mathrm{t}^{\mathrm{t}\rightarrow\mathrm{r}}(\mathbf{\Theta})
    =\log_2\Big(1+\frac{\rho_\mathrm{t}\varepsilon_v\varepsilon_{u_\mathrm{t}}
    |\hat{h}_\mathrm{t}(\mathbf{\Theta}^{\mathrm{t}})|^2}
    {\rho_\mathrm{r}\varepsilon_v\varepsilon_{u_\mathrm{r}}
    |\hat{h}_\mathrm{r}(\mathbf{\Theta}^{\mathrm{r}})|^2+\mathcal{E}}\Big),
\end{align}
\begin{align}\label{NOMA_3}
    R_\mathrm{r}^{\mathrm{t}\rightarrow\mathrm{r}}(\mathbf{\Theta})
    =\log_2\Big(1+\frac{\rho_\mathrm{r}\varepsilon_v\varepsilon_{u_\mathrm{r}}
    |\hat{h}_\mathrm{r}(\mathbf{\Theta}^{\mathrm{r}})|^2}{\mathcal{E}}\Big),
\end{align}
with
\begin{align}\label{NOMA_3_0}
    \mathcal{E}=\rho_\mathrm{t}\epsilon_\mathrm{t}(\mathbf{\Theta}^{\mathrm{t}})
    +\rho_\mathrm{r}\epsilon_\mathrm{r}(\mathbf{\Theta}^{\mathrm{r}})+\sigma_w^2,
\end{align}
where $\epsilon_\mathrm{i}(\mathbf{\Theta}^{\mathrm{i}})
=(1-\varepsilon_v\varepsilon_{u_\mathrm{i}})
|\hat{h}_\mathrm{i}(\mathbf{\Theta}^{\mathrm{i}})|^2
+\mathrm{C}_{\check{h}_\mathrm{i}(\mathbf{\Theta}^{\mathrm{i}})
\check{h}_\mathrm{i}(\mathbf{\Theta}^{\mathrm{i}})}$ represents the equivalent noise caused by the transceiver HWI and the channel estimation error. By contrast, when the AP detects the UE-R signal and then removes its effect from the composite received observation for the detection of the signal from the UE-T, let us denote this as the $\mathrm{r}\rightarrow\mathrm{t}$ SIC order. Then, the instantaneous achievable rate of the UE-T and that of the UE-R, denoted as $R_\mathrm{t}^{\mathrm{r}\rightarrow\mathrm{t}}(\mathbf{\Theta})$ and $R_\mathrm{r}^{\mathrm{r}\rightarrow\mathrm{t}}(\mathbf{\Theta})$, are given by
\begin{align}\label{NOMA_4}
    R_\mathrm{t}^{\mathrm{r}\rightarrow\mathrm{t}}(\mathbf{\Theta})
    =\log_2\Big(1+\frac{\rho_\mathrm{t}\varepsilon_v\varepsilon_{u_\mathrm{t}}
    |\hat{h}_\mathrm{t}(\mathbf{\Theta}^{\mathrm{t}})|^2}{\mathcal{E}}\Big),
\end{align}
\begin{align}\label{NOMA_5}
    R_\mathrm{r}^{\mathrm{r}\rightarrow\mathrm{t}}(\mathbf{\Theta})
    =\log_2\Big(1+\frac{\rho_\mathrm{r}\varepsilon_v\varepsilon_{u_\mathrm{r}}
    |\hat{h}_\mathrm{r}(\mathbf{\Theta}^{\mathrm{r}})|^2}
    {\rho_\mathrm{t}\varepsilon_v\varepsilon_{u_\mathrm{t}}
    |\hat{h}_\mathrm{t}(\mathbf{\Theta}^{\mathrm{t}})|^2+\mathcal{E}}\Big).
\end{align}
\end{theorem}
\begin{IEEEproof}
    See Appendix \ref{Appendix_B}.
\end{IEEEproof}

By employing the time-sharing strategy, where the time duration of $\beta$ is for the $\mathrm{t}\rightarrow\mathrm{r}$ SIC order and that of $(1-\beta)$ is for the $\mathrm{r}\rightarrow\mathrm{t}$ SIC order, the instantaneous achievable rate of the UE-T and that of the UE-R, denoted as $R_\mathrm{t}^{\text{NOMA}}(\mathbf{\Theta})$ and $R_\mathrm{r}^{\text{NOMA}}(\mathbf{\Theta})$ respectively, are
\begin{align}\label{NOMA_6}
    R_\mathrm{t}^{\text{NOMA}}(\mathbf{\Theta})
    =\beta{R}_\mathrm{t}^{\mathrm{t}\rightarrow\mathrm{r}}(\mathbf{\Theta})
    +(1-\beta){R}_\mathrm{t}^{\mathrm{r}\rightarrow\mathrm{t}}(\mathbf{\Theta}),
\end{align}
\begin{align}\label{NOMA_7}
    R_\mathrm{r}^{\text{NOMA}}(\mathbf{\Theta})
    =\beta{R}_\mathrm{r}^{\mathrm{t}\rightarrow\mathrm{r}}(\mathbf{\Theta})
    +(1-\beta){R}_\mathrm{r}^{\mathrm{r}\rightarrow\mathrm{t}}(\mathbf{\Theta}).
\end{align}
According to (\ref{NOMA_2}), (\ref{NOMA_3}), (\ref{NOMA_4}), (\ref{NOMA_5}), (\ref{NOMA_6}) and (\ref{NOMA_7}), the instantaneous sum-rate of the STAR-RIS aided NOMA uplink is given by
\begin{align}\label{NOMA_7_2}
    \notag&{R}_\text{sum}^{\text{NOMA}}(\mathbf{\Theta})\\
    \notag=&R_\mathrm{t}^{\text{NOMA}}(\mathbf{\Theta})
    +R_\mathrm{r}^{\text{NOMA}}(\mathbf{\Theta})\\
    \notag=&\beta\big({R}_\mathrm{t}^{\mathrm{t}\rightarrow\mathrm{r}}(\mathbf{\Theta})
    +{R}_\mathrm{r}^{\mathrm{t}\rightarrow\mathrm{r}}(\mathbf{\Theta})\big)
    +(1-\beta)\big({R}_\mathrm{t}^{\mathrm{r}\rightarrow\mathrm{t}}(\mathbf{\Theta})+
    {R}_\mathrm{r}^{\mathrm{r}\rightarrow\mathrm{t}}(\mathbf{\Theta})\big)\\
    =&\log_2\Big(1+\frac{\rho_\mathrm{t}\varepsilon_v\varepsilon_{u_\mathrm{t}}
    |\hat{h}_\mathrm{t}(\mathbf{\Theta}^{\mathrm{t}})|^2
    +\rho_\mathrm{r}\varepsilon_v\varepsilon_{u_\mathrm{r}}
    |\hat{h}_\mathrm{r}(\mathbf{\Theta}^{\mathrm{r}})|^2}{\mathcal{E}}\Big).
\end{align}
This shows that the sum-rate is independent of the time-sharing factor $\beta$. However, the time-sharing factor $\beta$ has an effect on the fairness of the achievable rate pair of the UE-T and UE-R.

According to (\ref{NOMA_7_2}), the ergodic sum-rate of the STAR-RIS aided uplink NOMA system, denoted as $R_\text{sum,erg}^{\text{NOMA}}(\mathbf{\Theta})$, is given by
\begin{align}\label{NOMA_7_3}
    \notag&R_\text{sum,erg}^{\text{NOMA}}(\mathbf{\Theta})\\
    \notag=&\mathbb{E}[R_\text{sum}^{\text{NOMA}}(\mathbf{\Theta})]\\
    =&\mathbb{E}\Big[\log_2\Big(1+\frac{\rho_\mathrm{t}\varepsilon_v\varepsilon_{u_\mathrm{t}}
    |\hat{h}_\mathrm{t}(\mathbf{\Theta}^{\mathrm{t}})|^2
    +\rho_\mathrm{r}\varepsilon_v\varepsilon_{u_\mathrm{r}}
    |\hat{h}_\mathrm{r}(\mathbf{\Theta}^{\mathrm{r}})|^2}{\mathcal{E}}\Big)\Big].
\end{align}
Since $R_\text{sum,erg}^{\text{NOMA}}(\mathbf{\Theta})$ is a concave function with respect to $|\hat{h}_\mathrm{t}(\mathbf{\Theta}^{\mathrm{t}})|^2$ and $|\hat{h}_\mathrm{r}(\mathbf{\Theta}^{\mathrm{r}})|^2$, we can formulate the upper bound of $R_\text{sum,erg}^{\text{NOMA}}(\mathbf{\Theta})$, denoted as $\ddot{R}_\text{sum,erg}^{\text{NOMA}}(\mathbf{\Theta})$, as follows:
\begin{align}\label{NOMA_7_4}
    \notag&\ddot{R}_\text{sum,erg}^{\text{NOMA}}(\mathbf{\Theta})\\
    \notag=&\log_2\Big(1+\frac{\rho_\mathrm{t}\varepsilon_v\varepsilon_{u_\mathrm{t}}
    \mathbb{E}[|\hat{h}_\mathrm{t}(\mathbf{\Theta}^{\mathrm{t}})|^2]
    +\rho_\mathrm{r}\varepsilon_v\varepsilon_{u_\mathrm{r}}
    \mathbb{E}[|\hat{h}_\mathrm{r}(\mathbf{\Theta}^{\mathrm{r}})|^2]}
    {\rho_\mathrm{t}\ddot{\epsilon}_\mathrm{t}(\mathbf{\Theta}^{\mathrm{t}})
    +\rho_\mathrm{r}\ddot{\epsilon}_\mathrm{r}(\mathbf{\Theta}^{\mathrm{r}})+\sigma_w^2}\Big)\\
    \notag=&\log_2\Big(1+\frac{1}
    {\rho_\mathrm{t}\ddot{\epsilon}_\mathrm{t}(\mathbf{\Theta}^{\mathrm{t}})
    +\rho_\mathrm{r}\ddot{\epsilon}_\mathrm{r}(\mathbf{\Theta}^{\mathrm{r}})+\sigma_w^2}\cdot\\
    \notag&\Big(\rho_\mathrm{t}\varepsilon_v\varepsilon_{u_\mathrm{t}}
    (|\mathbb{E}[h_\mathrm{t}(\mathbf{\Theta}^{\mathrm{t}})]|^2
    +\mathrm{C}_{\hat{h}_\mathrm{t}(\mathbf{\Theta}^{\mathrm{t}})
    \hat{h}_\mathrm{t}(\mathbf{\Theta}^{\mathrm{t}})})\\
    &+\rho_\mathrm{r}\varepsilon_v\varepsilon_{u_\mathrm{r}}
    (|\mathbb{E}[h_\mathrm{r}(\mathbf{\Theta}^{\mathrm{r}})]|^2
    +\mathrm{C}_{\hat{h}_\mathrm{r}(\mathbf{\Theta}^{\mathrm{r}})
    \hat{h}_\mathrm{r}(\mathbf{\Theta}^{\mathrm{r}})})\Big)\Big),
\end{align}
where $\ddot{\epsilon}_\mathrm{t}(\mathbf{\Theta}^{\mathrm{t}})
=(1-\varepsilon_v\varepsilon_{u_\mathrm{t}})(|\mathbb{E}[h_\mathrm{t}
(\mathbf{\Theta}^{\mathrm{t}})]|^2+\mathrm{C}_{\hat{h}_\mathrm{t}(\mathbf{\Theta}^{\mathrm{t}})
\hat{h}_\mathrm{t}(\mathbf{\Theta}^{\mathrm{t}})})+\mathrm{C}_{\check{h}_\mathrm{t}
(\mathbf{\Theta}^{\mathrm{t}})\check{h}_\mathrm{t}(\mathbf{\Theta}^{\mathrm{t}})}$ and $\ddot{\epsilon}_\mathrm{r}(\mathbf{\Theta}^{\mathrm{r}})
=(1-\varepsilon_v\varepsilon_{u_\mathrm{r}})(|\mathbb{E}[h_\mathrm{r}
(\mathbf{\Theta}^{\mathrm{r}})]|^2+\mathrm{C}_{\hat{h}_\mathrm{r}(\mathbf{\Theta}^{\mathrm{r}})
\hat{h}_\mathrm{r}(\mathbf{\Theta}^{\mathrm{r}})})+\mathrm{C}_{\check{h}_\mathrm{r}
(\mathbf{\Theta}^{\mathrm{r}})\check{h}_\mathrm{r}(\mathbf{\Theta}^{\mathrm{r}})}$.

For comparison, we consider an OMA scheme having a fraction $B$ of the time/frequency assigned to the UE-T and the remaining fraction $(1-B)$ of the time/frequency assigned to the UE-R. Then, the instantaneous achievable rate of the UE-T and that of the UE-R, denoted as $R_\mathrm{t}^{\text{OMA}}(\mathbf{\Theta})$ and $R_\mathrm{r}^{\text{OMA}}(\mathbf{\Theta})$, are given by~\cite{tse2005fundamentals}
\begin{align}\label{NOMA_8}
    R_\mathrm{t}^{\text{OMA}}(\mathbf{\Theta})
    =B\log_2\Big(1+\frac{\rho_\mathrm{t}\varepsilon_v\varepsilon_{u_\mathrm{t}}
    |\hat{h}_\mathrm{t}(\mathbf{\Theta}^{\mathrm{t}})|^2}
    {\rho_\mathrm{t}\epsilon_\mathrm{t}(\mathbf{\Theta}^{\mathrm{t}})
    +B\sigma_w^2}\Big),
\end{align}
\begin{align}\label{NOMA_9}
    R_\mathrm{r}^{\text{OMA}}(\mathbf{\Theta})
    =(1-B)\log_2\Big(1+\frac{\rho_\mathrm{r}\varepsilon_v\varepsilon_{u_\mathrm{r}}
    |\hat{h}_\mathrm{r}(\mathbf{\Theta}^{\mathrm{t}})|^2}
    {\rho_\mathrm{r}\epsilon_\mathrm{r}(\mathbf{\Theta}^{\mathrm{r}})
    +(1-B)\sigma_w^2}\Big).
\end{align}
Observe from (\ref{NOMA_8}) and (\ref{NOMA_9}) that varying $B$ from 0 to 1 finds all rate pairs that can be achieved by the OMA scheme. We denote the sum-rate of the OMA system as $R_\text{sum}^{\text{OMA}}(\mathbf{\Theta})$, which is given by
\begin{align}\label{NOMA_10}
    \notag R_\text{sum}^{\text{OMA}}(\mathbf{\Theta})
    =&R_\mathrm{t}^{\text{OMA}}(\mathbf{\Theta})
    +R_\mathrm{r}^{\text{OMA}}(\mathbf{\Theta})\\
    \notag=&B\log_2\Big(1+\frac{\rho_\mathrm{t}\varepsilon_v\varepsilon_{u_\mathrm{t}}
    |\hat{h}_\mathrm{t}(\mathbf{\Theta}^{\mathrm{t}})|^2}
    {\rho_\mathrm{t}\epsilon_\mathrm{t}(\mathbf{\Theta}^{\mathrm{t}})
    +B\sigma_w^2}\Big)\\
    &+(1-B)\log_2
    \Big(1+\frac{\rho_\mathrm{r}\varepsilon_v\varepsilon_{u_\mathrm{r}}
    |\hat{h}_\mathrm{r}(\mathbf{\Theta}^{\mathrm{r}})|^2}
    {\rho_\mathrm{r}\epsilon_\mathrm{r}(\mathbf{\Theta}^{\mathrm{r}})
    +(1-B)\sigma_w^2}\Big).
\end{align}

\begin{theorem}\label{theorem_3}
When the number of pilots is high enough, and the UE-T as well as the UE-R have the same hardware quality factor, i.e. $K\rightarrow\infty$ and $\varepsilon_{u_\mathrm{t}}=\varepsilon_{u_\mathrm{r}}$, the achievable sum-rate of the OMA scheme is not higher than that of the NOMA scheme, i.e.
\begin{align}\label{NOMA_11}
    R_\text{sum}^{\text{NOMA}}(\mathbf{\Theta})
    {\geq}R_\text{sum}^{\text{OMA}}(\mathbf{\Theta}),
\end{align}
where the equality is established only when the time/frequency fraction in the OMA scheme is $B=\frac{\rho_\mathrm{t}|\hat{h}_\mathrm{t}(\mathbf{\Theta}^{\mathrm{t}})|^2}
{\rho_\mathrm{t}|\hat{h}_\mathrm{t}(\mathbf{\Theta}^{\mathrm{t}})|^2
+\rho_\mathrm{r}|\hat{h}_\mathrm{r}(\mathbf{\Theta}^{\mathrm{r}})|^2}$.
\end{theorem}
\begin{IEEEproof}
    See Appendix \ref{Appendix_C}.
\end{IEEEproof}

\subsection{STAR-RIS Passive Beamforming Design}
Our objective is to design the STAR-RIS phase shift $\mathbf{\Theta}^{\mathrm{t}}$ and $\mathbf{\Theta}^{\mathrm{r}}$ for maximizing the ergodic sum-rate upper bound of the NOMA scheme. Thus, the optimization problem can be formulated as
\begin{align}\label{Beamforming_Design_1}
    \notag\text{(P1)}\quad &\max_{\bar{\theta}_{1}^{\mathrm{t}},\bar{\theta}_{2}^{\mathrm{t}},
    \cdots,\bar{\theta}_{N_\mathrm{t}}^{\mathrm{t}},
    \bar{\theta}_{1}^{\mathrm{r}},\bar{\theta}_{2}^{\mathrm{r}},
    \cdots,\bar{\theta}_{N_\mathrm{r}}^{\mathrm{r}}}\ \ddot{R}_\text{sum,erg}^{\text{NOMA}}(\mathbf{\Theta})\\
    \text{s.t.}&\quad \bar{\theta}^\mathrm{t}_{n_\mathrm{t}}\in[0,2\pi),\ \notag{n}_\mathrm{t}=1,2,\cdots,N_\mathrm{t},\\
    &\quad \bar{\theta}^\mathrm{r}_{n_\mathrm{r}}\in[0,2\pi),\ n_\mathrm{r}=1,2,\cdots,N_\mathrm{r}.
\end{align}

Since the STAR-RIS phase shift is designed based on the statistical CSI instead of the instantaneous CSI, it has considerably lower channel estimation overhead than the estimation of instantaneous CSI~\cite{pan2022overview}. Thus, the number of pilot sequences $K$ can be high enough to mitigate the channel estimation error effect on the system performance. As shown in (\ref{Channel_Estimation_10}) and (\ref{Channel_Estimation_13}), we have $\mathrm{C}_{\check{h}_\mathrm{t}(\mathbf{\Theta}^{\mathrm{t}})
\check{h}_\mathrm{t}(\mathbf{\Theta}^{\mathrm{t}})}=0$ and $\mathrm{C}_{\check{h}_\mathrm{r}(\mathbf{\Theta}^{\mathrm{r}})
\check{h}_\mathrm{r}(\mathbf{\Theta}^{\mathrm{r}})}=0$ when $K\rightarrow\infty$. Thus, the optimization problem (P1) can be formulated as follows.

\begin{theorem}\label{theorem_4}
When the number of pilot sequence is large enough, i.e. $K\rightarrow\infty$, the STAR-RIS phase shifts in the problem (P1) maximizing the ergodic sum-rate upper bound of the NOMA scheme are designed as
\begin{align}\label{Beamforming_Design_2_1}
    \notag\bar{\theta}^{\mathrm{t}}_{n_\mathrm{t}}=&\frac{2\pi}{\lambda}
    (\delta_{x}n_{\mathrm{t},x}(\sin\phi_\mathrm{t}\cos\varphi_\mathrm{t}
    -\sin\omega_\mathrm{t}\cos\varpi_\mathrm{t})\\
    &+\delta_{y}n_{\mathrm{t},y}(\cos\phi_\mathrm{t}-\cos\omega_\mathrm{t})),
\end{align}
\begin{align}\label{Beamforming_Design_2_2}
    \notag\bar{\theta}^{\mathrm{r}}_{n_\mathrm{r}}=&\frac{2\pi}{\lambda}
    (\delta_{x}n_{\mathrm{r},x}(\sin\phi_\mathrm{r}\cos\varphi_\mathrm{r}
    -\sin\omega_\mathrm{r}\cos\varpi_\mathrm{r})\\
    &+\delta_{y}n_{\mathrm{r},y}(\cos\phi_\mathrm{r}-\cos\omega_\mathrm{r})),
\end{align}
based on which $\sum_{n=1}^{N_\mathrm{t}}\bar{a}_{\mathrm{t},n}^*\bar{g}_{\mathrm{t},n}\mathrm{e}^
{\jmath\bar{\theta}_{n}^{\mathrm{t}}}=N_\mathrm{t}$ in $\mathbb{E}[h_\mathrm{t}(\mathbf{\Theta}^{\mathrm{t}})]$, $\mathbb{E}[|h_\mathrm{t}(\mathbf{\Theta}^{\mathrm{t}})|^2]$, $\mathrm{C}_{\hat{h}_\mathrm{t}(\mathbf{\Theta}^{\mathrm{t}})
\hat{h}_\mathrm{t}(\mathbf{\Theta}^{\mathrm{t}})}$, and $\mathrm{C}_{\check{h}_\mathrm{t}(\mathbf{\Theta}^{\mathrm{t}})
\check{h}_\mathrm{t}(\mathbf{\Theta}^{\mathrm{t}})}$, while $\sum_{n=1}^{N_\mathrm{r}}\bar{a}_{\mathrm{r},n}^*\bar{g}_{\mathrm{r},n}\mathrm{e}^
{\jmath\bar{\theta}_{n}^{\mathrm{r}}}=N_\mathrm{r}$ in $\mathbb{E}[h_\mathrm{r}(\mathbf{\Theta}^{\mathrm{r}})]$, $\mathbb{E}[|h_\mathrm{r}(\mathbf{\Theta}^{\mathrm{r}})|^2]$, $\mathrm{C}_{\hat{h}_\mathrm{r}(\mathbf{\Theta}^{\mathrm{r}})
\hat{h}_\mathrm{r}(\mathbf{\Theta}^{\mathrm{r}})}$, and $\mathrm{C}_{\check{h}_\mathrm{r}(\mathbf{\Theta}^{\mathrm{r}})
\check{h}_\mathrm{r}(\mathbf{\Theta}^{\mathrm{r}})}$.
\end{theorem}
\begin{IEEEproof}
    See Appendix \ref{Appendix_D}.
\end{IEEEproof}

\begin{figure*}[!t]
    \centering
    \subfloat[]{\begin{minipage}{0.33\linewidth}
        \centering
        \includegraphics[width=2.3in]{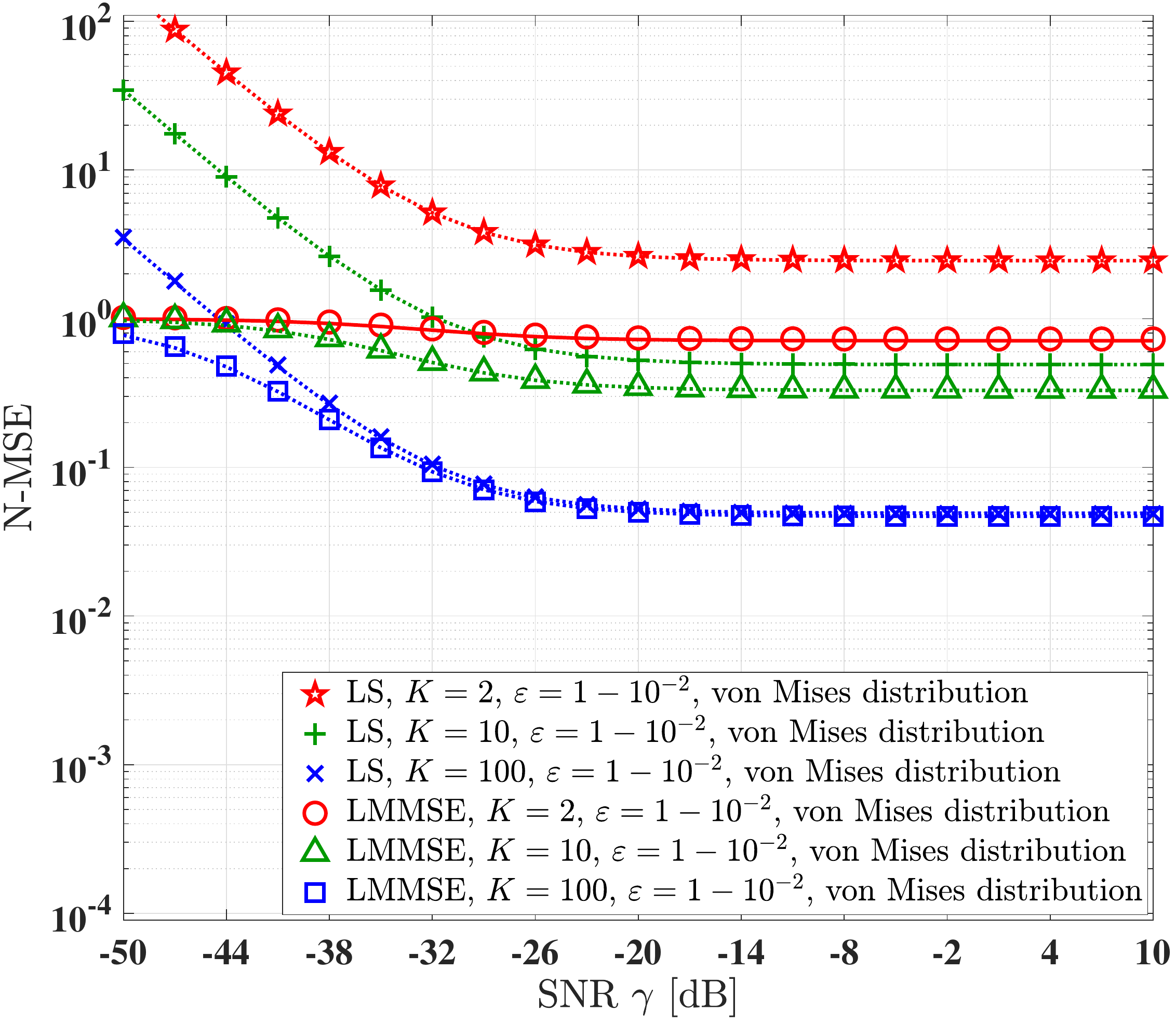}
    \end{minipage}}
    \subfloat[]{\begin{minipage}{0.33\linewidth}
        \centering
        \includegraphics[width=2.3in]{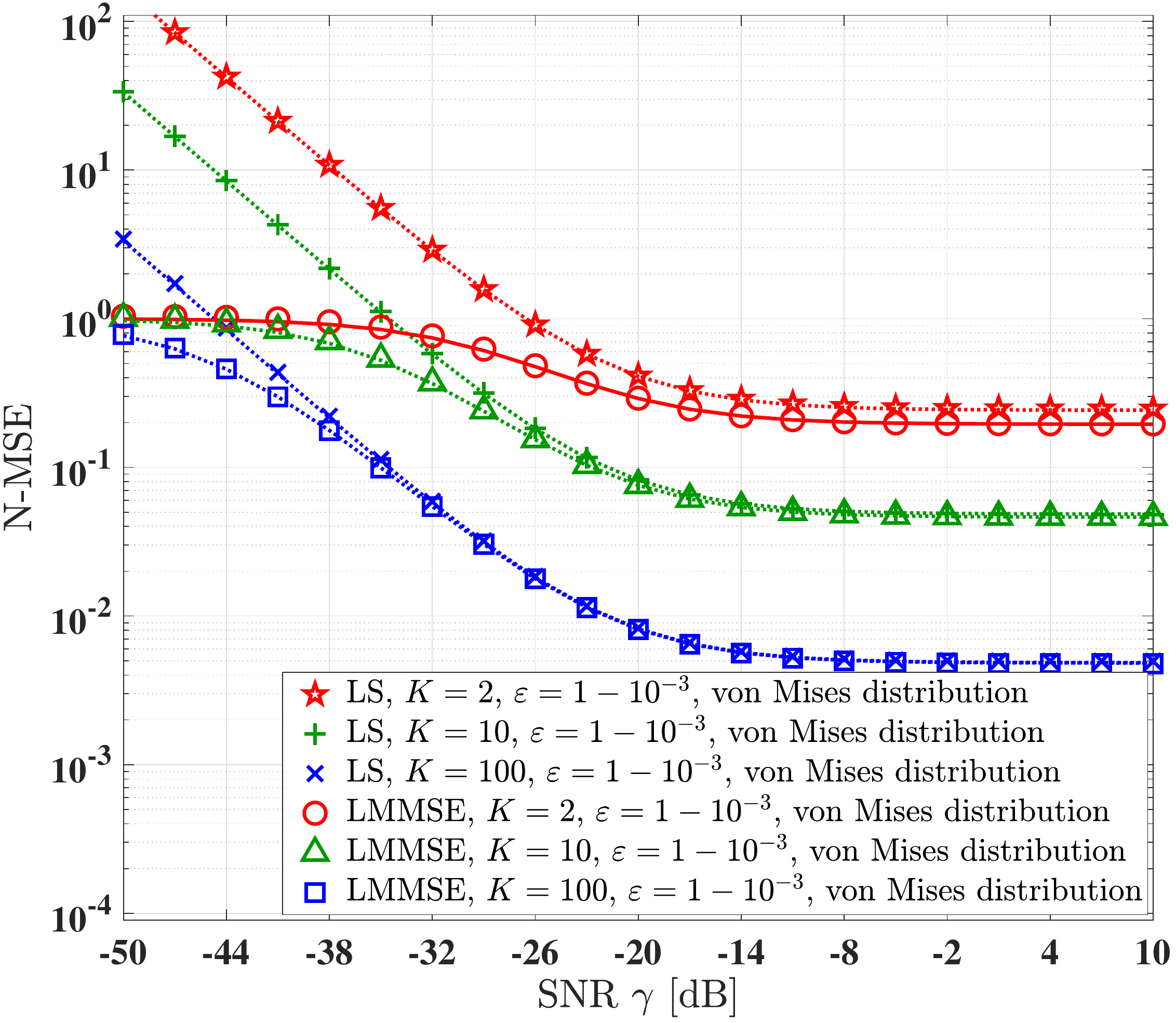}
    \end{minipage}}
    \subfloat[]{\begin{minipage}{0.33\linewidth}
        \centering
        \includegraphics[width=2.3in]{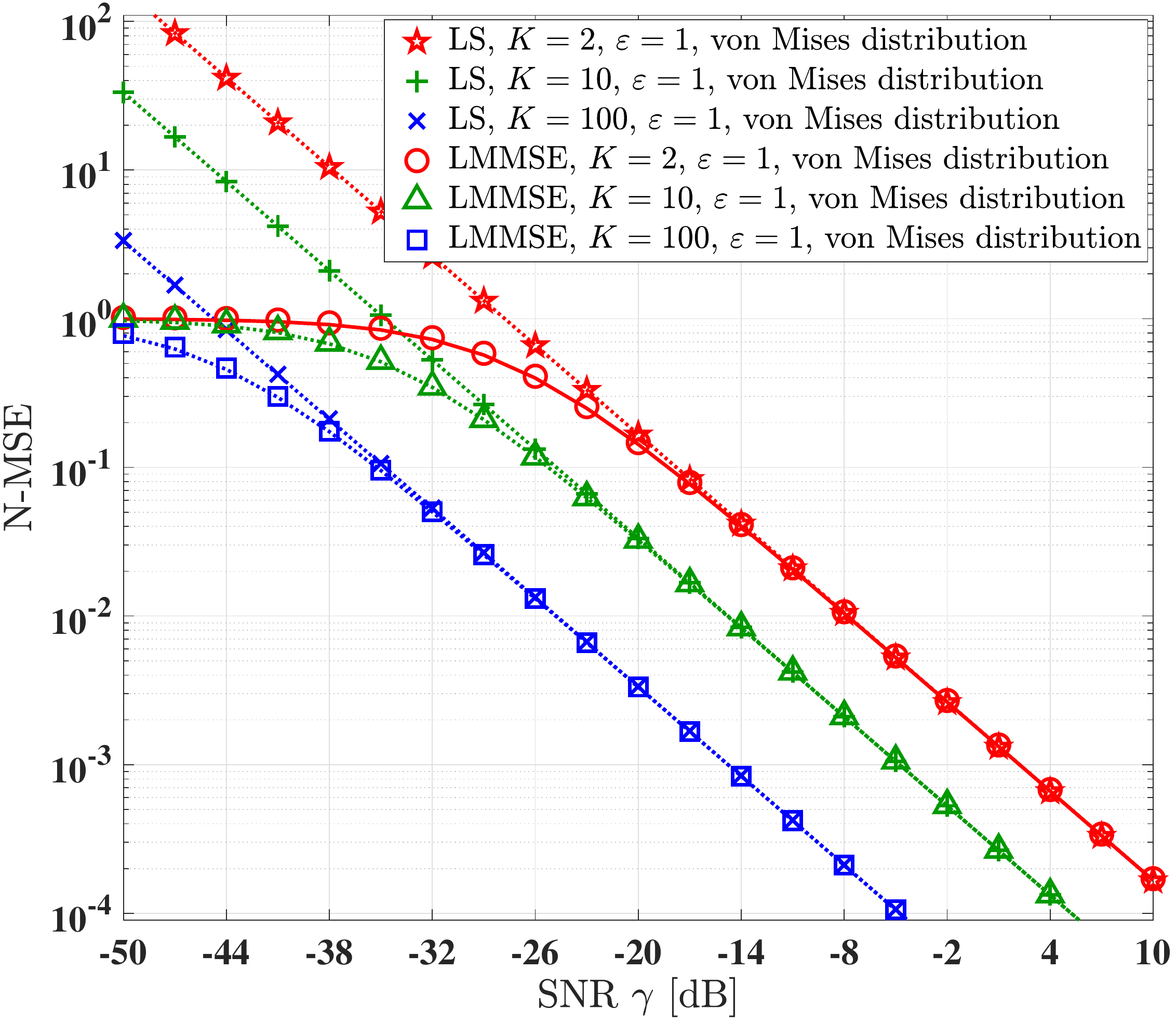}
    \end{minipage}}\\
    \subfloat[]{\begin{minipage}{0.33\linewidth}
        \centering
        \includegraphics[width=2.3in]{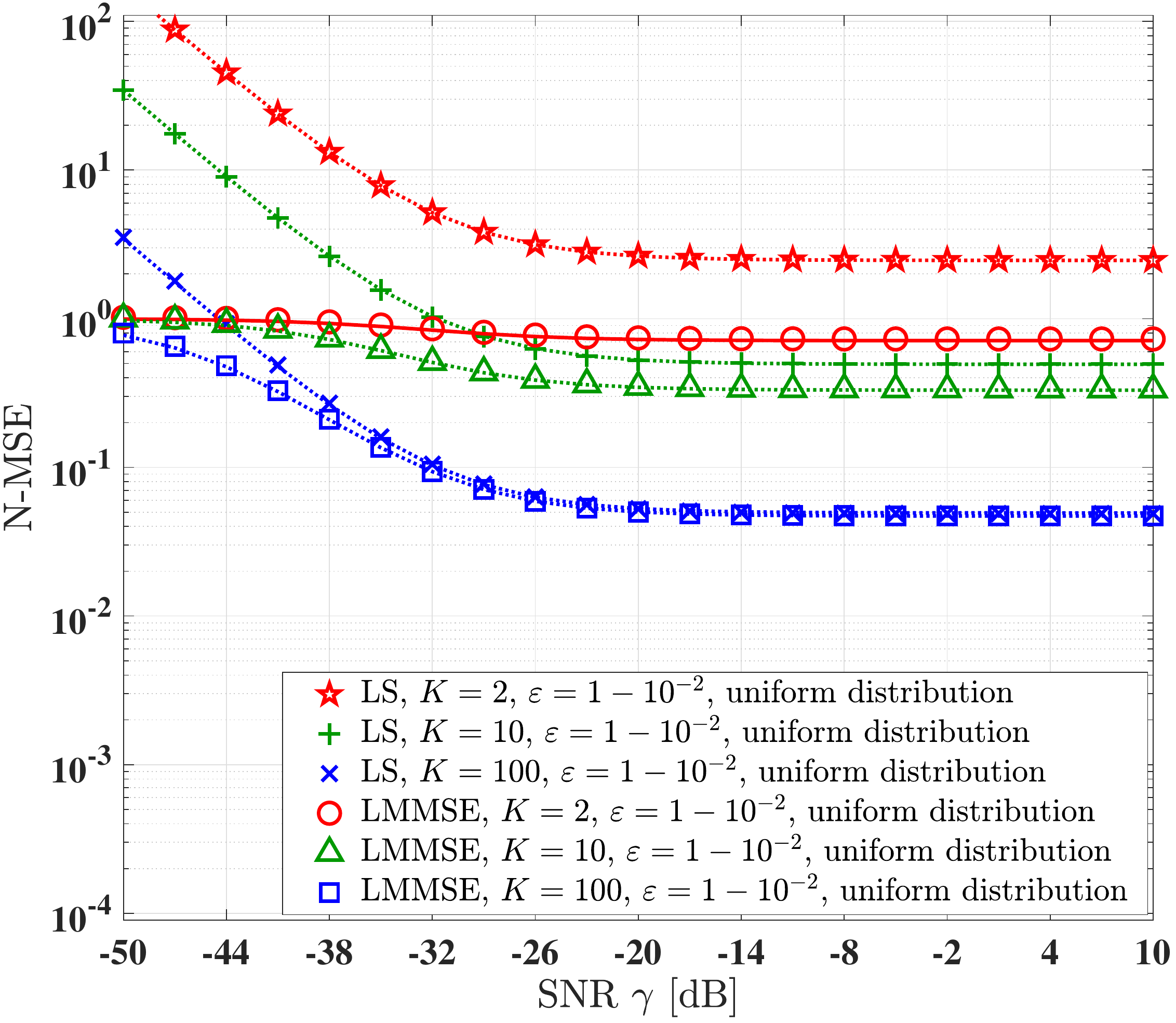}
    \end{minipage}}
    \subfloat[]{\begin{minipage}{0.33\linewidth}
        \centering
        \includegraphics[width=2.3in]{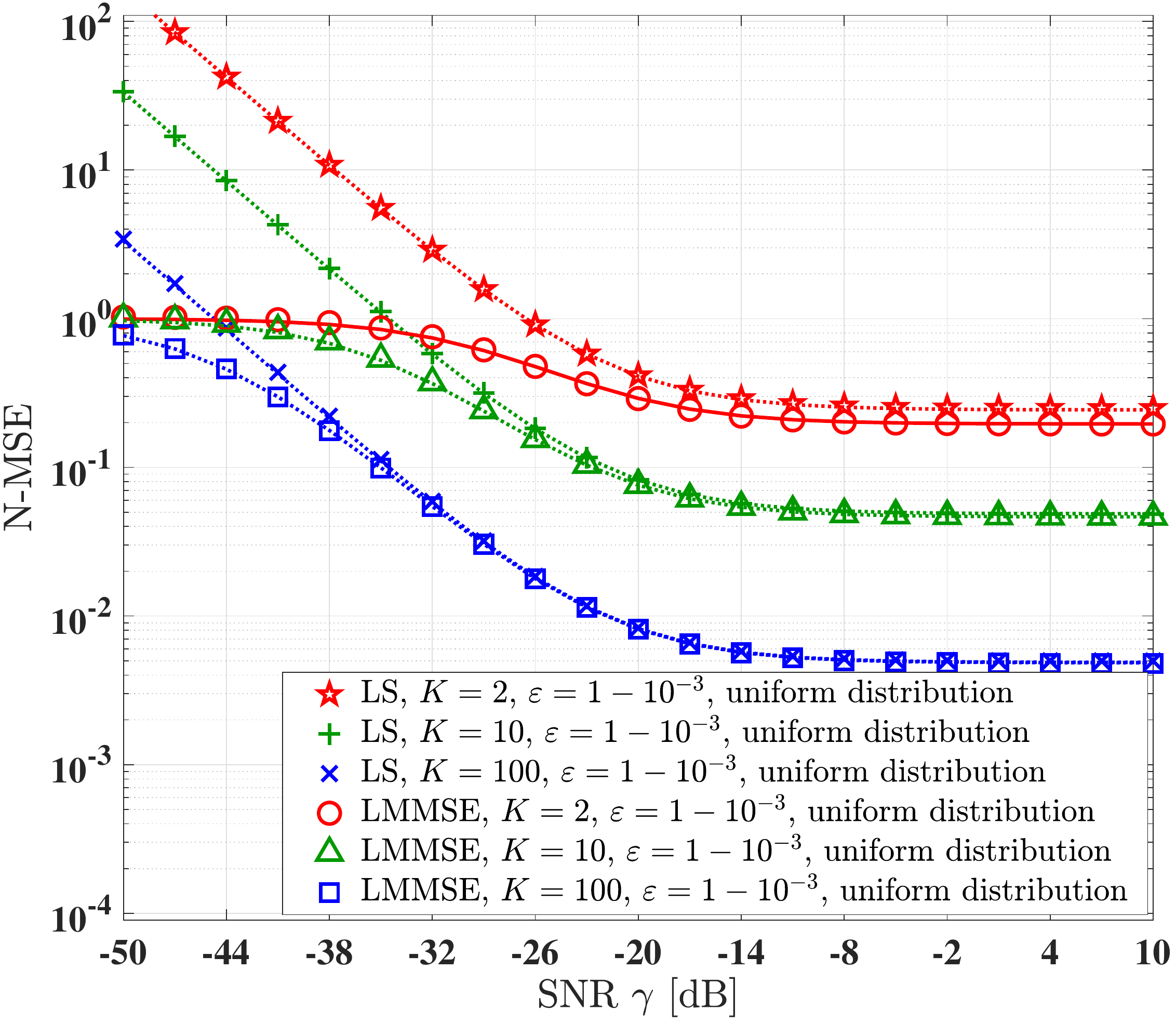}
    \end{minipage}}
    \subfloat[]{\begin{minipage}{0.33\linewidth}
        \centering
        \includegraphics[width=2.3in]{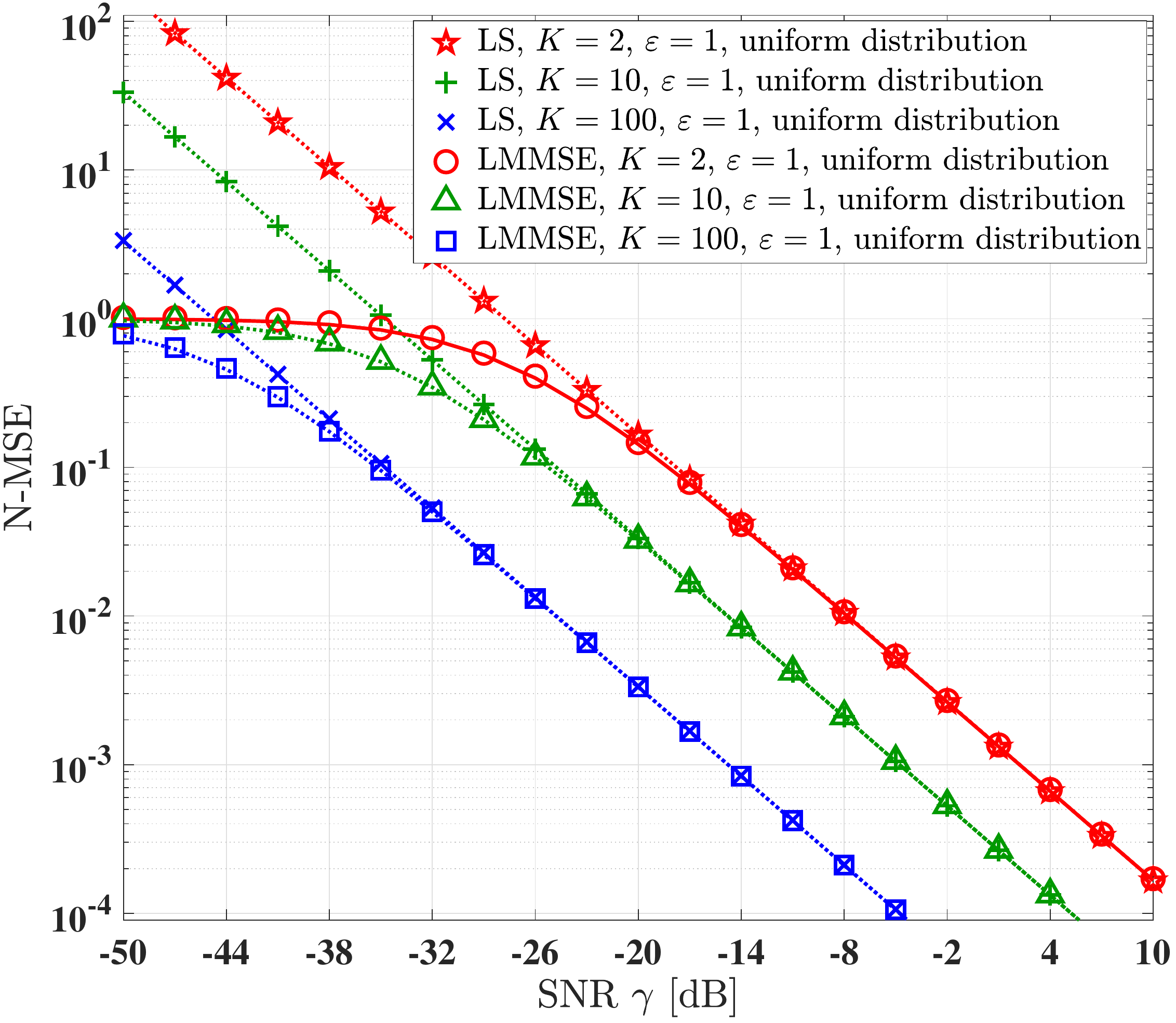}
    \end{minipage}}
    \caption{Comparison of the normalized mean square error versus the average received SNR $\gamma$ in the NOMA system with different transceiver hardware quality factors $\varepsilon$ and different pilot sequence length $K$, where the RIS phase noise follows the von Mises distribution and uniform distribution respectively, with the same power of $\sigma_\mathrm{p}^2=0.1$.}\label{Fig_simu_1}
\end{figure*}

\subsection{Effect of Imperfect SIC Decoding Algorithms}
The above analysis is based on the ideal SIC decoding algorithm. However, in practical NOMA system, the interference from the user which is decoded first cannot be completely removed for the user which is decoded later. Considering the effect of imperfect SIC decoding algorithm~\cite{mouni2021adaptive}, the instantaneous achievable rates of $R_\mathrm{t}^{\mathrm{t}\rightarrow\mathrm{r}}(\mathbf{\Theta})$ and $R_\mathrm{r}^{\mathrm{t}\rightarrow\mathrm{r}}(\mathbf{\Theta})$ are expressed as
\begin{align}\label{Imperfect_SIC_1}
    R_\mathrm{t}^{\mathrm{t}\rightarrow\mathrm{r}}(\mathbf{\Theta})
    =\log_2\Big(1+\frac{\rho_\mathrm{t}\varepsilon_v\varepsilon_{u_\mathrm{t}}
    |\hat{h}_\mathrm{t}(\mathbf{\Theta}^{\mathrm{t}})|^2}
    {\rho_\mathrm{r}\varepsilon_v\varepsilon_{u_\mathrm{r}}
    |\hat{h}_\mathrm{r}(\mathbf{\Theta}^{\mathrm{r}})|^2+\mathcal{E}}\Big),
\end{align}
\begin{align}\label{Imperfect_SIC_2}
    R_\mathrm{r}^{\mathrm{t}\rightarrow\mathrm{r}}(\mathbf{\Theta})
    =\log_2\Big(1+\frac{\rho_\mathrm{r}\varepsilon_v\varepsilon_{u_\mathrm{r}}
    |\hat{h}_\mathrm{r}(\mathbf{\Theta}^{\mathrm{r}})|^2}
    {\eta\rho_\mathrm{t}\varepsilon_v\varepsilon_{u_\mathrm{t}}
    |\hat{h}_\mathrm{t}(\mathbf{\Theta}^{\mathrm{t}})|^2+\mathcal{E}}\Big),
\end{align}
and  the instantaneous achievable rates of $R_\mathrm{t}^{\mathrm{r}\rightarrow\mathrm{t}}(\mathbf{\Theta})$ and $R_\mathrm{r}^{\mathrm{r}\rightarrow\mathrm{t}}(\mathbf{\Theta})$ are expressed as
\begin{align}\label{Imperfect_SIC_3}
    R_\mathrm{t}^{\mathrm{r}\rightarrow\mathrm{t}}(\mathbf{\Theta})
    =\log_2\Big(1+\frac{\rho_\mathrm{t}\varepsilon_v\varepsilon_{u_\mathrm{t}}
    |\hat{h}_\mathrm{t}(\mathbf{\Theta}^{\mathrm{t}})|^2}
    {\eta\rho_\mathrm{r}\varepsilon_v\varepsilon_{u_\mathrm{r}}
    |\hat{h}_\mathrm{r}(\mathbf{\Theta}^{\mathrm{r}})|^2+\mathcal{E}}\Big),
\end{align}
\begin{align}\label{Imperfect_SIC_4}
    R_\mathrm{r}^{\mathrm{r}\rightarrow\mathrm{t}}(\mathbf{\Theta})
    =\log_2\Big(1+\frac{\rho_\mathrm{r}\varepsilon_v\varepsilon_{u_\mathrm{r}}
    |\hat{h}_\mathrm{r}(\mathbf{\Theta}^{\mathrm{r}})|^2}
    {\rho_\mathrm{t}\varepsilon_v\varepsilon_{u_\mathrm{t}}
    |\hat{h}_\mathrm{t}(\mathbf{\Theta}^{\mathrm{t}})|^2+\mathcal{E}}\Big),
\end{align}
where $\eta\in[0,1]$ represents the SIC imperfection coefficient resulting from implementation issues such as complexity scaling and error propagation~\cite{do2019impacts}. A value of $\eta=0$ implies that the interference can be completely removed for the following user information recovery, i.e. perfect SIC.

According to (\ref{Imperfect_SIC_1}), (\ref{Imperfect_SIC_2}), (\ref{Imperfect_SIC_3}) and (\ref{Imperfect_SIC_4}), the instantaneous sum-rate of the STAR-RIS aided NOMA uplink is given by
\begin{align}\label{Imperfect_SIC_5}
    \notag&{R}_\text{sum}^{\text{NOMA}}(\mathbf{\Theta})\\
    \notag=&R_\mathrm{t}^{\text{NOMA}}(\mathbf{\Theta})
    +R_\mathrm{r}^{\text{NOMA}}(\mathbf{\Theta})\\
    \notag=&\beta\big({R}_\mathrm{t}^{\mathrm{t}\rightarrow\mathrm{r}}(\mathbf{\Theta})
    +{R}_\mathrm{r}^{\mathrm{t}\rightarrow\mathrm{r}}(\mathbf{\Theta})\big)
    +(1-\beta)\big({R}_\mathrm{t}^{\mathrm{r}\rightarrow\mathrm{t}}(\mathbf{\Theta})+
    {R}_\mathrm{r}^{\mathrm{r}\rightarrow\mathrm{t}}(\mathbf{\Theta})\big)\\
    \notag=&\log_2\Big(1+\frac{\rho_\mathrm{t}\varepsilon_v\varepsilon_{u_\mathrm{t}}
    |\hat{h}_\mathrm{t}(\mathbf{\Theta}^{\mathrm{t}})|^2
    +\rho_\mathrm{r}\varepsilon_v\varepsilon_{u_\mathrm{r}}
    |\hat{h}_\mathrm{r}(\mathbf{\Theta}^{\mathrm{r}})|^2}
    {\mathcal{E}}\Big)\\
    &-\Delta{R}(\mathbf{\Theta}),
\end{align}
where $\Delta{R}(\mathbf{\Theta})$ is the sum-rate degradation imposed by the imperfect SIC algorithm compared to that of the perfect SIC algorithm in (\ref{NOMA_7_2}), given by
\begin{align}\label{Imperfect_SIC_6}
    \notag\Delta{R}(\mathbf{\Theta})
    =&\beta\Big(\log_2\Big(1+\frac{\eta\rho_\mathrm{t}\varepsilon_v\varepsilon_{u_\mathrm{t}}
    |\hat{h}_\mathrm{t}(\mathbf{\Theta}^{\mathrm{t}})|^2}
    {\mathcal{E}}\Big)\\
    \notag&-\log_2\Big(1+\frac{\eta\rho_\mathrm{t}\varepsilon_v\varepsilon_{u_\mathrm{t}}
    |\hat{h}_\mathrm{t}(\mathbf{\Theta}^{\mathrm{t}})|^2}
    {\rho_\mathrm{r}\varepsilon_v\varepsilon_{u_\mathrm{r}}
    |\hat{h}_\mathrm{r}(\mathbf{\Theta}^{\mathrm{r}})|^2+\mathcal{E}}\Big)\Big)\\
    \notag&+(1-\beta)\Big(\log_2\Big(1+\frac{\eta\rho_\mathrm{r}\varepsilon_v\varepsilon_{u_\mathrm{r}}
    |\hat{h}_\mathrm{r}(\mathbf{\Theta}^{\mathrm{r}})|^2}{\mathcal{E}}\Big)\\
    &-\log_2\Big(1+\frac{\eta\rho_\mathrm{r}\varepsilon_v\varepsilon_{u_\mathrm{r}}
    |\hat{h}_\mathrm{r}(\mathbf{\Theta}^{\mathrm{r}})|^2}
    {\rho_\mathrm{t}\varepsilon_v\varepsilon_{u_\mathrm{t}}
    |\hat{h}_\mathrm{t}(\mathbf{\Theta}^{\mathrm{t}})|^2+\mathcal{E}}\Big)\Big).
\end{align}
According to (49) and (50), it can be shown that in contrast to the perfect SIC algorithm, the achievable sum-rate of the imperfect SIC algorithm is determined by the decoding order fraction $\beta$.

\begin{theorem}\label{theorem_5}
To maximize the achievable sum-rate, the decoding order fraction $\beta$ is designed as follows:
\begin{align}\label{Imperfect_SIC_7}
    \beta=\left\{
            \begin{array}{ll}
              0, & \rho_\mathrm{t}\varepsilon_{u_\mathrm{t}}
                    |\hat{h}_\mathrm{t}(\mathbf{\Theta}^{\mathrm{t}})|^2
                   >\rho_\mathrm{r}\varepsilon_{u_\mathrm{r}}
                    |\hat{h}_\mathrm{r}(\mathbf{\Theta}^{\mathrm{r}})|^2 \\
              1, & \rho_\mathrm{t}\varepsilon_{u_\mathrm{t}}
                    |\hat{h}_\mathrm{t}(\mathbf{\Theta}^{\mathrm{t}})|^2
                   <\rho_\mathrm{r}\varepsilon_{u_\mathrm{r}}
                    |\hat{h}_\mathrm{r}(\mathbf{\Theta}^{\mathrm{r}})|^2
            \end{array}
          \right..
\end{align}
\end{theorem}
\begin{IEEEproof}
    See Appendix \ref{Appendix_E}.
\end{IEEEproof}

\section{Numerical and Simulation Results}\label{Numerical_and_Simulation_Results}
In this section, the theoretical and simulation results of the N-MSE and the achievable rate are presented. Unless otherwise specified, the simulation parameters are given in Table \ref{Table_Simulation}, the transceiver hardware quality factors satisfies $\varepsilon=\varepsilon_v=\varepsilon_\mathrm{t}=\varepsilon_\mathrm{r}$, the number of RIS elements satisfies $N=N_\mathrm{t}=N_\mathrm{r}$, the pilot transmit power and the UE transmit power follow $P_\mathrm{t}=P_\mathrm{r}=\rho_\mathrm{t}=\rho_\mathrm{r}$. For channel estimation, the average received signal-to-noise-ratio (SNR) is defined as $\gamma=\frac{P_\mathrm{t}\varrho_a\varrho_\mathrm{t}}{\sigma_w^2}
=\frac{P_\mathrm{r}\varrho_a\varrho_\mathrm{r}}{\sigma_w^2}$ since $P_\mathrm{t}=P_\mathrm{r}$ and $\varrho_\mathrm{t}=\varrho_\mathrm{r}$. For achievable rate, the average received SNR is defined as $\gamma=\frac{\rho_\mathrm{t}\varrho_a\varrho_\mathrm{t}}{\sigma_w^2}
=\frac{\rho_\mathrm{r}\varrho_a\varrho_\mathrm{r}}{\sigma_w^2}$ since $\rho_\mathrm{t}=\rho_\mathrm{r}$ and $\varrho_\mathrm{t}=\varrho_\mathrm{r}$. We utilize lines, e.g. `$--$' and `$-\cdot-\cdot$', to represent theoretical analysis, and utilize markers, e.g. `$\square$', and `$\blacklozenge$', to represent simulation results.

\begin{table}[!b]
\vspace{-0em}
\footnotesize
\setlength{\abovecaptionskip}{1em}
\setlength{\belowcaptionskip}{0em}
\begin{center}
\caption{Simulation Parameters.}\label{Table_Simulation}
\begin{tabular}{*{2}{l}}
\toprule
    \textbf{Parameters} & \textbf{Values} \\
\midrule
\midrule
    Cartesian coordinates of the AP & (0m, -80m, 20m) \\
\midrule
    Cartesian coordinates of the UE-T & (0m, 20m, 5m) \\
\midrule
    Cartesian coordinates of the UE-R & (0m, -20m, 5m) \\
\midrule
    Cartesian coordinates of the RIS & (0m, 0m, 15m) \\
\midrule
    Rician factors between the RIS and the AP & $\kappa_a=0\text{dB}$\\
\midrule
    Rician factors between the UEs and the RIS & $\kappa_\text{t}=\kappa_\text{r}=0\text{dB}$\\
\midrule
    Path loss at the reference distance of 1 meter & $\varrho_0=-30\text{dB}$ \\
\midrule
    Path loss exponent between the RIS and the AP & $\alpha_a=2.4$ \\
\midrule
    Path loss exponent between the UEs and the RIS & $\alpha_\mathrm{t}=\alpha_\mathrm{r}=2.542$ \\
\midrule
    Noise power & $\sigma_w^2=-100\text{dBm}$ \\
\midrule
    Distance between adjacent RIS elements & $\delta_x=\delta_y=\frac{\lambda}{2}$\\
\midrule
    Number of RIS elements & $N_\mathrm{t}=N_\mathrm{r}=20\times20$\\
\bottomrule
\end{tabular}
\end{center}
\vspace{-0em}
\end{table}

\begin{figure}[!t]
    \centering
    \includegraphics[width=2.3in]{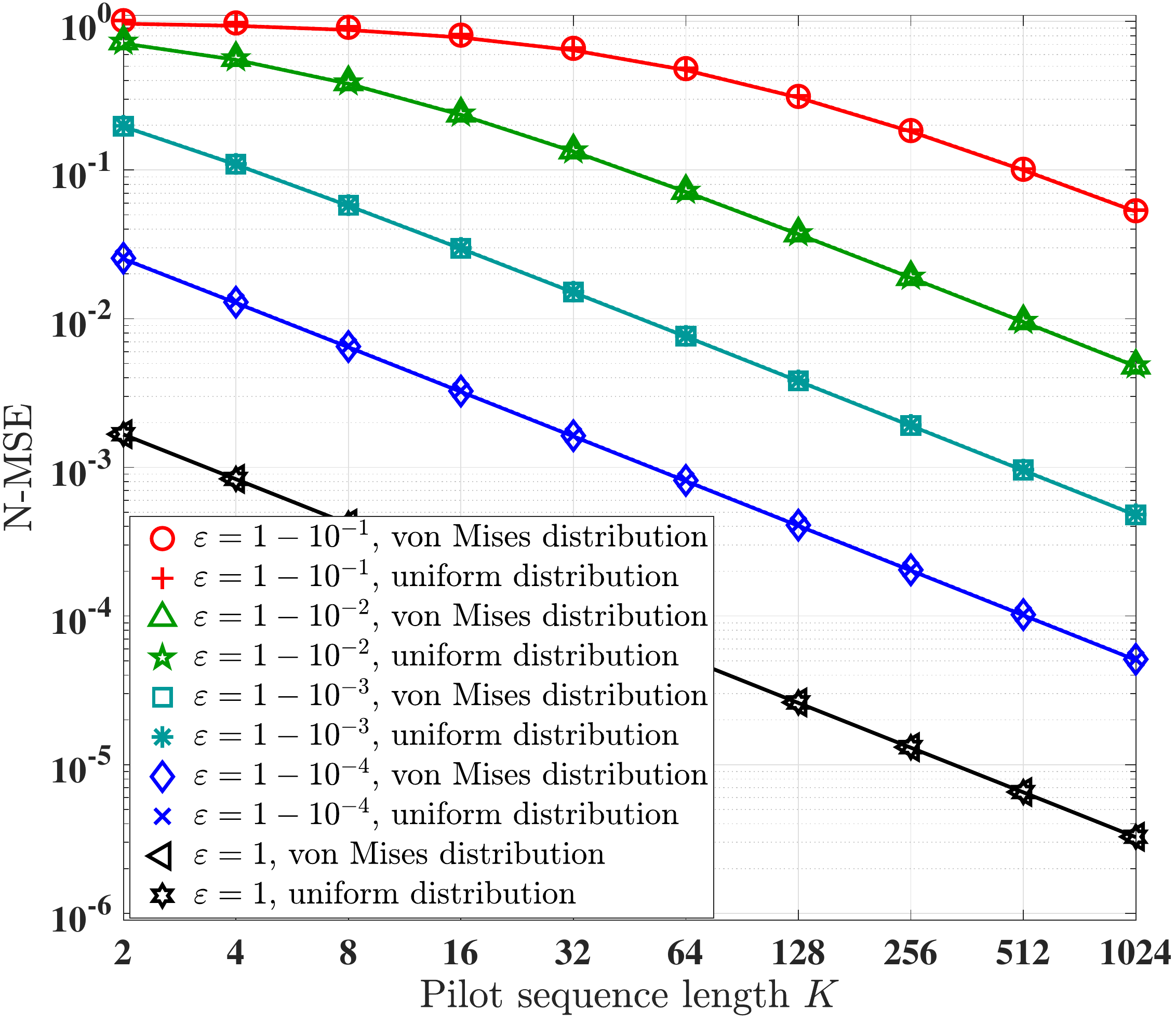}
    \caption{Comparison of the normalized mean square error based on the LMMSE scheme versus the pilot sequence length $K$ in the NOMA system with different transceiver hardware quality factors $\varepsilon$.}\label{Fig_simu_2}
\end{figure}

\subsection{Channel Estimation Performance}
Fig. \ref{Fig_simu_1} compares the normalized mean square error versus the average received SNR $\gamma$ in the NOMA system, while employing the LMMSE estimator based on (\ref{Channel_Estimation_15}) and the LS estimator for different transceiver hardware quality factors $\varepsilon$ and different pilot sequence lengths $K$, where the RIS phase noise follows the von Mises distribution and uniform distribution respectively, with the same power of $\sigma_\mathrm{p}^2=0.1$. It shows that when the hardware quality factor $\varepsilon<1$, the N-MSE tends to a constant value even when $\gamma\rightarrow\infty$, which can be illustrated in (\ref{Channel_Estimation_17}). Furthermore, increasing the pilot sequence length $K$ effectively improves the N-MSE performance. Almost the same N-MSE performance can be achieved, when the RIS phase noise follows the von Mises distribution and uniform distribution with the identical phase noise power. Furthermore, for any values of hardware quality factors, the LMMSE estimator achieves better N-MSE performance than the LS estimator in the low SNR region due to the employment of the first-order and second-order statistical information of the links.

Fig. \ref{Fig_simu_2} shows the normalized mean square error performance of the STAR-RIS assisted NOMA uplink, when employing the LMMSE scheme versus the pilot sequence length $K$, while considering different transceiver hardware quality factors $\varepsilon$. In these simulation results, we consider the average received SNR $\gamma=0\mathrm{dB}$, and the RIS phase noise power $\sigma_\mathrm{p}^2=0.1$ following both the von Mises distribution and the uniform distribution. Our results show that the N-MSE tends to 0 with the increase of the pilot sequence length $K$. Furthermore, the N-MSE performance degradation caused by hardware impairments can be compensated upon increasing the pilot sequence length.

\begin{figure*}[!t]
    \centering
    \subfloat[]{\begin{minipage}{0.33\linewidth}
        \centering
        \includegraphics[width=2.3in]{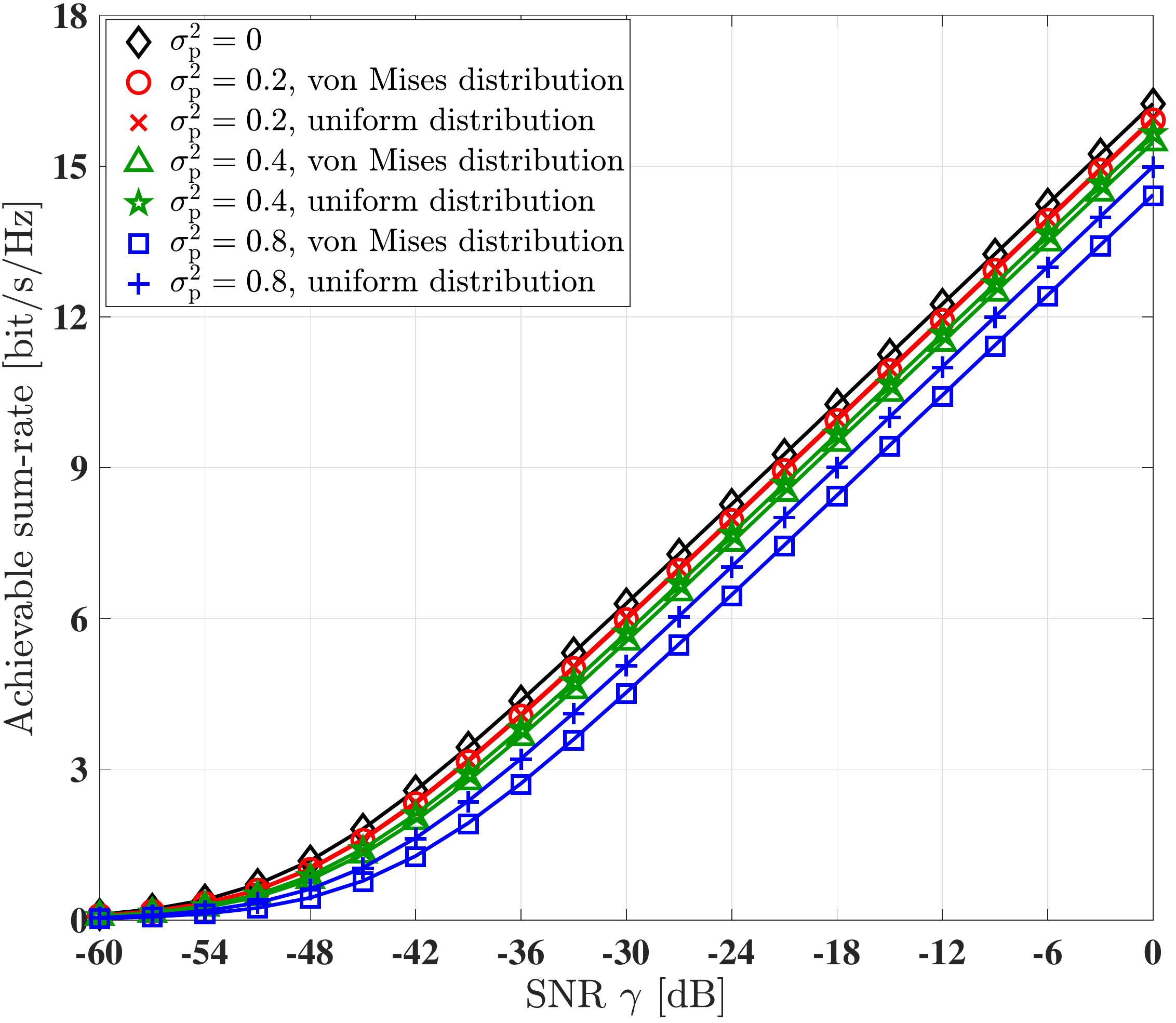}
    \end{minipage}}
    \subfloat[]{\begin{minipage}{0.33\linewidth}
        \centering
        \includegraphics[width=2.3in]{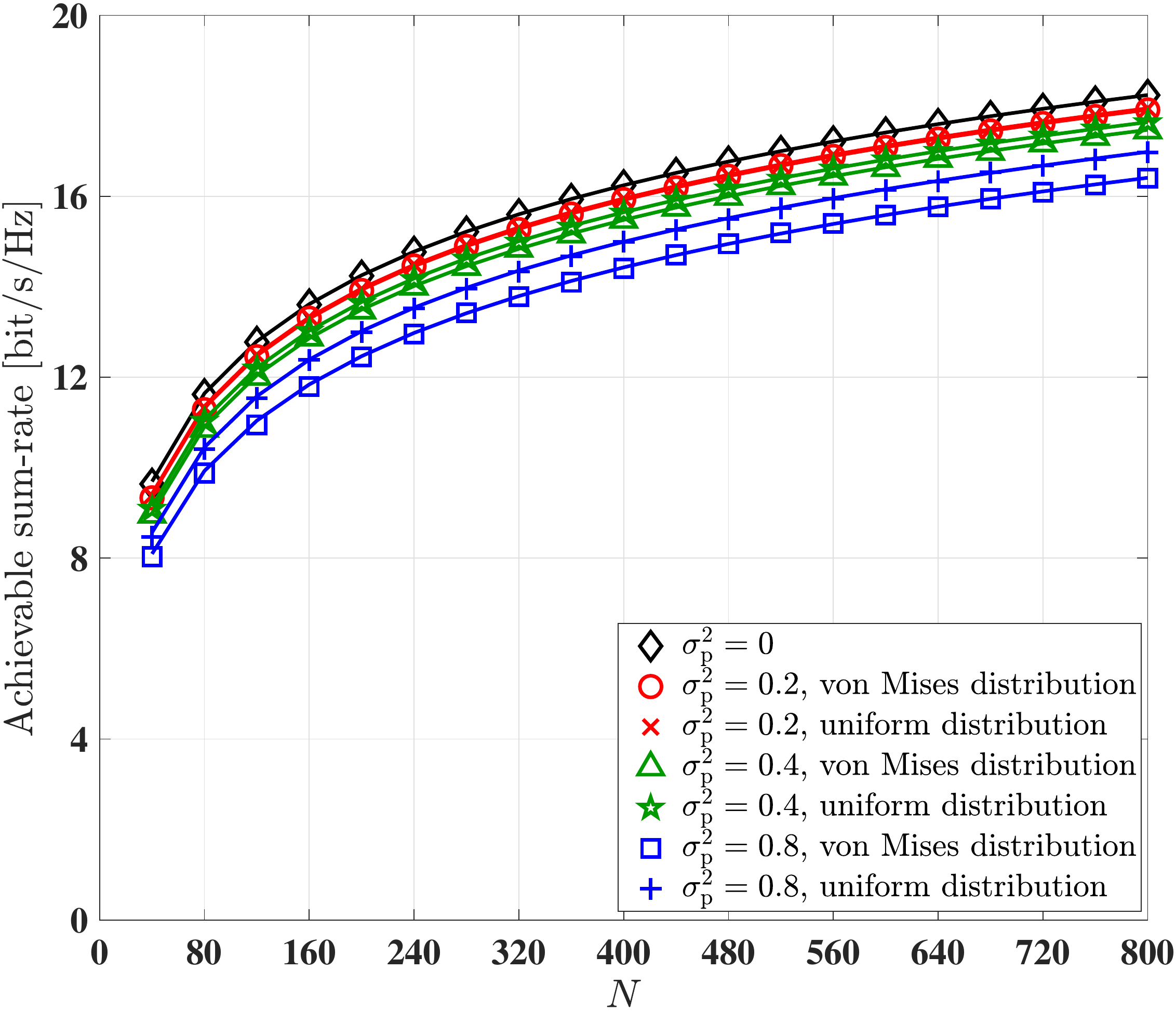}
    \end{minipage}}
    \subfloat[]{\begin{minipage}{0.33\linewidth}
        \centering
        \includegraphics[width=2.3in]{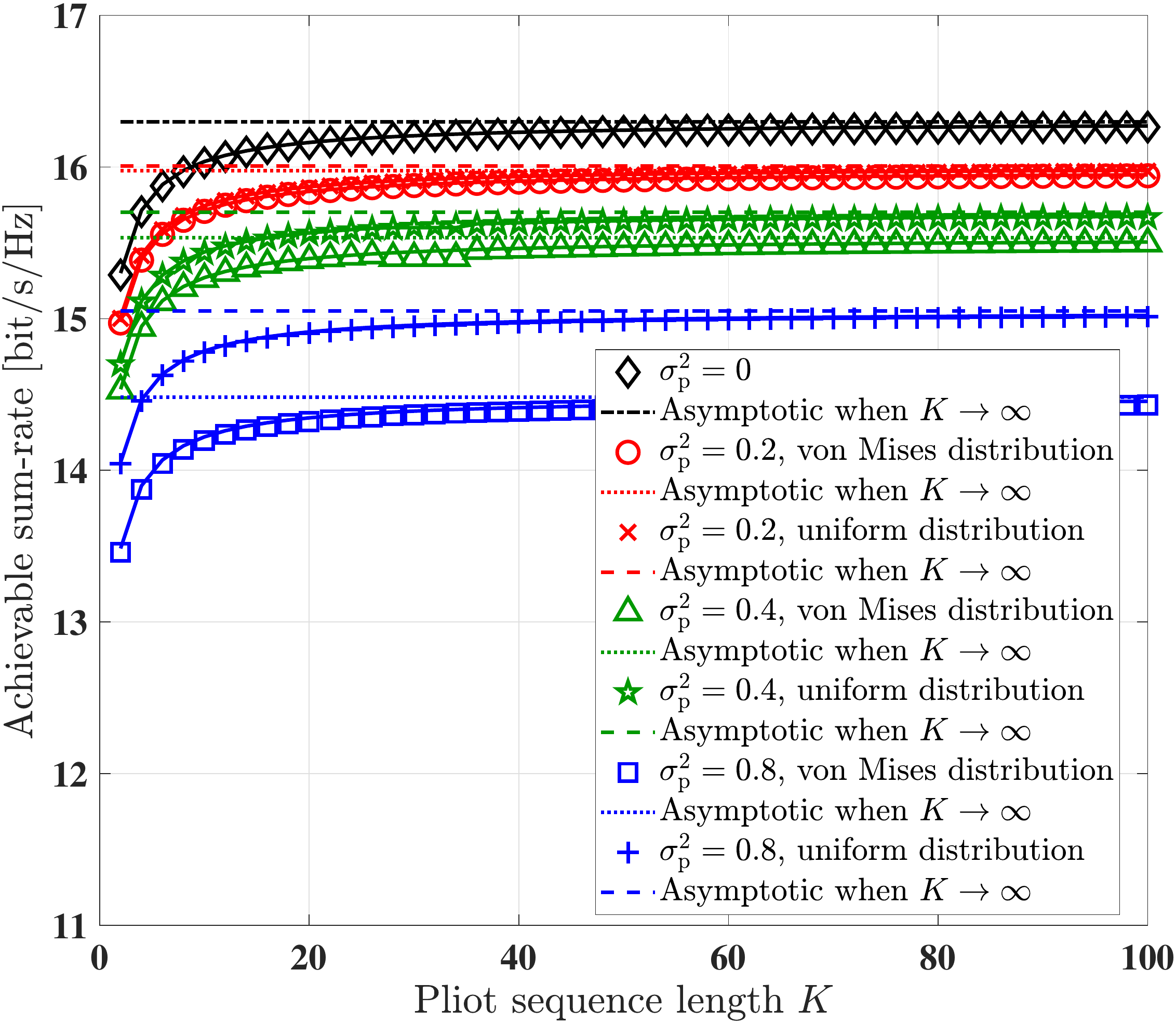}
    \end{minipage}}
    \caption{Theoretical analysis and simulation results of the ergodic sum-rate in the NOMA system with different RIS phase noise power $\sigma_\mathrm{p}^2$.}\label{Fig_simu_4}
\end{figure*}

\subsection{Ergodic Rate Performance}
\begin{figure}[!t]
    \centering
    \subfloat[]{\begin{minipage}{1\linewidth}
        \centering
        \includegraphics[width=2.3in]{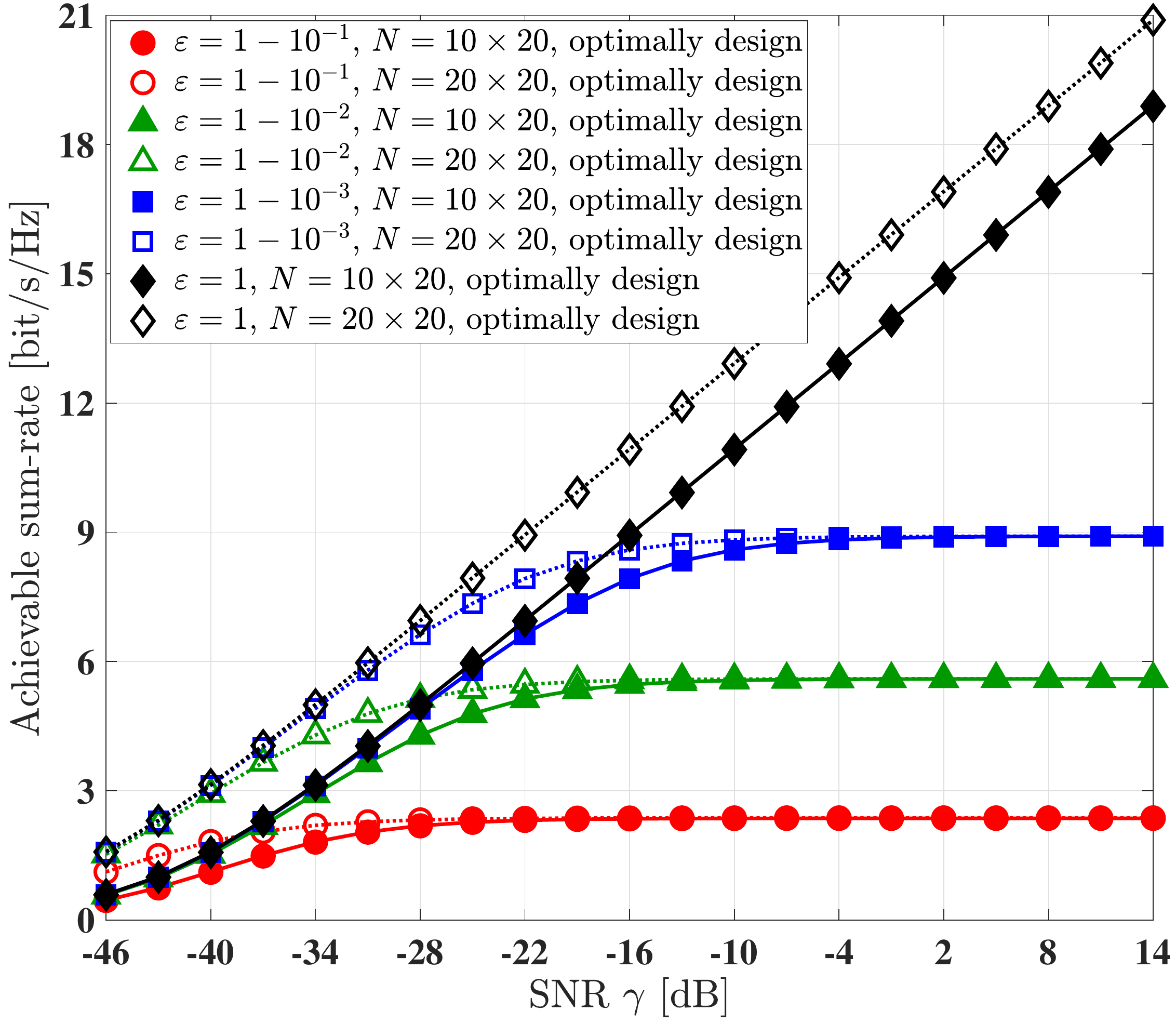}
    \end{minipage}}\\
    \subfloat[]{\begin{minipage}{1\linewidth}
        \centering
        \includegraphics[width=2.3in]{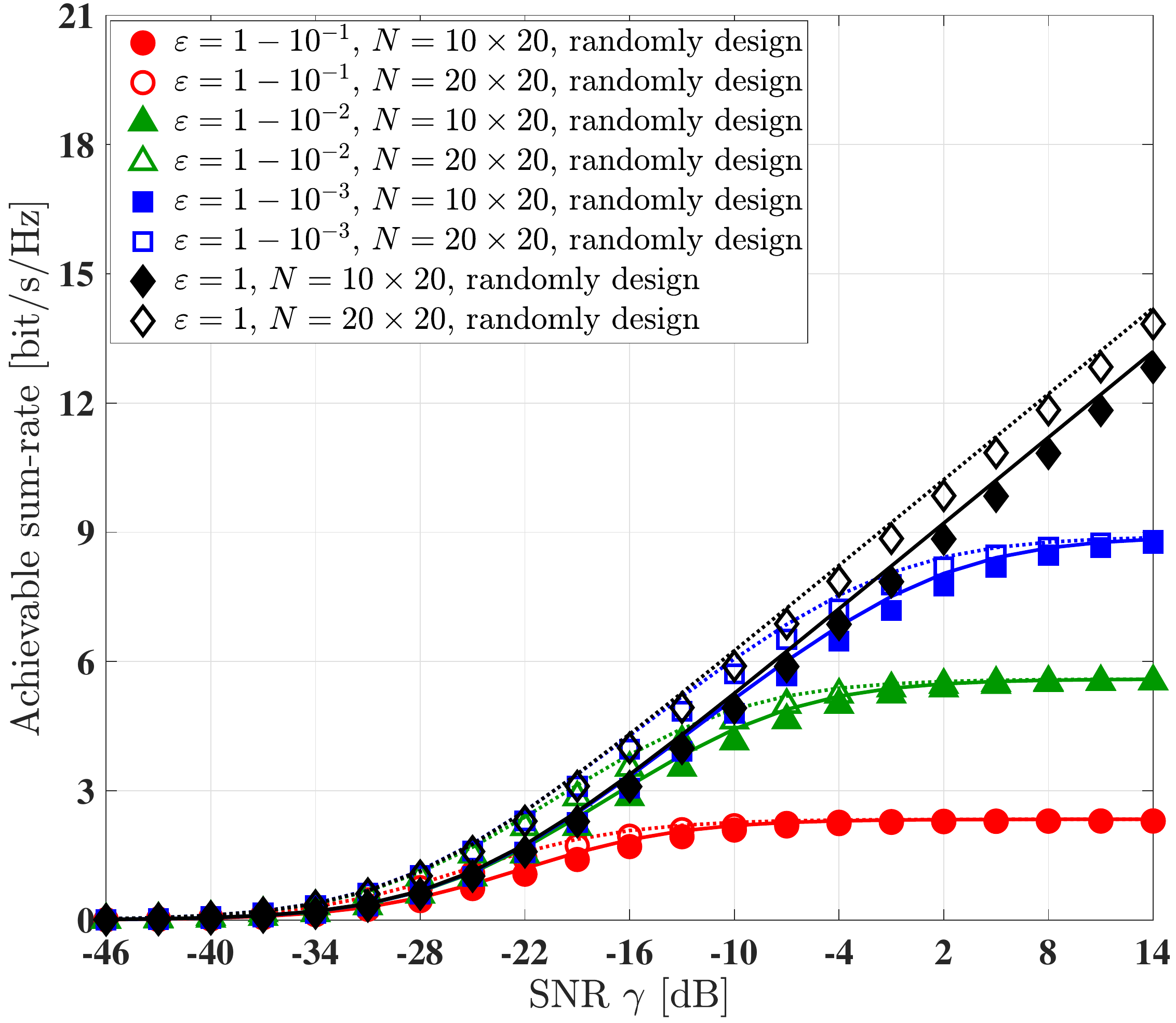}
    \end{minipage}}
    \caption{Comparison of the ergodic sum-rate versus the average received SNR $\gamma$ in the NOMA system with different transceiver hardware quality factors $\varepsilon$ and different number of RIS elements with $N=N_\mathrm{t}=N_\mathrm{r}$, where the RIS phase shift is optimally designed and randomly designed, respectively.}\label{Fig_simu_5}
\end{figure}

\begin{figure*}[!t]
    \centering
    \subfloat[]{\begin{minipage}{0.33\linewidth}
        \centering
        \includegraphics[width=2.3in]{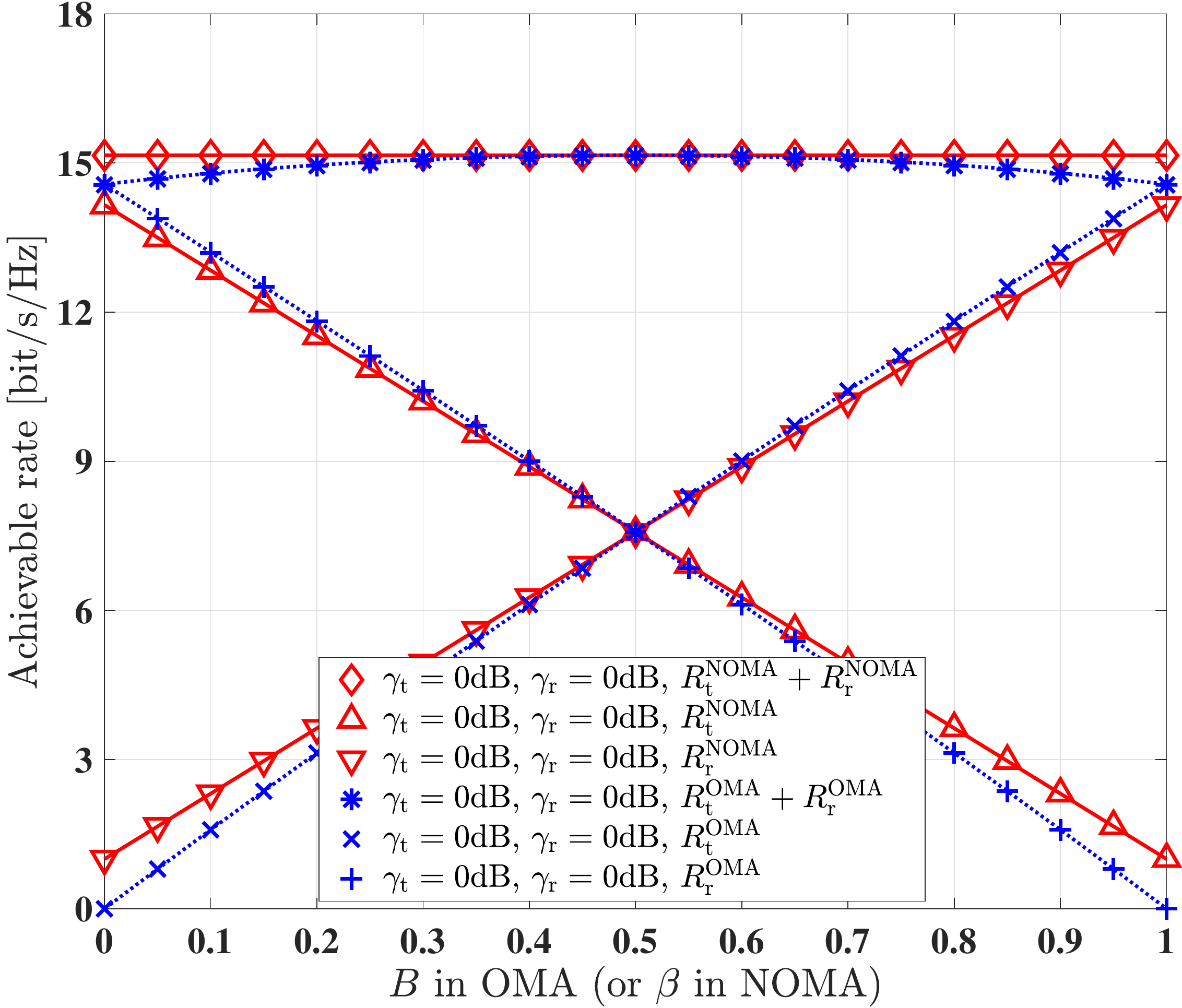}
    \end{minipage}}
    \subfloat[]{\begin{minipage}{0.33\linewidth}
        \centering
        \includegraphics[width=2.3in]{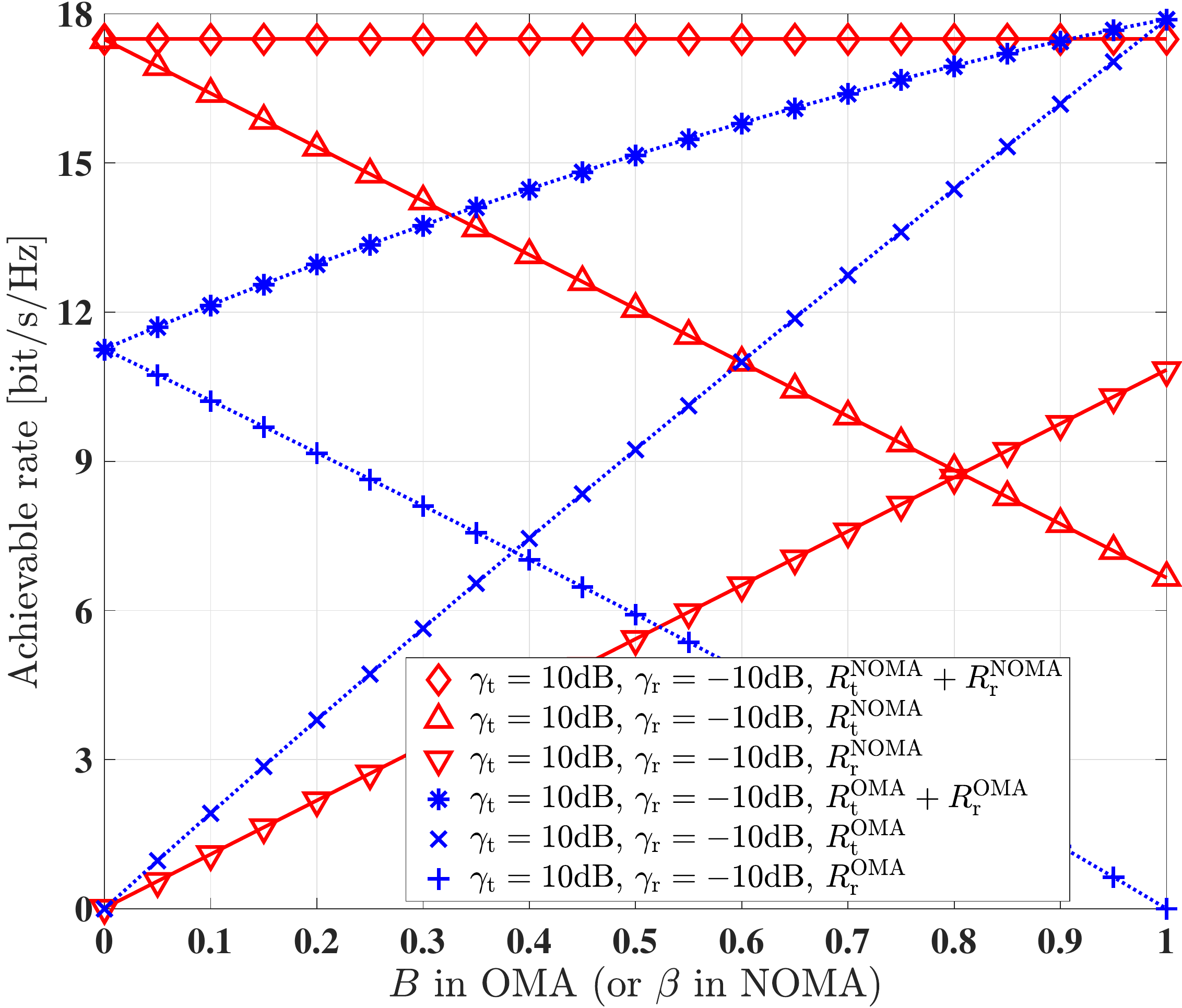}
    \end{minipage}}
    \subfloat[]{\begin{minipage}{0.33\linewidth}
        \centering
        \includegraphics[width=2.3in]{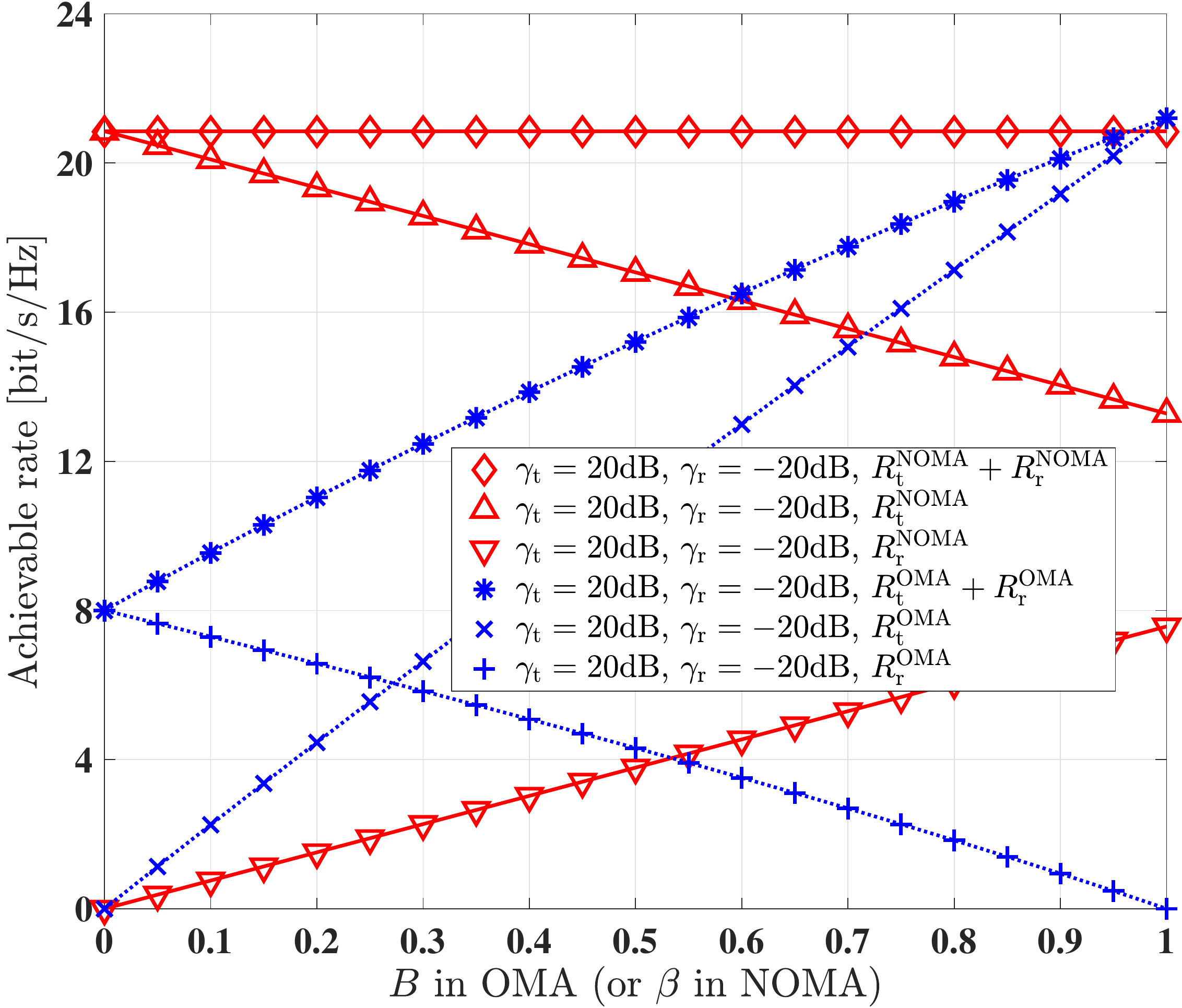}
    \end{minipage}}
    \caption{Comparison of the achievable rate versus the time/frequency fraction $B$ in the OMA scheme or the decoding order fraction $\beta$ in the NOMA system with different received SNR.}\label{Fig_simu_6}
\end{figure*}

\begin{figure*}[!t]
    \centering
    \subfloat[]{\begin{minipage}{0.33\linewidth}
        \centering
        \includegraphics[width=2.3in]{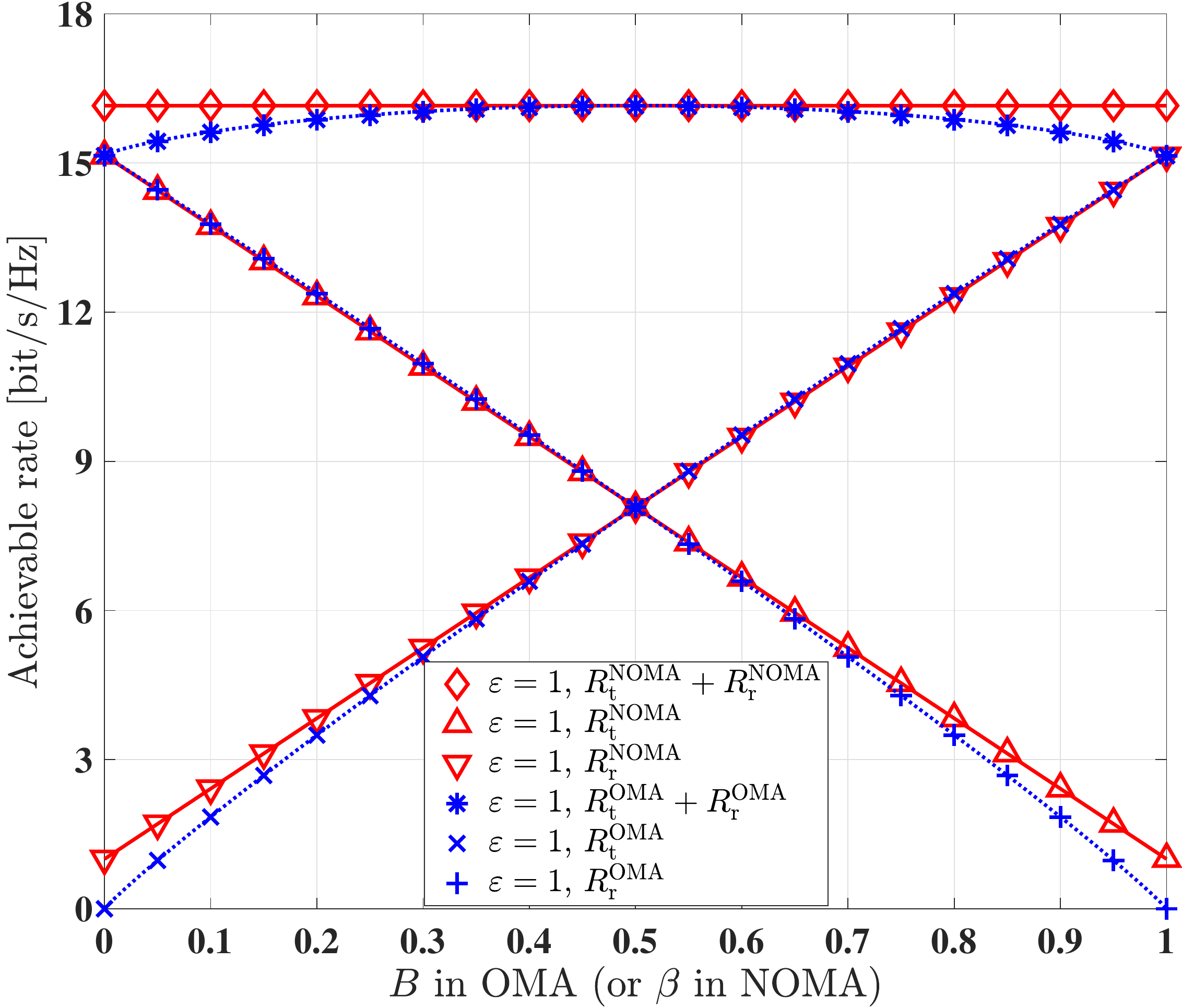}
    \end{minipage}}
    \subfloat[]{\begin{minipage}{0.33\linewidth}
        \centering
        \includegraphics[width=2.3in]{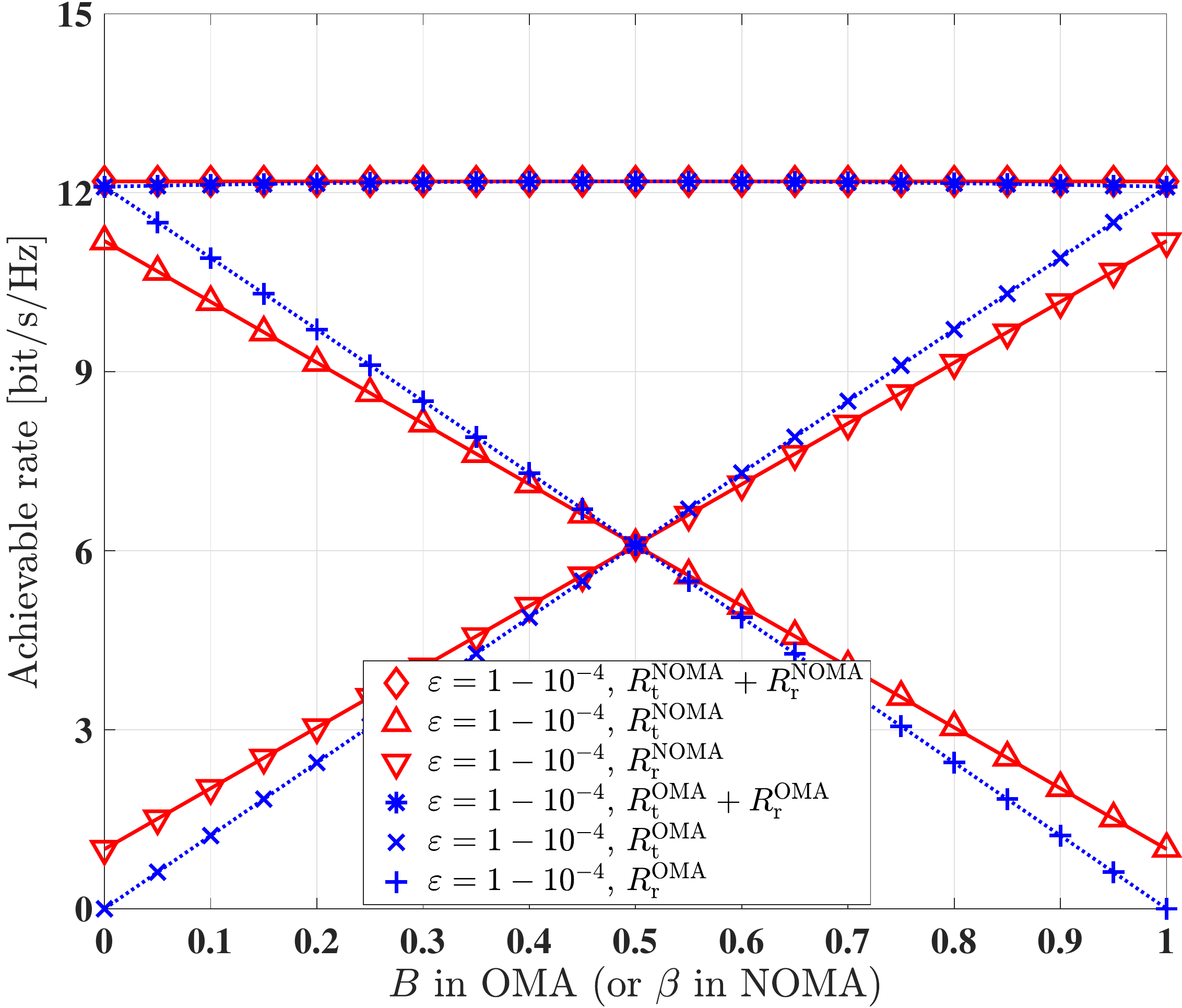}
    \end{minipage}}
    \subfloat[]{\begin{minipage}{0.33\linewidth}
        \centering
        \includegraphics[width=2.3in]{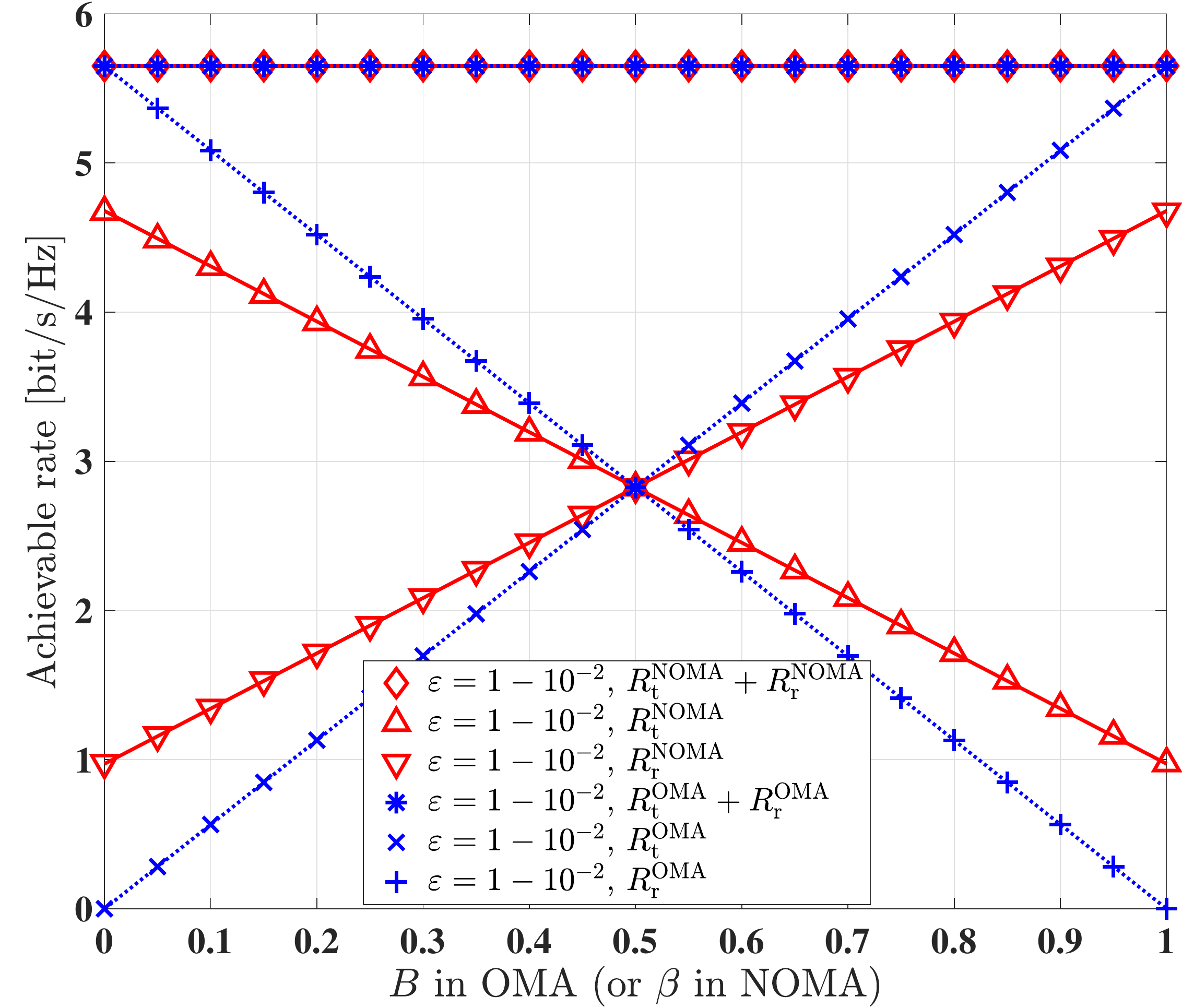}
    \end{minipage}}
    \caption{Comparison of the achievable rate versus the time/frequency fraction $B$ in the OMA scheme or the decoding order fraction $\beta$ in the NOMA system with different transceiver hardware qualities, where the received SNR of UE-T and UE-R are $\gamma_\mathrm{t}=0\mathrm{dB}$ and $\gamma_\mathrm{r}=0\mathrm{dB}$, respectively.}\label{Fig_simu_7}
\end{figure*}

\begin{figure*}[!t]
    \centering
    \subfloat[]{\begin{minipage}{0.33\linewidth}
        \centering
        \includegraphics[width=2.3in]{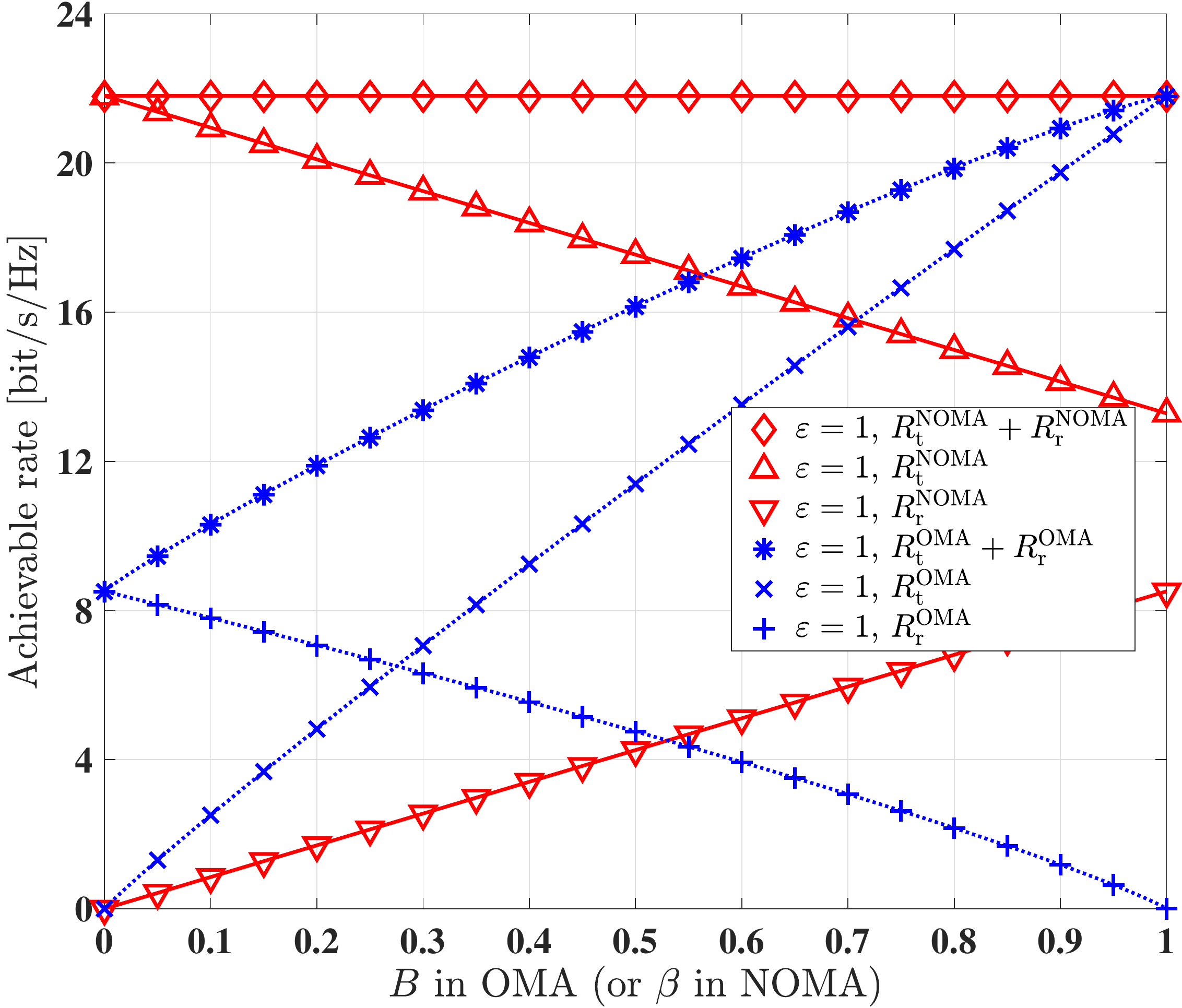}
    \end{minipage}}
    \subfloat[]{\begin{minipage}{0.33\linewidth}
        \centering
        \includegraphics[width=2.3in]{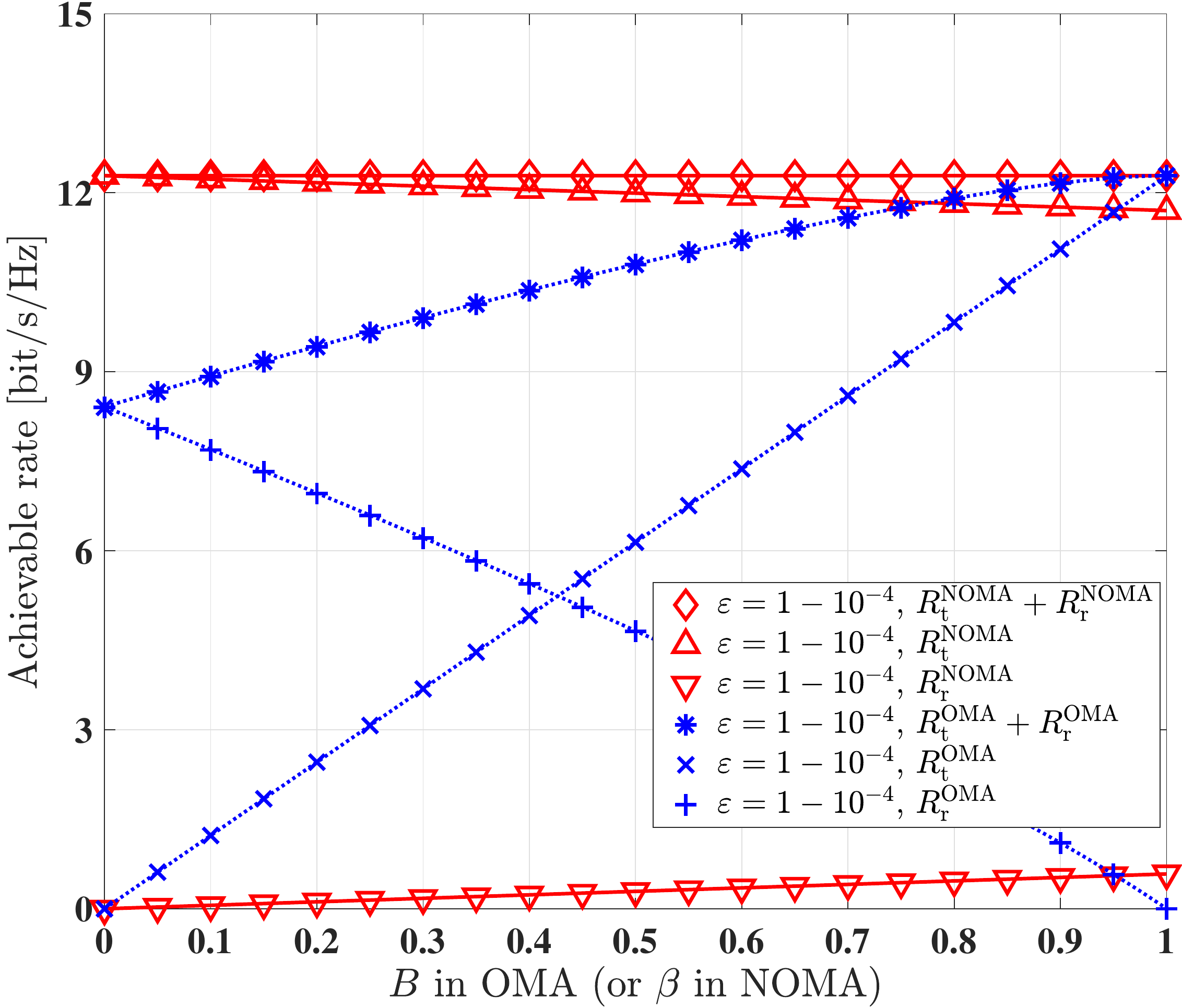}
    \end{minipage}}
    \subfloat[]{\begin{minipage}{0.33\linewidth}
        \centering
        \includegraphics[width=2.3in]{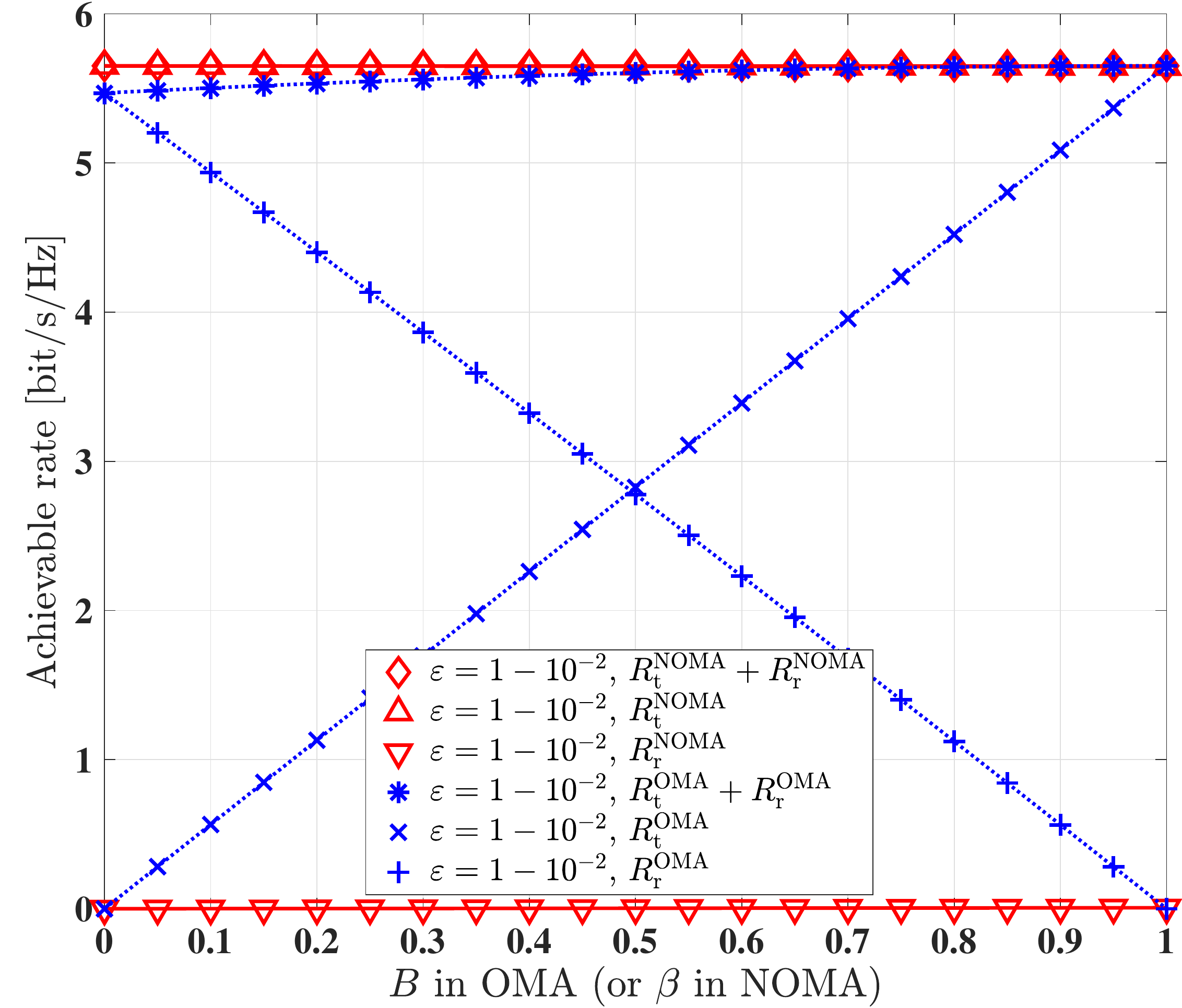}
    \end{minipage}}
    \caption{Comparison of the achievable rate versus the time/frequency fraction $B$ in the OMA scheme or the decoding order fraction $\beta$ in the NOMA system with different transceiver hardware qualities, where the received SNR of UE-T and UE-R are $\gamma_\mathrm{t}=20\mathrm{dB}$ and $\gamma_\mathrm{r}=-20\mathrm{dB}$, respectively.}\label{Fig_simu_8}
\end{figure*}

In Fig. \ref{Fig_simu_4} to \ref{Fig_simu_8} we present the ergodic sum-rate, including our simulation results based on (\ref{NOMA_7_3}) and the theoretical upper bound based on (\ref{NOMA_7_4}), for the NOMA system considered based on a perfect SIC decoding algorithm, where the CSI is inferred by the LMMSE estimator.

Fig. \ref{Fig_simu_4} shows the effect of the RIS phase noise on the ergodic sum-rate performance, when considering ideal transceiver hardware, and the RIS phase shift is optimized based on the statistical CSI according to (\ref{Beamforming_Design_2_1}) and (\ref{Beamforming_Design_2_2}). Fig. \ref{Fig_simu_4}(a) compares the ergodic sum-rate versus the average received SNR $\gamma$ in the NOMA system having different RIS phase noise powers $\sigma_\mathrm{p}^2$, when the pilot sequence length is $K=50$. Fig. \ref{Fig_simu_4}(a) shows that there is some ergodic sum-rate performance loss upon increasing the RIS phase noise power $\sigma_\mathrm{p}^2$. Furthermore, the von Mises-distributed RIS phase noise has higher performance degradation than its uniformly-distributed counterpart at the same phase noise power. Fig. \ref{Fig_simu_4}(b) compares the ergodic sum-rate versus the number of RIS elements $N$ in the NOMA system, where the pilot sequence length is $K=50$, and the average received SNR is $\gamma=0\mathrm{dB}$. It can be seen in Fig. \ref{Fig_simu_4}(b) that the sum-rate performance degradation caused by the RIS phase noise can be compensated upon deploying more RIS elements. For example, the NOMA system with the RIS phase noise power of $\sigma_\mathrm{p}^2=0.8$ following the von Mises distribution can achieve the same sum-rate performance as that having a RIS phase noise power of $\sigma_\mathrm{p}^2=0$ by an approximately twice the number of RIS elements. Fig. \ref{Fig_simu_4}(c) compares the ergodic sum-rate, as well as its asymptotic trend for $K\rightarrow\infty$, versus the pilot sequence length $K$ in the NOMA system, with the average received SNR is $\gamma=0\mathrm{dB}$. Fig. \ref{Fig_simu_4}(c) shows that the sum-rate degradation caused by channel estimation can be eliminated upon increasing the pilot sequence length. For example, the ergodic sum-rate can approach its upper bound when the pilot sequence length is approximately 30.

Fig. \ref{Fig_simu_5} compares the ergodic sum-rate versus the average received SNR $\gamma$ in the NOMA system for different transceiver hardware quality factors and for diverse RIS configurations, where the number of pilots is $K=50$ and the RIS phase noise power is $\sigma_\mathrm{p}^2=0$. It can be seen that upon doubling the number of RIS elements, 6dB performance improvement can be achieved when the RIS phase shift is optimally designed. This reduces to 3dB performance improvement, when the RIS phase shift is randomly designed. However, when the transceiver hardware quality factors obey $\varepsilon<1$ and the average SNR is $\gamma\rightarrow\infty$, the same ergodic sum-rate can be achieved in the case of optimal RIS phase shift design and random RIS phase shift design. This is seemed to be due to the fact that the achievable rate is limited by the transceiver hardware quality in the high-SNR region, which can be explained with the aid of (\ref{NOMA_7_2}).

In Fig. \ref{Fig_simu_6} to Fig. \ref{Fig_simu_8}, the achievable rate performance, including the sum-rate and rate pair fairness of the NOMA system and the OMA system is compared.

Fig. \ref{Fig_simu_6} compares the achievable ergodic rate of UE-T $R_\mathrm{t}$ and that of UE-R $R_\mathrm{r}$, as well as their sum-rate $R_\mathrm{t}+R_\mathrm{r}$, in both the NOMA and OMA system for different transceiver hardware quality factors associated with different received SNR, where the pilot sequence length is $K=2$, the transceiver hardware quality is ideal and the RIS phase noise follows a uniform distribution having the power of $\sigma_\mathrm{p}^2=0.1$. On the x-axis, the time-frequency fraction $B$ is for the OMA system and the SIC decoding order ratio $\beta$ is for the NOMA system. Specifically, in Fig. \ref{Fig_simu_6}(a), the received SNR of UE-T is $\gamma_\mathrm{t}=0\mathrm{dB}$ and that of UE-R is $\gamma_\mathrm{r}=0\mathrm{dB}$, which means that the UE-T and UE-R have the same channel gain. By contrast, in Fig. \ref{Fig_simu_6}(b), the received SNR of UE-T is $\gamma_\mathrm{t}=10\mathrm{dB}$ and that of UE-R is $\gamma_\mathrm{r}=-10\mathrm{dB}$, while in Fig. \ref{Fig_simu_6}(c), the received SNR of UE-T is $\gamma_\mathrm{t}=20\mathrm{dB}$ and that of UE-R is $\gamma_\mathrm{r}=-20\mathrm{dB}$. It can be seen in Fig. \ref{Fig_simu_6} that the achievable sum-rate of the NOMA system remains constant under different decoding order fraction $\beta$. By contrast, the optimal sum-rate of the OMA system is achieved when the time/frequency fraction $B=0.5$ in the case of identical channel gain for the UE-T and the UE-R, and when $B\approx1$ in the case of higher channel gain for the UE-T compared to the UE-R. Furthermore, Fig. \ref{Fig_simu_6}(a) shows that when $\gamma_\mathrm{t}=\gamma_\mathrm{r}$, the optimal sum-rate of the OMA system is the same as that of the NOMA. By contrast, Fig. \ref{Fig_simu_6}(b) and Fig. \ref{Fig_simu_6}(c) show that when $\gamma_\mathrm{t}>\gamma_\mathrm{r}$, the optimal sum-rate of the OMA system can be slightly higher than that of the NOMA system when the time/frequency fraction is $B\approx1$. However, in terms of the rate pair fairness, when $\gamma_\mathrm{t}=\gamma_\mathrm{r}$, the OMA system and the NOMA system have the same performance for the time/frequency fraction of $B=0.5$ in the OMA system and the decoding order fraction of $\beta=0.5$ in the NOMA system. By contrast, when $\gamma_\mathrm{t}>\gamma_\mathrm{r}$, the NOMA system has better fairness than that of the OMA system. Specifically, when $\gamma_\mathrm{t}=10\mathrm{dB}$ and $\gamma_\mathrm{r}=-10\mathrm{dB}$, the two users achieve the same rate of $R_\mathrm{t}^{\mathrm{OMA}}=R_\mathrm{r}^{\mathrm{OMA}}\approx7.2$ bit/s/Hz when $B\approx0.39$ in the OMA system, while they achieve the same rate of $R_\mathrm{t}^{\mathrm{NOMA}}=R_\mathrm{r}^{\mathrm{NOMA}}\approx8.7$ bit/s/Hz when $\beta\approx0.8$ in the NOMA system. When $\gamma_\mathrm{t}=20\mathrm{dB}$ and $\gamma_\mathrm{r}=-20\mathrm{dB}$, the two users achieve the same rate of $R_\mathrm{t}^{\mathrm{OMA}}=R_\mathrm{r}^{\mathrm{OMA}}\approx6.1$ bit/s/Hz when $B\approx0.27$ in the OMA system, while the optimal rate pair in the NOMA system is achieved when $\beta=1$ in which $R_\mathrm{t}^{\mathrm{NOMA}}\approx13.3$ bit/s/Hz and $R_\mathrm{r}^{\mathrm{NOMA}}\approx7.6$ bit/s/Hz.

Fig. \ref{Fig_simu_7} and Fig. \ref{Fig_simu_8} compare the achievable ergodic rate of UE-T $R_\mathrm{t}$ and that of UE-R $R_\mathrm{r}$, as well as their sum-rate $R_\mathrm{t}+R_\mathrm{r}$, in both the NOMA and OMA system for different transceiver hardware qualities, while considering perfect channel knowledge. The RIS phase noise follows uniform distribution having the power of $\sigma_\mathrm{p}^2=0.1$. In Fig. \ref{Fig_simu_7}, the UE-T and the UE-R have identical channel gain, i.e. $\gamma_\mathrm{t}=\gamma_\mathrm{r}=0\mathrm{dB}$, where it can be seen that the OMA system and the NOMA system can get the same performance in terms of both the achievable sum-rate and the rate pair fairness. In Fig. \ref{Fig_simu_8}, the UE-T has higher channel gain than the UE-R associated with $\gamma_\mathrm{t}=20\mathrm{dB}$ and $\gamma_\mathrm{r}=-20\mathrm{dB}$. Fig. \ref{Fig_simu_8} shows that in terms of the achievable sum-rate, the OMA system can get the same performance as the NOMA system, when the time/frequency fraction is $B\approx1$, which can be theoretically derived as $B=\frac{\gamma_\mathrm{t}}{\gamma_\mathrm{t}+\gamma_\mathrm{r}}=\frac{10000}{10001}$, when considering all transceiver hardware qualities. By contrast, the performance of the rate pair fairness depends on the transceiver hardware qualities. Specifically, when the transceiver hardware quality is ideal, i.e. $\varepsilon=1$, the NOMA system outperforms the OMA systems, since both users achieve the same rate of $R_\mathrm{t}^{\mathrm{OMA}}=R_\mathrm{r}^{\mathrm{OMA}}\approx6.5$ bit/s/Hz when $B\approx0.28$ in the OMA system, while the UE-T and the UE-R achieve the rate of $R_\mathrm{t}^{\mathrm{NOMA}}\approx13.3$ bit/s/Hz and $R_\mathrm{r}^{\mathrm{OMA}}\approx8.5$ bit/s/Hz when $\beta\approx1$ in the NOMA system. However, when the transceiver hardware quality is non-ideal, the OMA system achieves better rate pair fairness
than the NOMA system, albeit at the cost of some sum-rate performance degradation. For example, in Fig. \ref{Fig_simu_8}(b), when the transceiver hardware quality factor is $\varepsilon=1-10^{-4}$, the OMA system has better rate pair fairness than the NOMA system, since the optimal rate pair for the OMA system is $R_\mathrm{t}^{\mathrm{OMA}}=R_\mathrm{r}^{\mathrm{OMA}}\approx5.2$ bit/s/Hz when $B\approx0.43$, while that for the NOMA system is $R_\mathrm{t}^{\mathrm{NOMA}}\approx11.7$ bit/s/Hz and $R_\mathrm{r}^{\mathrm{NOMA}}\approx0.58$ bit/s/Hz when $\beta=1$. However, in this case, the sum-rate of the OMA system, i.e. $R_\mathrm{t}^{\mathrm{OMA}}+R_\mathrm{r}^{\mathrm{OMA}}\approx10.4$ bit/s/Hz, is lower than that of the NOMA system, i.e. $R_\mathrm{t}^{\mathrm{NOMA}}+R_\mathrm{r}^{\mathrm{NOMA}}\approx12.3$ bit/s/Hz.

\begin{figure}[!t]
    \centering
    \includegraphics[width=2.3in]{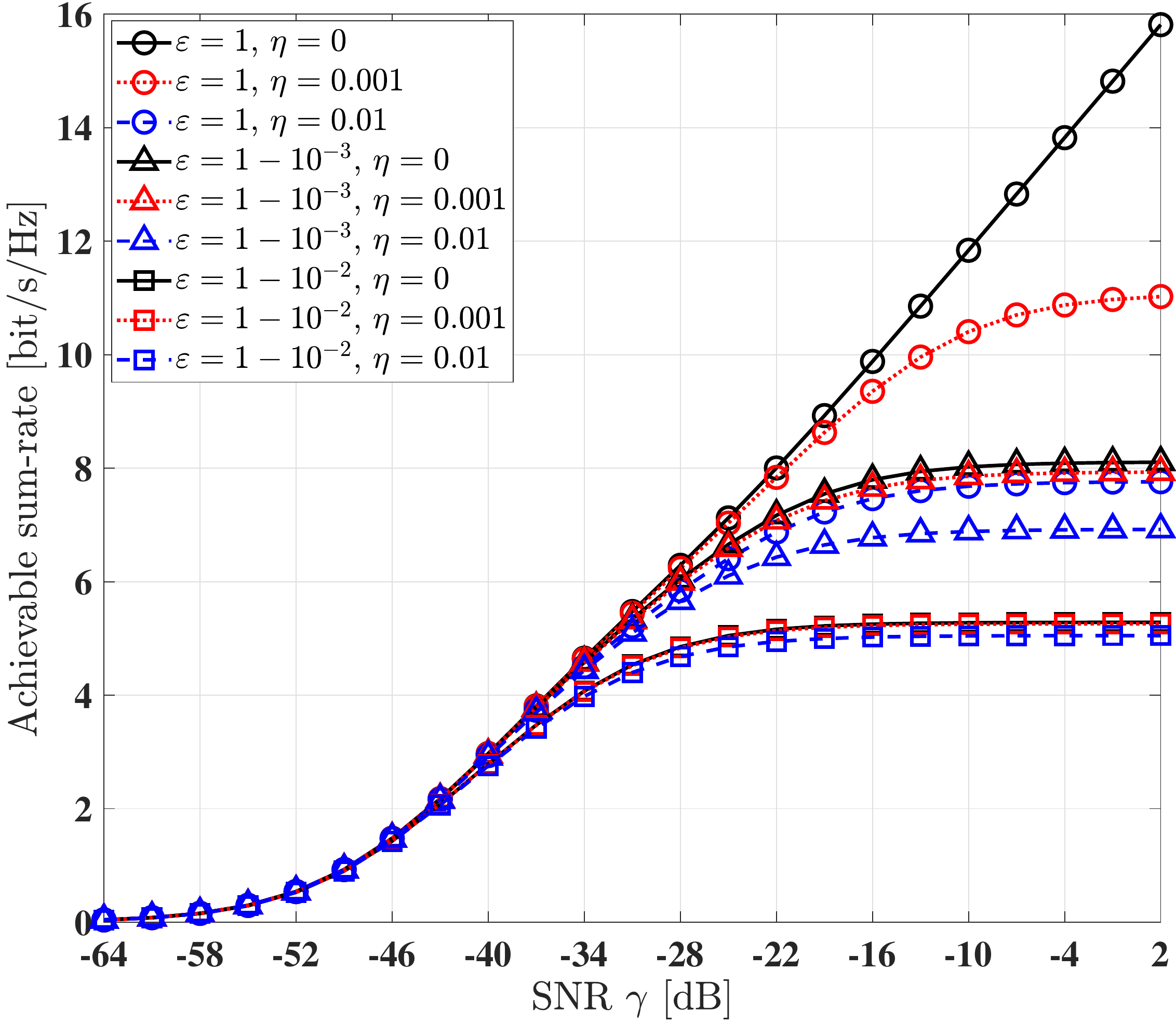}
    \caption{Simulation results of the achievable sum-rate versus the average received SNR $\gamma$ in the NOMA system based on imperfect SIC decoding algorithm with different transceiver hardware qualities $\varepsilon$ and different SIC imperfection coefficients $\eta$.}\label{Fig_simu_9}
\end{figure}

Fig. \ref{Fig_simu_9} presents the simulation results of the achievable sum-rate versus the average received SNR $\gamma$ in the NOMA system based on imperfect SIC decoding for different transceiver hardware quality $\varepsilon$ and different SIC imperfection coefficients $\eta$, where the CSI is inferred by the LMMSE estimators for a pilot sequence length of $K=2$, and the RIS phase noise power $\sigma_\mathrm{p}^2=0.1$ following the uniform distribution. It shows that when the transceiver hardware quality is low, i.e. $\varepsilon=1-10^{-2}$, the effect of the imperfect SIC on the sum-rate performance is limited. By contrast, when the transceiver hardware quality is high, i.e. $\varepsilon=1-10^{-3}$ or $\varepsilon=1$, imperfect SIC decoding has a substantial effect on the achievable sum rate. Furthermore, when the SIC imperfection coefficients obey $\eta>0$, the achievable sum-rate tends to a constant value as $\gamma\rightarrow\infty$, even though the transceiver hardware is perfect.

\section{Conclusions}\label{Conclusion}
We theoretically analyzed the ergodic sum-rate of the STAR-RIS assisted uplink NOMA scheme relying on SIC detection in the face of imperfect CSI, as well as in the presence of RIS and transceiver HWI. Our theoretical analysis and simulation results showed that the CSI accuracy can be improved by increasing the pilot sequence length, and deploying more RIS elements can compensate the achievable sum-rate performance degradation caused the RIS phase noise. However, the transceiver HWI results in performance floor at high transmit power region in terms of both the channel estimation and the achievable sum-rate. Besides, transceiver HWIs also have significant side effect on the rate pair fairness in the STAR-RIS aided uplink NOMA systems.

\appendices
\section{Proof of Theorem \ref{theorem_1}}\label{Appendix_A}
Firstly, when the RIS phase noise obeys $\tilde{\theta}^{\mathrm{t}}_{n_\mathrm{t}}\sim\mathcal{VM}(0,\varsigma_\text{p})$, upon referring to~\cite{hillen2017moments}, we have $\mathbb{E}[\text{e}^{\jmath\tilde{\theta}^{\mathrm{t}}_{n_\mathrm{t}}}]
=\frac{I_{1}(\varsigma_\text{p})}{I_{0}(\varsigma_\text{p})}$. Secondly, when the RIS phase noise obeys $\tilde{\theta}^{\mathrm{t}}_{n_\mathrm{t}}\sim\mathcal{UF}(-\iota_\mathrm{p},
\iota_\mathrm{p})$, the $m$th-order moment of $\tilde{\theta}^{\mathrm{t}}_{n_\mathrm{t}}$, namely $\mathbb{E}[(\tilde{\theta}^{\mathrm{t}}_{n_\mathrm{t}})^m]$, is equal to 0 when $m$ is odd and it is equal to $\frac{1}{m+1}\iota_\mathrm{p}^m$ when $m$ is even. Thus, we can show that $\mathbb{E}[\text{e}^{\jmath\tilde{\theta}^{\mathrm{t}}_{n_\mathrm{t}}}]
=\sum_{m=0}^{\infty}\frac{(-1)^m}{(2m)!}\mathbb{E}
[(\tilde{\theta}^{\mathrm{t}}_{n_\mathrm{t}})^{2m}]
\notag=\sum_{m=0}^{\infty}\frac{(-1)^m\iota_\mathrm{p}^{2m}}{(2m+1)!}
=\frac{\sin(\iota_\mathrm{p})}{\iota_\mathrm{p}}$.
Hence, we have $\mathbb{E}[\text{e}^{\jmath\tilde{\theta}^{\mathrm{t}}_{n_\mathrm{t}}}]=\xi$, with $\xi=\frac{I_{1}(\varsigma_\text{p})}{I_{0}(\varsigma_\text{p})}$ when $\tilde{\theta}^{\mathrm{t}}_{n_\mathrm{t}}\sim\mathcal{VM}(0,\varsigma_\text{p})$ or $\xi=\frac{\sin(\iota_\mathrm{p})}{\iota_\mathrm{p}}$ when $\tilde{\theta}^{\mathrm{t}}_{n_\mathrm{t}}\sim\mathcal{UF}(-\iota_\mathrm{p},\iota_\mathrm{p})$.

The mean and the second moment of the equivalent channel $h_\mathrm{t}(\mathbf{\Theta}^{\mathrm{t}})$ is given by
\begin{align}\label{Appendix_A_2}
    \notag\mathbb{E}[h_\mathrm{t}(\mathbf{\Theta}^{\mathrm{t}})]
    =&\sqrt{\varrho_\mathrm{t}\varrho_a}\mathbb{E}[\mathbf{a}_{\mathrm{t}}^{\text{H}}
    \mathbf{\Theta}^{\mathrm{t}}\mathbf{g}_{\text{t}}]\\
    \notag=&\sqrt{\frac{\varrho_\mathrm{t}\varrho_a\kappa_{\mathrm{t}}\kappa_a}
    {(1+\kappa_{\mathrm{t}})(1+\kappa_a)}}\sum_{n=1}^{N_\mathrm{t}}
    \mathbb{E}[\bar{a}_{\mathrm{t},n}^*\bar{g}_{\mathrm{t},n}\mathrm{e}^
    {\jmath(\bar{\theta}_{n}^{\mathrm{t}}+\tilde{\theta}_{n}^{\mathrm{t}})}]\\
    \overset{(a)}=&\sqrt{\frac{\varrho_\mathrm{t}\varrho_a\kappa_{\mathrm{t}}\kappa_a}
    {(1+\kappa_{\mathrm{t}})(1+\kappa_a)}}\xi
    \sum_{n=1}^{N_\mathrm{t}}\bar{a}_{\mathrm{t},n}^*\bar{g}_{\mathrm{t},n}\mathrm{e}^
    {\jmath\bar{\theta}_{n}^{\mathrm{t}}},
\end{align}
\begin{align}\label{Appendix_A_3}
    \notag&\mathbb{E}[|h_\mathrm{t}(\mathbf{\Theta}^{\mathrm{t}})|^2]\\
    \notag=&\varrho_\mathrm{t}\varrho_a\mathbb{E}[|\mathbf{a}_{\mathrm{t}}^{\text{H}}
    \mathbf{\Theta}^{\mathrm{t}}\mathbf{g}_{\text{t}}|^2]\\
    =&\frac{\varrho_\mathrm{t}\varrho_a(\kappa_{\mathrm{t}}\kappa_a\xi^2
    |\sum\nolimits_{n=1}^{N_\mathrm{t}}\bar{a}_{\mathrm{t},n}^*\bar{g}_{\mathrm{t},n}
    \mathrm{e}^{\jmath\bar{\theta}_{n}^{\mathrm{t}}}|^2
    +(\kappa_{\mathrm{t}}+\kappa_a+1)N_\mathrm{t})}
    {(1+\kappa_{\mathrm{t}})(1+\kappa_a)},
\end{align}
where (a) is based on $\mathbb{E}[\text{e}^{\jmath\tilde{\theta}^{\mathrm{t}}_{n_\mathrm{t}}}]=\xi$.  According to (\ref{Appendix_A_2}) and (\ref{Appendix_A_3}), we can express the variance of the equivalent channels $h_\mathrm{t}(\mathbf{\Theta}^{\mathrm{t}})$ as
\begin{align}\label{Appendix_A_4}
    \notag\mathrm{C}_{h_\mathrm{t}(\mathbf{\Theta}^{\mathrm{t}})h_\mathrm{t}
    (\mathbf{\Theta}^{\mathrm{t}})}
    =&\mathbb{E}[|h_\mathrm{t}(\mathbf{\Theta}^{\mathrm{t}})|^2]
    -|\mathbb{E}[h_\mathrm{t}(\mathbf{\Theta}^{\mathrm{t}})]|^2\\
    =&\frac{\varrho_\mathrm{t}\varrho_a(\kappa_{\mathrm{t}}+\kappa_a+1)N_\mathrm{t}}
    {(1+\kappa_{\mathrm{t}})(1+\kappa_a)}.
\end{align}

Similarly, we can formulate the mean, the second moment and the variance of the equivalent channel $h_\mathrm{r}(\mathbf{\Theta}^{\mathrm{r}})$ in (\ref{Channel_Estimation_1_2}), (\ref{Channel_Estimation_1_2_2}) and (\ref{Channel_Estimation_2_2}), respectively.

\section{Proof of Theorem \ref{theorem_2}}\label{Appendix_B}
Based on the SIC algorithm, in the $\mathrm{t}\rightarrow\mathrm{r}$ SIC order, $R_\mathrm{t}^{\text{NOMA,t}}$ and $R_\mathrm{r}^{\text{NOMA,t}}$ are given by
\begin{align}\label{Appendix_B_1}
    R_\mathrm{t}(\mathbf{\Theta}^{\mathrm{t}})
    =\log_2\Big(\frac{\rho_\mathrm{t}\varepsilon_v\varepsilon_{u_\mathrm{t}}
    |\hat{h}_\mathrm{t}(\mathbf{\Theta}^{\mathrm{t}})|^2}
    {\mathbb{E}[|y(\mathbf{\Theta})|^2]
    -\rho_\mathrm{t}\varepsilon_v\varepsilon_{u_\mathrm{t}}
    |\hat{h}_\mathrm{t}(\mathbf{\Theta}^{\mathrm{t}})|^2}\Big),
\end{align}
\begin{align}\label{Appendix_B_2}
    \notag&R_\mathrm{r}(\mathbf{\Theta}^{\mathrm{r}})\\
    =&\log_2\Big(\frac{\rho_\mathrm{r}\varepsilon_v\varepsilon_{u_\mathrm{r}}
    |\hat{h}_\mathrm{r}(\mathbf{\Theta}^{\mathrm{r}})|^2}
    {\mathbb{E}[|y(\mathbf{\Theta})|^2]
    -\rho_\mathrm{t}\varepsilon_v\varepsilon_{u_\mathrm{t}}
    |\hat{h}_\mathrm{t}(\mathbf{\Theta}^{\mathrm{t}})|^2
    -\rho_\mathrm{r}\varepsilon_v\varepsilon_{u_\mathrm{r}}
    |\hat{h}_\mathrm{r}(\mathbf{\Theta}^{\mathrm{r}})|^2}\Big),
\end{align}
where $\mathbb{E}[|y(\mathbf{\Theta})|^2]$ can be derived as
\begin{align}\label{Appendix_B_3}
    \notag\mathbb{E}[|y(\mathbf{\Theta})|^2]
    =&\rho_\mathrm{t}(|\hat{h}_\mathrm{t}(\mathbf{\Theta}^{\mathrm{t}})|^2
    +\mathrm{C}_{\check{h}_\mathrm{t}(\mathbf{\Theta}^{\mathrm{t}})
    \check{h}_\mathrm{t}(\mathbf{\Theta}^{\mathrm{t}})})\\
    &+\rho_\mathrm{r}(|\hat{h}_\mathrm{r}(\mathbf{\Theta}^{\mathrm{r}})|^2
    +\mathrm{C}_{\check{h}_\mathrm{r}(\mathbf{\Theta}^{\mathrm{r}})
    \check{h}_\mathrm{r}(\mathbf{\Theta}^{\mathrm{r}})})
    +\sigma_w^2.
\end{align}
According to (\ref{Appendix_B_1}), (\ref{Appendix_B_2}), (\ref{Appendix_B_3}) and some further manipulations, we arrive at (\ref{NOMA_2}) and (\ref{NOMA_3}). In the $\mathrm{r}\rightarrow\mathrm{t}$ SIC order, $R_\mathrm{t}^{\mathrm{r}\rightarrow\mathrm{t}}(\mathbf{\Theta})$ and $R_\mathrm{r}^{\mathrm{r}\rightarrow\mathrm{t}}(\mathbf{\Theta})$ in (\ref{NOMA_4}) and (\ref{NOMA_5}) can be similarly derived.

\section{Proof of Theorem \ref{theorem_3}}\label{Appendix_C}
As shown in (\ref{Channel_Estimation_10}) and (\ref{Channel_Estimation_13}), we have $\mathrm{C}_{\check{h}_\mathrm{t}(\mathbf{\Theta}^{\mathrm{t}})
\check{h}_\mathrm{t}(\mathbf{\Theta}^{\mathrm{t}})}=0$ and $\mathrm{C}_{\check{h}_\mathrm{r}(\mathbf{\Theta}^{\mathrm{r}})
\check{h}_\mathrm{r}(\mathbf{\Theta}^{\mathrm{r}})}=0$ when $K\rightarrow\infty$. When the UE-T and the UE-R have the same hardware quality factor, we can set $\varepsilon_{u_\mathrm{t}}=\varepsilon_{u_\mathrm{r}}=\varepsilon_u$. Thus, we have $\epsilon_\mathrm{t}(\mathbf{\Theta}^{\mathrm{t}})
=(1-\varepsilon_v\varepsilon_u)|\hat{h}_\mathrm{t}(\mathbf{\Theta}^{\mathrm{t}})|^2$ and $\epsilon_\mathrm{r}(\mathbf{\Theta}^{\mathrm{r}})
=(1-\varepsilon_v\varepsilon_u)|\hat{h}_\mathrm{r}(\mathbf{\Theta}^{\mathrm{r}})|^2$. In this case, according to (\ref{NOMA_10}), we can get the achievable sum-rate of the OMA scheme as
\begin{align}\label{Appendix_C_1}
    \notag&{R}_\text{sum}^{\text{OMA}}(\mathbf{\Theta})\\
    \notag=&B\log_2\Big(1+\frac{\rho_\mathrm{t}\varepsilon_v\varepsilon_u
    |\hat{h}_\mathrm{t}(\mathbf{\Theta}^{\mathrm{t}})|^2}
    {\rho_\mathrm{t}(1-\varepsilon_v\varepsilon_u)
    |\hat{h}_\mathrm{t}(\mathbf{\Theta}^{\mathrm{t}})|^2+B\sigma_w^2}\Big)\\
    &+(1-B)\log_2\Big(1+\frac{\rho_\mathrm{r}\varepsilon_v\varepsilon_u
    |\hat{h}_\mathrm{r}(\mathbf{\Theta}^{\mathrm{r}})|^2}
    {\rho_\mathrm{r}(1-\varepsilon_v\varepsilon_u)
    |\hat{h}_\mathrm{r}(\mathbf{\Theta}^{\mathrm{r}})|^2+(1-B)\sigma_w^2}\Big).
\end{align}
To get the maximum value of ${R}_\text{sum}^{\text{OMA}}(\mathbf{\Theta})$, we firstly derive the partial derivative of ${R}_\text{sum}^{\text{OMA}}(\mathbf{\Theta})$ with respect to $B$ as shown in (\ref{Appendix_C_2}).
\begin{figure*}[!t]
\begin{align}\label{Appendix_C_2}
    \notag\frac{\partial{R}_\text{sum}^{\text{OMA}}(\mathbf{\Theta})}{\partial B}
    =&\log_2\mathrm{e}\cdot\Big(\log_2\Big(1+\frac{\rho_\mathrm{t}\varepsilon_v\varepsilon_u
    |\hat{h}_\mathrm{t}(\mathbf{\Theta}^{\mathrm{t}})|^2}
    {\rho_\mathrm{t}(1-\varepsilon_v\varepsilon_u)
    |\hat{h}_\mathrm{t}(\mathbf{\Theta}^{\mathrm{t}})|^2+B\sigma_w^2}\Big)
    -\log_2\Big(1+\frac{\rho_\mathrm{r}\varepsilon_v\varepsilon_u
    |\hat{h}_\mathrm{r}(\mathbf{\Theta}^{\mathrm{r}})|^2}
    {\rho_\mathrm{r}(1-\varepsilon_v\varepsilon_u)
    |\hat{h}_\mathrm{r}(\mathbf{\Theta}^{\mathrm{r}})|^2
    +(1-B)\sigma_w^2}\Big)\Big)\\
    \notag&-\log_2\mathrm{e}\cdot\Big(\frac{\rho_\mathrm{t}\varepsilon_v\varepsilon_u
    |\hat{h}_\mathrm{t}(\mathbf{\Theta}^{\mathrm{t}})|^2B\sigma_w^2}
    {(\rho_\mathrm{t}|\hat{h}_\mathrm{t}(\mathbf{\Theta}^{\mathrm{t}})|^2
    +B\sigma_w^2)(\rho_\mathrm{t}(1-\varepsilon_v\varepsilon_u)
    |\hat{h}_\mathrm{t}(\mathbf{\Theta}^{\mathrm{t}})|^2
    +B\sigma_w^2)}\\
    &-\frac{\rho_\mathrm{r}\varepsilon_v\varepsilon_u
    |\hat{h}_\mathrm{r}(\mathbf{\Theta}^{\mathrm{r}})|^2(1-B)\sigma_w^2}
    {(\rho_\mathrm{r}|\hat{h}_\mathrm{r}(\mathbf{\Theta}^{\mathrm{r}})|^2
    +(1-B)\sigma_w^2)(\rho_\mathrm{r}(1-\varepsilon_v\varepsilon_u)
    |\hat{h}_\mathrm{r}(\mathbf{\Theta}^{\mathrm{r}})|^2
    +(1-B)\sigma_w^2)}\Big).
\end{align}
\hrulefill
\end{figure*}
According to (\ref{Appendix_C_2}), we can get $\frac{\partial{R}_\text{sum}^{\text{OMA}}(\mathbf{\Theta})}{\partial B}=0$ when
$B=\frac{\rho_\mathrm{t}|\hat{h}_\mathrm{t}(\mathbf{\Theta}^{\mathrm{t}})|^2}
{\rho_\mathrm{t}|\hat{h}_\mathrm{t}(\mathbf{\Theta}^{\mathrm{t}})|^2+
\rho_\mathrm{r}|\hat{h}_\mathrm{r}(\mathbf{\Theta}^{\mathrm{r}})|^2}$. This means that $R_\text{sum}^{\text{OMA}}(\mathbf{\Theta})$ can be maximized when $B=\frac{\rho_\mathrm{t}|\hat{h}_\mathrm{t}(\mathbf{\Theta}^{\mathrm{t}})|^2}
{\rho_\mathrm{t}|\hat{h}_\mathrm{t}(\mathbf{\Theta}^{\mathrm{t}})|^2+
\rho_\mathrm{r}|\hat{h}_\mathrm{r}(\mathbf{\Theta}^{\mathrm{r}})|^2}$. Then, by substituting $B=\frac{\rho_\mathrm{t}|\hat{h}_\mathrm{t}(\mathbf{\Theta}^{\mathrm{t}})|^2}
{\rho_\mathrm{t}|\hat{h}_\mathrm{t}(\mathbf{\Theta}^{\mathrm{t}})|^2+
\rho_\mathrm{r}|\hat{h}_\mathrm{r}(\mathbf{\Theta}^{\mathrm{r}})|^2}$ into (\ref{Appendix_C_1}), we can get the maximum value of $R_\text{sum}^{\text{OMA}}(\mathbf{\Theta})$, denoted by $\ddot{R}_\text{sum}^{\text{OMA}}(\mathbf{\Theta})$, as
\begin{align}\label{Appendix_C_3}
    \notag&\ddot{R}_\text{sum}^{\text{OMA}}(\mathbf{\Theta})\\
    \notag=&\log_2
    \Big(1+\\
    \notag&\frac{\rho_\mathrm{t}\varepsilon_v\varepsilon_u
    |\hat{h}_\mathrm{t}(\mathbf{\Theta}^{\mathrm{t}})|^2
    +\rho_\mathrm{r}\varepsilon_v\varepsilon_u
    |\hat{h}_\mathrm{r}(\mathbf{\Theta}^{\mathrm{r}})|^2}
    {\rho_\mathrm{t}(1-\varepsilon_v\varepsilon_u)
    |\hat{h}_\mathrm{t}(\mathbf{\Theta}^{\mathrm{t}})|^2
    +\rho_\mathrm{r}(1-\varepsilon_v\varepsilon_u)
    |\hat{h}_\mathrm{r}(\mathbf{\Theta}^{\mathrm{r}})|^2+\sigma_w^2}\Big)\\
    \notag\overset{(a)}=&\log_2
    \Big(1+\frac{\rho_\mathrm{t}\varepsilon_v\varepsilon_u
    |\hat{h}_\mathrm{t}(\mathbf{\Theta}^{\mathrm{t}})|^2
    +\rho_\mathrm{r}\varepsilon_v\varepsilon_u
    |\hat{h}_\mathrm{r}(\mathbf{\Theta}^{\mathrm{r}})|^2}{\mathcal{E}}\Big)\\
    =&R_\text{sum}^{\text{NOMA}}(\mathbf{\Theta}),
\end{align}
where (a) is true for $K\rightarrow\infty$ and $\varepsilon_u=\varepsilon_{u_\mathrm{t}}=\varepsilon_{u_\mathrm{r}}$. Based on (\ref{Appendix_C_3}), we can have $R_\text{sum}^{\text{NOMA}}(\mathbf{\Theta}){\geq}R_\text{sum}^{\text{OMA}}(\mathbf{\Theta})$, where the equality is established only when the time/frequency fraction in the OMA scheme is $B=\frac{\rho_\mathrm{t}|\hat{h}_\mathrm{t}(\mathbf{\Theta}^{\mathrm{t}})|^2}
{\rho_\mathrm{t}|\hat{h}_\mathrm{t}(\mathbf{\Theta}^{\mathrm{t}})|^2+
\rho_\mathrm{r}|\hat{h}_\mathrm{r}(\mathbf{\Theta}^{\mathrm{r}})|^2}$.

\section{Proof of Theorem \ref{theorem_4}}\label{Appendix_D}
According to (\ref{Channel_Estimation_1_1}), (\ref{Channel_Estimation_1_2_1}), (\ref{Channel_Estimation_9}), (\ref{Channel_Estimation_10}), (\ref{Channel_Estimation_12}), (\ref{Channel_Estimation_13}) and (\ref{NOMA_7_4}), when $K\rightarrow\infty$, we can get
\begin{align}\label{Appendix_D_1}
    \notag&\ddot{R}_\text{sum,erg}^{\text{NOMA}}(\mathbf{\Theta})\\
    =&\log_2\Big(1+\frac{\sum_{\mathrm{i}\in\{\mathrm{t},\mathrm{r}\}}
    \rho_\mathrm{i}\varepsilon_v\varepsilon_{u_\mathrm{i}}
    \mathbb{E}[|h_\mathrm{i}(\mathbf{\Theta}^{\mathrm{i}})|^2]}
    {\sum\limits_{\mathrm{i}\in\{\mathrm{t},\mathrm{r}\}}
    \rho_\mathrm{i}(1-\varepsilon_v\varepsilon_{u_\mathrm{i}})
    \mathbb{E}[|h_\mathrm{i}(\mathbf{\Theta}^{\mathrm{i}})|^2]+\sigma_w^2}\Big).
\end{align}
According to (\ref{Channel_Estimation_1_2_1}) and (\ref{Appendix_D_1}), we can formulate the derivative of $\ddot{R}_\text{sum,erg}^{\text{NOMA}}
(\mathbf{\Theta})$ with respect to
$\left|\sum_{n=1}^{N_\mathrm{t}}\bar{a}_{\mathrm{t},n}^*\bar{g}_{\mathrm{t},n}
\mathrm{e}^{\jmath\bar{\theta}_{n}^{\mathrm{t}}}\right|^2$ as
\begin{align}\label{Appendix_D_2}
    \notag\frac{\partial\ddot{R}_\text{sum,erg}^{\text{NOMA}}(\mathbf{\Theta})}
    {\partial\Big|\sum\limits_{n=1}^{N_\mathrm{t}}\bar{a}_{\mathrm{t},n}^*\bar{g}_{\mathrm{t},n}
    \mathrm{e}^{\jmath\bar{\theta}_{n}^{\mathrm{t}}}\Big|^2}
    =&\frac{\log_2\mathrm{e}\cdot\rho_\mathrm{t}\varepsilon_v\varepsilon_{u_\mathrm{t}}
    \frac{\partial\mathbb{E}[|h_\mathrm{t}(\mathbf{\Theta}^{\mathrm{t}})|^2]}
    {\partial\big|\sum_{n=1}^{N_\mathrm{t}}\bar{a}_{\mathrm{t},n}^*\bar{g}_{\mathrm{t},n}
    \mathrm{e}^{\jmath\bar{\theta}_{n}^{\mathrm{t}}}\big|^2}}
    {\sum\limits_{\mathrm{i}\in\{\mathrm{t},\mathrm{r}\}}
    \rho_\mathrm{i}(1-\varepsilon_v\varepsilon_{u_\mathrm{i}})
    \mathbb{E}[|h_\mathrm{i}(\mathbf{\Theta}^{\mathrm{i}})|^2]
    +\sigma_w^2}\\
    >&0.
\end{align}
Similarly, we can formulate the derivative of $\ddot{R}_\text{sum,erg}^{\text{NOMA}}
(\mathbf{\Theta})$ with respect to
$\big|\sum_{n=1}^{N_\mathrm{r}}\bar{a}_{\mathrm{r},n}^*\bar{g}_{\mathrm{r},n}
\mathrm{e}^{\jmath\bar{\theta}_{n}^{\mathrm{r}}}\big|^2$ as
\begin{align}\label{Appendix_D_3}
    \notag\frac{\partial\ddot{R}_\text{sum,erg}^{\text{NOMA}}(\mathbf{\Theta})}
    {\partial\Big|\sum\limits_{n=1}^{N_\mathrm{r}}\bar{a}_{\mathrm{r},n}^*\bar{g}_{\mathrm{r},n}
    \mathrm{e}^{\jmath\bar{\theta}_{n}^{\mathrm{r}}}\Big|^2}
    =&\frac{\log_2\mathrm{e}\cdot\rho_\mathrm{r}\varepsilon_v\varepsilon_{u_\mathrm{r}}
    \frac{\partial\mathbb{E}[|h_\mathrm{r}(\mathbf{\Theta}^{\mathrm{r}})|^2]}
    {\partial\big|\sum_{n=1}^{N_\mathrm{r}}\bar{a}_{\mathrm{r},n}^*\bar{g}_{\mathrm{r},n}
    \mathrm{e}^{\jmath\bar{\theta}_{n}^{\mathrm{r}}}\big|^2}}
    {\sum\limits_{\mathrm{i}\in\{\mathrm{t},\mathrm{r}\}}
    \rho_\mathrm{i}(1-\varepsilon_v\varepsilon_{u_\mathrm{i}})
    \mathbb{E}[|h_\mathrm{i}(\mathbf{\Theta}^{\mathrm{i}})|^2]
    +\sigma_w^2}\\
    >&0.
\end{align}

Therefore, the optimization problem (P1) is equivalent to the following two sub-problems.
\begin{align}\label{Appendix_D_4}
    \notag\text{(P2a)}\quad &\max_{\bar{\theta}_{1}^{\mathrm{t}},\bar{\theta}_{2}^{\mathrm{t}},
    \cdots,\bar{\theta}_{N_\mathrm{t}}^{\mathrm{t}}}\ \mathbb{E}[|h_\mathrm{t}(\mathbf{\Theta}^{\mathrm{t}})|^2]\\
    \text{s.t.}&\quad \bar{\theta}^\mathrm{t}_{n_\mathrm{t}}\in[0,2\pi),\ {n}_\mathrm{t}=1,2,\cdots,N_\mathrm{t}.
\end{align}
\begin{align}\label{Appendix_D_5}
    \notag\text{(P2b)}\quad &\max_{\bar{\theta}_{1}^{\mathrm{r}},\bar{\theta}_{2}^{\mathrm{r}},
    \cdots,\bar{\theta}_{N_\mathrm{r}}^{\mathrm{r}}}\ \mathbb{E}[|h_\mathrm{r}(\mathbf{\Theta}^{\mathrm{r}})|^2]\\
    \text{s.t.}&\quad \bar{\theta}^\mathrm{r}_{n_\mathrm{r}}\in[0,2\pi),\ n_\mathrm{r}=1,2,\cdots,N_\mathrm{r}.
\end{align}
Based on (\ref{Channel_Model_3_1}), (\ref{Channel_Model_4_1}), (\ref{Appendix_D_4}) and (\ref{Appendix_D_5}), the optimal STAR-RIS phase shift is designed according to (\ref{Beamforming_Design_2_1}) and (\ref{Beamforming_Design_2_2}).

\section{Proof of Theorem \ref{theorem_5}}\label{Appendix_E}
According to (\ref{Imperfect_SIC_5}) and (\ref{Imperfect_SIC_6}), when $\rho_\mathrm{t}\varepsilon_{u_\mathrm{t}}|\hat{h}_\mathrm{t}(\mathbf{\Theta}^{\mathrm{t}})|^2
>\rho_\mathrm{r}\varepsilon_{u_\mathrm{r}}|\hat{h}_\mathrm{r}(\mathbf{\Theta}^{\mathrm{r}})|^2$, we can get $\frac{\partial{R}_\text{sum}^{\text{NOMA}}(\mathbf{\Theta})}{\partial\beta}$ as
\begin{align}\label{Appendix_E_1}
    \notag&\frac{\partial{R}_\text{sum}^{\text{NOMA}}(\mathbf{\Theta})}{\partial\beta}\\
    \notag=&\log_2\Big(1+\frac{\eta\rho_\mathrm{r}\varepsilon_v\varepsilon_{u_\mathrm{r}}
    |\hat{h}_\mathrm{r}(\mathbf{\Theta}^{\mathrm{r}})|^2}{\mathcal{E}}\Big)\\
    \notag&+\log_2\Big(1+\frac{\eta\rho_\mathrm{t}\varepsilon_v\varepsilon_{u_\mathrm{t}}
    |\hat{h}_\mathrm{t}(\mathbf{\Theta}^{\mathrm{t}})|^2}
    {\rho_\mathrm{r}\varepsilon_v\varepsilon_{u_\mathrm{r}}
    |\hat{h}_\mathrm{r}(\mathbf{\Theta}^{\mathrm{r}})|^2+\mathcal{E}}\Big)\\
    \notag&-\log_2\Big(1+\frac{\eta\rho_\mathrm{r}\varepsilon_v\varepsilon_{u_\mathrm{r}}
    |\hat{h}_\mathrm{r}(\mathbf{\Theta}^{\mathrm{r}})|^2}
    {\rho_\mathrm{t}\varepsilon_v\varepsilon_{u_\mathrm{t}}
    |\hat{h}_\mathrm{t}(\mathbf{\Theta}^{\mathrm{t}})|^2+\mathcal{E}}\Big)\\
    \notag&-\log_2\Big(1+\frac{\eta\rho_\mathrm{t}\varepsilon_v\varepsilon_{u_\mathrm{t}}
    |\hat{h}_\mathrm{t}(\mathbf{\Theta}^{\mathrm{t}})|^2}{\mathcal{E}}\Big)\\
    \notag=&\log_2\Big(\eta\big(\rho_\mathrm{t}\varepsilon_v\varepsilon_{u_\mathrm{t}}
    |\hat{h}_\mathrm{t}(\mathbf{\Theta}^{\mathrm{t}})|^2\big)
    \big(\rho_\mathrm{r}\varepsilon_v\varepsilon_{u_\mathrm{r}}
    |\hat{h}_\mathrm{r}(\mathbf{\Theta}^{\mathrm{r}})|^2\big)\\
    \notag&+\mathcal{E}\big(\rho_\mathrm{t}\varepsilon_v\varepsilon_{u_\mathrm{t}}
    |\hat{h}_\mathrm{t}(\mathbf{\Theta}^{\mathrm{t}})|^2
    +\eta\rho_\mathrm{r}\varepsilon_v\varepsilon_{u_\mathrm{r}}
    |\hat{h}_\mathrm{r}(\mathbf{\Theta}^{\mathrm{r}})|^2+\mathcal{E}\big)\Big)\\
    \notag&-\log_2\Big(\eta\big(\rho_\mathrm{t}\varepsilon_v\varepsilon_{u_\mathrm{t}}
    |\hat{h}_\mathrm{t}(\mathbf{\Theta}^{\mathrm{t}})|^2\big)
    \big(\rho_\mathrm{r}\varepsilon_v\varepsilon_{u_\mathrm{r}}
    |\hat{h}_\mathrm{r}(\mathbf{\Theta}^{\mathrm{r}})|^2\big)\\
    \notag&+\mathcal{E}\big(\rho_\mathrm{r}\varepsilon_v\varepsilon_{u_\mathrm{r}}
    |\hat{h}_\mathrm{r}(\mathbf{\Theta}^{\mathrm{r}})|^2
    +\eta\rho_\mathrm{t}\varepsilon_v\varepsilon_{u_\mathrm{t}}
    |\hat{h}_\mathrm{t}(\mathbf{\Theta}^{\mathrm{t}})|^2+\mathcal{E}\big)\Big)\\
    \notag&+\log_2\Big(\rho_\mathrm{r}\varepsilon_v\varepsilon_{u_\mathrm{r}}
    |\hat{h}_\mathrm{r}(\mathbf{\Theta}^{\mathrm{r}})|^2
    +\eta\rho_\mathrm{t}\varepsilon_v\varepsilon_{u_\mathrm{t}}
    |\hat{h}_\mathrm{t}(\mathbf{\Theta}^{\mathrm{t}})|^2+\mathcal{E}\Big)\\
    \notag&-\log_2\Big(\rho_\mathrm{t}\varepsilon_v\varepsilon_{u_\mathrm{t}}
    |\hat{h}_\mathrm{t}(\mathbf{\Theta}^{\mathrm{t}})|^2
    +\eta\rho_\mathrm{r}\varepsilon_v\varepsilon_{u_\mathrm{r}}
    |\hat{h}_\mathrm{r}(\mathbf{\Theta}^{\mathrm{r}})|^2+\mathcal{E}\Big)\\
    \overset{(a)}<&0.
\end{align}
where (a) is based on $\rho_\mathrm{t}\varepsilon_v\varepsilon_{u_\mathrm{t}}
|\hat{h}_\mathrm{t}(\mathbf{\Theta}^{\mathrm{t}})|^2
>\rho_\mathrm{r}\varepsilon_v\varepsilon_{u_\mathrm{r}}
|\hat{h}_\mathrm{r}(\mathbf{\Theta}^{\mathrm{r}})|^2$. Upon exploiting (\ref{Appendix_E_1}), we can show that $\frac{\partial{R}_\text{sum}^{\text{NOMA}}(\mathbf{\Theta})}{\partial\beta}<0$. Thus, the achievable sum-rate is maximized, when $\beta=0$. Similarly, when $\rho_\mathrm{t}\varepsilon_{u_\mathrm{t}}|\hat{h}_\mathrm{t}(\mathbf{\Theta}^{\mathrm{t}})|^2
<\rho_\mathrm{r}\varepsilon_{u_\mathrm{r}}|\hat{h}_\mathrm{r}(\mathbf{\Theta}^{\mathrm{r}})|^2$, we can get $\frac{\partial{R}_\text{sum}^{\text{NOMA}}(\mathbf{\Theta})}{\partial\beta}>0$, which means that the achievable sum-rate is maximized when $\beta=1$.

\bibliographystyle{IEEEtran}
\bibliography{IEEEabrv,TAMS}
\end{document}